\documentclass[useAMS]{mn2e}
\usepackage{graphicx}
\usepackage{supertabular}
\usepackage{aalongtable}

\newcommand{\kms}{km\,s$^{-1}$}

\title[MDI of Ap Stars I. Observations]{Stokes $IQUV$ Magnetic Doppler Imaging of Ap stars\\ I. ESPaDOnS and NARVAL Observations\thanks{
Based on observations obtained at the Canada-France-Hawaii Telescope (CFHT) which is operated by the National Research Council of Canada, 
the Institut National des Sciences de l'Univers of the Centre National de la Recherche Scientifique of France,  and the University of Hawaii.  Also based on observations obtained at the Bernard Lyot Telescope (TBL, Pic du Midi, France) of the Midi-Pyr\'en\'ees Observatory, 
which is operated by the Institut National des Sciences de l'Univers of the Centre National de la Recherche Scientifique of France. }
}

\author[J.Silvester et al.]
{J. Silvester$^{1,2}$, G.A. Wade$^{2}$, O. Kochukhov$^{3}$, S. Bagnulo$^{4}$, C.P. Folsom$^{4}$, 
\newauthor {D. Hanes$^{1}$}\\
$^{1}$Department of Physics, Engineering Physics \& Astronomy, Queen's University, Kingston, Ontario, Canada, K7L 3N6\\
$^{2}$Department of Physics, Royal Military College of Canada, P.O. Box 17000, Station `Forces', Kingston, Ontario, Canada, K7K 7B4\\
$^{3}$Department of Astronomy and Space Physics, Uppsala University, 751 20, Uppsala, Sweden.\\
$^{4}$Armagh Observatory, College Hill, Armagh BT61 9DG, Northern Ireland, U.K \\ 
}  

\begin{document}

\date{Accepted . Received }

\pagerange{\pageref{firstpage}--\pageref{lastpage}} \pubyear{2012}

\maketitle

\label{firstpage}

\begin{abstract}
In this paper we describe and evaluate new spectral line polarisation observations obtained with the goal of mapping the surfaces of magnetic Ap stars in great detail. One hundred complete or partial Stokes $IQUV$ sequences, corresponding to 297 individual polarised spectra, have been obtained for 7 bright Ap stars using the ESPaDOnS and NARVAL high resolution spectropolarimeters. The targets span a range of mass from approximately 1.8 to 3.4~$M_\odot$, a range of rotation period from $2.56$ to $6.80$ days, and a range of maximum longitudinal magnetic field strength from 0.3 to over 4 kG. For 3 of the 7 stars, we have obtained dense phase coverage sampling the entire rotational cycle. These datasets are suitable for immediate magnetic and chemical abundance surface mapping using Magnetic Doppler Imaging (MDI). For the 4 remaining stars, partial phase coverage has been obtained, and additional observations will be required in order to map the surfaces of these stars. The median signal-to-noise ratio of the reduced observations is over 700 per 1.8~\kms\ pixel.  Spectra of all stars show Stokes $V$ Zeeman signatures in essentially all individual lines, and most stars show clear Stokes $QU$ signatures in many individual spectral lines. The observations provide a vastly improved data set compared to previous generations of observations in terms of signal-to-noise ratio, resolving power and measurement uncertainties.  Measurement of the longitudinal magnetic field demonstrates that the data are internally consistent within computed uncertainties typically at the 50 to 100$\sigma$ level.  Data are also shown to be in excellent agreement with published observations and in qualitative agreement with the predictions of published surface structure models. In addition to providing the foundation for the next generation of surface maps of Ap stars, this study establishes the performance and stability of the ESPaDOnS and NARVAl high-resolution spectropolarimeters during the period 2006-2010.

\end{abstract}


\section{Introduction}
The classification Ap identifies a (main sequence) A or B type star which displays peculiar chemical abundances, usually combined with an observable magnetic field. Although other classes of chemically peculiar stars exist (e.g. Am stars, Hg-Mn stars, He-weak stars), these stars have been demonstrated to lack strong, organised magnetic fields at their surfaces (e.g. Shorlin et al. 2002, Wade et al 2006, Makaganiuk et al. 2011).  Ap stars appear to be the only class of middle main-sequence stars for which, in all cases, an observable magnetic field is present (Auri\`erie et al. 2007).  

Since their discovery by Babcock in 1947, the magnetic fields of Ap stars have been established through observation to have important global dipole components with polar strengths ranging from hundreds to tens of thousands of gauss. The symmetry axis of the dipole component is almost always significantly tilted relative the stellar rotation axis. In addition, Ap stars generally spin much more slowly than non-peculiar stars of similar masses (Stepie\'n 2000), and as they spin they exhibit line profile variations attributed to rotational modulation of patchy, non-axisymmetric lateral and vertical distributions of chemical abundance in their photospheres. The distributions of abundance vary significantly from element to element: some are distributed relatively uniformly, while others show strong contrast; some are distributed in relatively simple patterns, while others show complex distributions. While it is generally accepted that the fundamental mechanism responsible for the chemical peculiarities is microscopic chemical diffusion (as described by Michaud 1970), the origin of chemical patchiness, and the relationship to the magnetic field, is poorly understood.

The earliest studies of the magnetic field geometries of Ap stars interpreted the rotational variations of their longitudinal magnetic fields in the context of Stibbs' Oblique Rotator Model assuming a simple magnetic dipole field (e.g. Babcock 1947, 1951; Stibbs 1950). However, with the acquisition of increasingly sophisticated diagnostic data (the mean surface field (or mean field modulus), and high-resolution line profiles), it became clear that the large-scale field topologies exhibited important departures from the simple dipolar model (e.g. Preston \& Sturch 1967, Preston 1969, 1970, Landstreet 1970, 1988, 1989).

Leroy and collaborators (Landolfi et al. 1993, Leroy et al. 1993, 1994, 1995ab, 1996, Bagnulo et al. 1995, Wade et al. 1996) systematically studied Ap stars using broadband linear polarisation measurements and models, constraining the transverse component of the magnetic field. Importantly, they found that differences between the observed linear polarisation variations and those predicted by the simple dipole model could not be fully explained by abundance inhomogeneities alone (Leroy et al. 1996). With this work, they established a modified dipolar model with a trend toward an outward expansion of the field lines over some parts of the magnetic equator, and showed the potential of linear polarisation for diagnosing small-scale structure of the magnetic fields of Ap stars. Thus, the observations and modeling undertaken during the latter half of the 20th century allowed progress from a simple view of the magnetic fields of Ap stars to a relatively sophisticated picture in which fields were known to show both global-scale and local-scale departures from a simple dipole. 

Leroy et al. (1996) commented that high-resolution spectropolarimetry represented the next step in furthering the study of the magnetic field geometry of Ap stars.  Four years later, Wade et al. (2000a) published the first compendium of phase-resolved high-resolution spectropolarimetric observations of Ap stars in both circular and linear polarisation. Using the MuSiCoS spectropolarimeter, $R=35,000$ Stokes $IQUV$ spectra with a median S/N of 300 (per 4.6~\kms\ pixel) were obtained for 14 Ap stars. While the quality of the spectra was sufficiently good to show the shape and phase variation of all Stokes parameters in mean Least-Squares Deconvolved (LSD) line profiles, measurement in individual spectral lines was restricted to a few particularly strong lines, principally those of Fe~{\sc ii} multiplet 42. Nevertheless, the Stokes profiles of 53 Cam were used by Bagnulo et al. (2001) and Kochukhov et al. (2004) to evaluate published magnetic models developed based on less sophisticated data.  Those authors found that models based on so-called "magnetic observables" (e.g Bagnulo 2000) led to derivation of surface magnetic field characteristics that were not consistent with the detailed Stokes profiles, and that both circular and especially linear polarisation profiles were required for realistic reconstruction of the field.  

Following these conclusions, Kochukhov et al. (2002) for $\alpha^2$~CVn and Lueftinger et al. (2010) for HD 24712 employed the new Magnetic Doppler Imaging technique (MDI), described by Piskunov \& Kochukhov (2002) and Kochukhov \& Piskunov (2002), to construct high resolution maps of the surface vector magnetic field maps using Stokes $IV$ observations and by preferring a global low-order multipolar field structure. Maps using Linear polarisation profiles (Stokes $Q$ and $U$) made by Kochukhov et al. (2004) and Kochukhov \& Wade (2010) for the Ap stars 53 Cam and $\alpha^2$~CVn. These maps were distinguished from earlier models in that they were computed directly from the observed polarised line profiles, making no {\em a priori} assumptions regarding the large-scale or small-scale topology of the field. The MDI surface magnetic field maps of both stars revealed that their magnetic topologies depart significantly from low-order multipoles. In particular, both studies concluded that while the global topology of the magnetic field was reasonably smooth, the strength of the field was quite patchy, indicating complex structure on relatively small scales. Simultaneous mapping of the distributions of the surface chemical abundances of several elements was also performed, allowing a comparison between the local field properties and local photospheric chemistry. 

It is important to note that the observational material used in the MDI studies of 53 Cam and $\alpha^2$ CVn represented the best data sets obtained from several years of MuSiCoS observations. In those spectra the uniquely valuable Stokes $Q$ and $U$ Zeeman signatures were only clearly detectable in 3 strong lines, with a significance (i.e. amplitude divided by error bar) of 5 or less.  
The relatively low signal-to-noise ratio and resolving power achievable with the MuSiCoS instrument led to some ambiguity in the field reconstruction, and limited the useful sample of stars to those with bright apparent magnitudes, strong fields and sharp lines. Because MDI exploits the indirect resolution of the stellar disc due to stellar rotation, this means that those stars best suited to reconstruction (relatively rapidly-rotating stars, with consequentially weaker Stokes profiles) were inaccessible to MuSiCoS. As a result, only an extremely limited range of stellar properties (rotation, mass, temperature, magnetic field, etc.) which may influence the phenomena of interest could be studied using the MuSiCoS data. 
  
To address outstanding questions surrounding the detailed magnetic structure of Ap stars and the effect of the magnetic field on atmospheric chemical transport processes, we have acquired new higher-resolution and signal-to-noise Stokes $IQUV$ spectra of a small sample of well-studied magnetic Ap stars using the new generation of high-resolution spectropolarimeters. In this paper we describe the observations obtained. We demonstrate the stability of the instrumentation during the 5 years of observation by evaluating the internal and external agreement of the data. We illustrate the quality of the observed Stokes profiles, comparing with MuSiCoS results and demonstrating that they represent a qualitative step forward in our ability to diagnose the magnetic structure of Ap stars.

\section{Targets}

\begin{table*}
\caption{Stars discussed in this paper, along with ancillary data: HD designation, other name, $V$ magnitude, spectral type, projected rotational velocity, rotational period, radius, inclination, the number of longitudinal field measurements obtained, the median longitudinal field uncertainty obtained, the number of net linear polarisation measurements obtained. Projected rotational velocities ($v\sin i$) have been measured in this study, as discussed in the analysis section.  The periods were obtained from various sources as described in the Sect 6. Masses are those reported by Kochukhov and Bagnulo (2006) and the average longitudinal magnetic fields are taken from the catalogue of stellar effective magnetic fields (Bychkov et al. 2003), using values obtained from Least-Squares Deconvolved profiles if available. Temperatures are those reported by Kochukhov and Bagnulo (2006) and stellar radii as reported in Leone et al. (2000) for HD 32633, Pasinetti Fracassini et al. (2001) for HD 40312, Kochukhov and Wade (2010) for HD 112413 and Wade (1997) for the remaining stars.  Inclinations are taken from  Stepie\'n (1989) for HD 32633, Rice, Holmgren and Bohlender (2004) for HD 40312 and Kochukhov and Wade (2010) for HD 112413 and from Leroy et al. (1996) for the remaining stars.} 
\begin{tabular}{llllllllllllll}
\hline
\noalign{\smallskip}    
HD & Name & $V$  & Spectral & $T_{\rm eff}$ & $M$ &$v\sin i$& $P_{\rm rot}$ & $R$   & $i$ &  \#            & \#       &Median &$\langle B_\ell\rangle^{rms}$   \\
   &      &      & type     &  (K) & ($M_\odot$)  &  (\kms) & (days)   &   ($R_\odot$) &  ($\degr$)  & $B_\ell$ &  net $Q/U$  &  $\sigma_B$ (G)& (kG) \\
\noalign{\smallskip}
\hline
4778   &        & 6.1  &   A0p   & 10000 & 2.29 & $ 36\pm 2$    &  $2 \fd 56171$  & 1.83 & 70 & 7 &   13  & 20  & 1.02  \\
32633  &        & 7.1 &  B9p & 12800  & 3.10 &$19 \pm 2 $  & $6\fd 43000$& 2.4 & 76  &21 & 40  &30  & 2.85\\ 
40312  & $\theta$ Aur & 2.6 &  A0p & 10200& 3.41 &$ 53 \pm 1 $ & $3\fd 61860$& 2.2 & 54 &7 & 14 &14  & 0.21 \\
62140  & 49 Cam & 6.5 & F0p & 7700 & 1.77 &$24\pm 2$ & $ 4\fd 28679$& 1.95 & 90 &19  & 38   &21    & 1.11   \\
71866  &        & 6.8 & A0p & 8800  & 2.24& $15 \pm 2$ & $6\fd 80022$& 2.0 & 110 &14    &       28  &27 & 1.54 \\
112413 & $\alpha^2$ CVn & 2.9 &  A0p & 11500  & 2.98 &$17 \pm 1$  & $5\fd 46939$& 2.49 & 120 &27 & 48 &55 & 0.53 \\
118022 & 78 Vir & 4.9 & A1p & 9100 & 2.16  & $13 \pm 1$ & $3\fd 7220$& 2.1 & 25 &5 & 10 &12  & 0.63        \\
\hline
\hline
\end{tabular}
\label{stars}
\end{table*}

\begin{figure}
\begin{center}
   \includegraphics[width=0.50\textwidth]{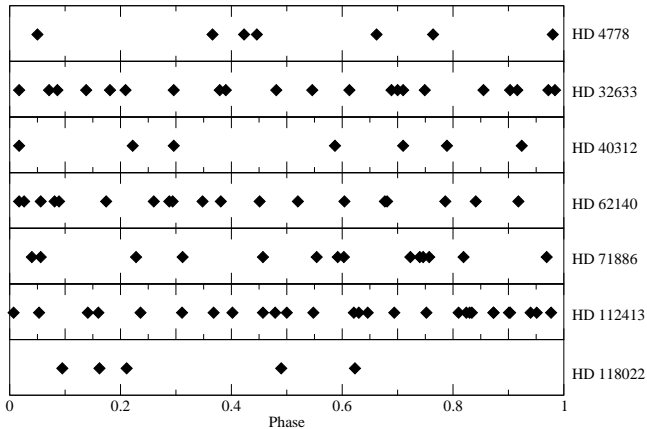}
   \caption{Phase coverage obtained for the target stars. Symbols denote rotational phases at which observations were obtained.}
\label{phase}
\end{center}
\end{figure}

Targets selected for this study are bright Ap stars demonstrated to exhibit strong Stokes profiles by Wade et al. (2000a). We attempted to select targets spanning a large range of stellar physical properties as well as field strengths and geometries. Target stars were generally required to have the following characteristics to be suitable for this study:

\begin{itemize}
\item 
Well-determined rotation periods -  This study requires phase-resolved timeseries observations. All targets must therefore have well-determined, unambiguous rotation periods so that each phase is observed correctly.
\item 
Suitable projected rotational velocity - The projected rotational velocity must be neither too rapid (rapid rotators $>$ 50 ~\kms\  typically have shallow Stokes profiles which are challenging to detect and interpret) nor too slow ($<$ 2 ~\kms\ whilst MDI can be applied to stars with such small rotational velocities, the advantage of Doppler tomography is lost).
\item
Strong magnetic fields and variability - Because MDI relies on both the shape and variation of line profiles to determine the geographic location of magnetic and chemical features, target stars should display strong and variable Stokes profiles.
\end{itemize}

Ultimately, 7 targets were selected for monitoring. The target list is shown in Table \ref{stars}.

\section{Observations obtained with ESPaDonS and NARVAL}

Both the ESPaDOnS and NARVAL instruments consist of a bench mounted cross-dispersed \'echelle spectrograph, fibre-fed from a Cassegrain-mounted polarimeter unit.   These instruments are designed to overcome the limitations encountered with MuSiCoS, with improved resolution ($R = \lambda / \Delta \lambda \simeq 65000$), sensitivity (approximately 15-20\% throughput) and wavelength coverage from 369-1048 nm (with gaps at 922.4 to 923.4 nm, 960.8 to 963.6 nm and 1002.6 to 1007.4 nm). The ESPaDOnS (\'Echelle SpectroPolarimetric Device for the Observation of Stars) spectropolarimeter is installed at the 3.6 m Canada-France-Hawaii Telescope (CFHT), and the NARVAL spectropolarimeter is installed at the 2 m Bernard Lyot telescope at Pic du Midi observatory. ESPaDOnS and NARVAL are essentially identical instruments, with NARVAL constructed based on the experience of ESPaDOnS.

The polarimetric unit (at the Cassegrain focus) allows two orthogonal states of a given polarisation (circular or linear) to be are recorded throughout the entire spectral range. The polarimeters of ESPaDOnS and NARVAL are of a similar design to the "Sempol" visitor polarimeter (Donati et al. 2003) used on the AAT (Anglo-Australian Telescope). The polarimeter is split into two parts, the upper part is for guiding and calibration, containing the guiding camera (a commercial FLI MaxCam series CCD camera is used in NARVAL and as of semester 2011A a QSI Imaging CCD camera is used in ESPaDOnS), an atmospheric dispersion corrector and calibration wheel.  The lower part contains the Fresnel rhomb retarders which are used to perform the polarimetric analysis. These optics are in 4 different drawers which contain the following (in order): a half-wave rhomb (consisting of a pair of quarter-wave rhombs), a quarter-wave rhomb, a second half-wave rhomb, and finally a Fabry-Perot wheel on one side and a Wollaston/wedge plate slide on the other side. To enhance the achromaticity of the phase delay, and to keep it at $90\pm 0.5\degr$ across the whole optical domain, the rhombs are coated with a thin layer of MgF$_{\rm 2}$, yielding a performance significantly better than achromatic crystalline plates which can vary by about $20\degr$ from the required quarter-wave retardance. Another advantage of Fresnel rhombs is that they do not produce detectable spectral ripples due to the fact that the fringe spacing is on the order of the pixel size. The half-wave rhombs can rotate about the optical axis by a specified angle.  The final component, the Wollaston prism, consisting of two orthogonal calcite prisms that are cemented together, acts as a polarising beamsplitter. The two beams of light from the beamsplitter are transmitted by some 30 m of optical fibre to the spectrograph. ESPaDOnS includes a fiber agitator, which shakes the optical fiber to remove modal noise that may be present.

The ESPaDOnS spectrograph unit consists of a double set of high-reflectance collimators cut from a single 680 mm parabolic mirror, with a focal length of 1500 mm. The grating is a 79 gr/mm monolithic grating with a dimensions of 200 by 400 mm. The camera lens is a fully dioptric $f/2$ 388 mm focal length lens, with a 210 mm free diameter (7 lenses in 4 blocks, one block being a 220 mm quadruplet). For cross-dispersing, a high dispersion prism made of a train of 2 identical PBL25Y prisms with an $35\degr$ apex and 220 mm cross section is used. Up until the 2011A semester the detector used with ESPaDOnS was a grade 1 EEV detector with 2K by 4.5K 0.0135 mm square pixels (known as EEV1 at CFHT). This was replaced in 2011A with a new E2V detector (named Olapa at CFHT). 

The spectrograph unit is mounted on an optical bench, which is housed in a thermal enclosure found in the inner Coud\'e room at CFHT.

This configuration yields full spectral coverage of the optical domain (from grating order 61 centred at 372 nm to grating order 22 centred at 1029 nm) in a single exposure. In polarimetric mode this should, in principle, achieve a resolution in excess of 65,000, but due to a charge transfer efficiency issue with the EEV1 CCD detector, the true resolution varies from approximately 68,000 in the blue to 61,000 in the red. The peak throughput of the spectrograph (with CCD detector) is about 40\% to 45\%, bringing the total instrument peak efficiency to a level of about 15\% to 20\%.

The configuration of NARVAL is much the same with the exception of an 2.8 arcsec aperture pupil (versus 1.6 arcsec for ESPaDOnS). NARVAL does however benefit from better spectrograph thermal stability (by approximately a factor of 10) than ESPaDOnS, due to the use of a double thermal layer enclosure.

\begin{figure}
\begin{center}
 \includegraphics[width=0.54\textwidth]{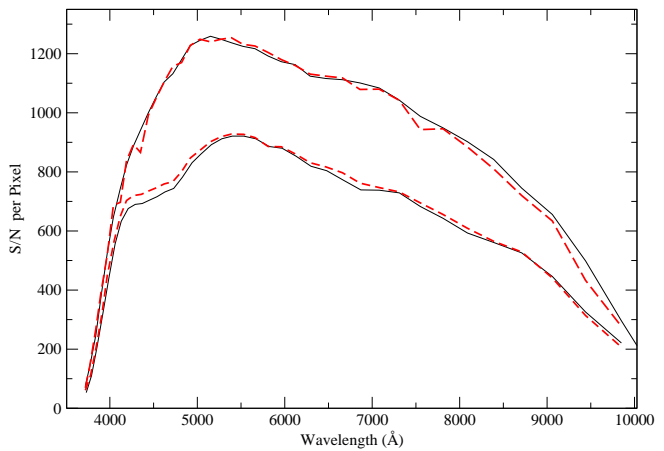}
\caption{Signal-to-noise ratio (per 1.8~\kms\ spectral pixel) as a function of wavelength for two observations of $\alpha^{2}$ CVn obtained with ESPaDOnS (top curves, 02 March 07 (airmass = 1.104) and 13 Jan 09 (airmass = 1.058) both total exposures of 120s) and two obtained with NARVAL (lower curves, 20 Dec 06 (airmass = 1.250) and 11 Jan 09 (airmass = 1.005) both total exposures of 240s). Solid vs dashed lines indicate observations taken on respective nights.}
\label{snvary}
\end{center}
\end{figure}

\begin{figure}
\begin{center}
 \includegraphics[width=0.54\textwidth]{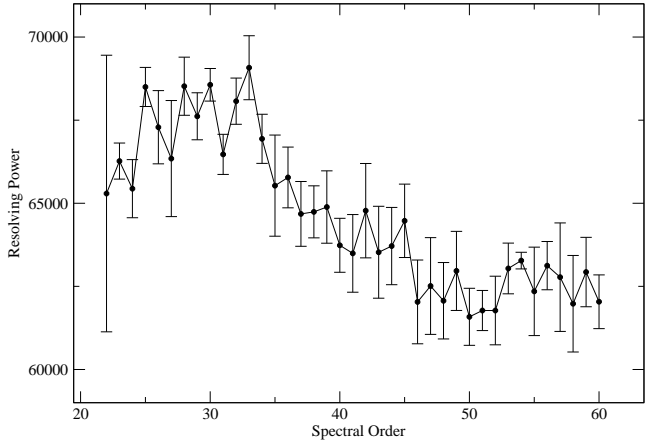}
\caption{The mean spectral resolving power as a function of spectral order for both ESPaDOnS and NARVAL for a selection of observing nights (3 observations from each instrument spanning 12 months). Error bars represent the standard deviation of the resolving power of the various observations within each spectral order. }
\label{reschange}
\end{center}
\end{figure}

The resolving power and signal-to-noise ratio of both instruments vary with wavelength in a predictable manner, as illustrated in Fig. \ref{snvary}, which shows the signal-to-noise ratio as a function of wavelength for four observations of $\alpha^{2}$ CVn (two acquired with ESPaDOnS, and two with NARVAL).  The variation in the spectral resolving power $R$ as a function of spectral order for a selection of observing nights is shown in Fig. \ref{reschange}. This figure illustrates that the characteristics of the instruments do not vary significantly on the timescales relevant for this project (nights to years). Variability in atmospheric conditions (e.g. seeing) may well be a dominant contributor to the scatter in resolving power.

In terms of observations the following steps occur; the two output beams from the Wollaston prism, which have been analysed into the two components of circular polarisation (using the quarter-wave rhomb) or linear polarisation (using the half-wave rhomb, are then carried by the pair of optical fibres to the spectrograph where two interleaved spectra are formed. The $I$ component of the stellar Stokes vector is formed by adding the two corresponding spectra, while the $V$, $Q$ or $U$ polarisation component is obtained from the ratio method as discussed by Bagnulo et al. (2009). To minimise systematic errors due to small misalignments, differences in transmission, effects of seeing, etc., one complete observation of a star consists of four successive sub-exposures; for the second and third, the waveplate settings are changed so as to exchange the positions of the two analysed spectra on the CCD. This same procedure is used for all polarisation spectra, Stokes $V$, $Q$ and $U$.

Calibration of the instrument uses a combination of thorium/argon and thorium/neon lamps, with the lamp calibrations taken at the beginning and at the end of each night for a primary wavelength calibration.  Telluric lines are then later used to perform a second wavelength calibration during the reduction process using libre-ESpRIT. Filters are used to minimise the blooming on the chip at the red end of the spectrum. Two tungsten lamps are utilised for the flat fields frames, with one low intensity lamp being used with a red filter and the other lamp being higher intensity and used with a blue filter.

The reduction of observations is carried out at the observatories using the dedicated software package Libre-ESpRIT, which yields both the $I$ spectrum and the $V$ circular polarisation spectrum and/or $QU$ linear polarisation spectra of each star observed. It is important to note that libre-ESpRIT automatically finds and removes continuum polarisation.  In this work each reduced spectrum is normalised order-by-order using a FORTRAN code specifically optimised to fit the continuum of these stars.

A diagnostic null spectrum called the $N$ spectrum, computed by combining the four sub-exposures in such a way as to have real polarisation cancel out, is also calculated by Libre-ESpRIT (again, see Bagnulo et al. 2009 for the definition of the null spectrum). The $N$ spectrum tests the system for spurious polarisation signals. The result of the reduction and normalisation procedure are continuum-normalised, one-dimensional spectra in the form of wavelength, $I/I_{\rm c}$, $V/I_{\rm c}$, $Q/I_{\rm c}$, $U/I_{\rm c}$, two independent normalised $N$ spectra $N_{\rm 1}/I_{\rm c}$ and $N_{\rm 2}/I_{\rm c}$, and an error bar (computed by propagating photon uncertainties through the reduction procedure), tabulated pixel-by-pixel.

Weak signatures are visible in LSD $N$ profiles associated with some Stokes $Q$ and $U$ spectra. Experiments conducted in order to diagnose and mitigate polarisation crosstalk (see Barrick et al., in preparation) indicate that these signatures are related to this phenomenon. However, data acquired during early NARVAL runs in 2006 exhibit substantially stronger signatures. These signatures are reported by TBL staff to result from problems with the coatings on the $\lambda/2$ rhombs in use in 2006. The rhombs were subsequently replaced, and no similar strong signatures are detected in later observations.

Based on our comparisons of Stokes $Q$ and $U$ profiles obtained with the two instruments during the course of our observing program, it is clear that these $N$ signatures are not diagnostic of any detectable contamination of the associated $Q$ and $U$ profiles.

In this study we have obtained 100 polarimetric sequences corresponding to 297 individual polarised spectra. The observations were initially obtained in classical observing mode, and later in service mode at both telescopes. In total 48 sequences were obtained with ESPaDOnS and 52 with NARVAL.  The log of observations is reported in Table \ref{oblog}, and the achieved phase coverage of the stellar targets is illustrated in Fig. \ref{phase}. The resulting reduced spectra are illustrated in Fig. \ref{atlas} which demonstrates the quality of the data,  with Stokes $VQU$ signatures seen in many individual lines.

\section{Crosstalk}
During the commissioning of ESPaDOnS in 2004 it was found that the instrument exhibited crosstalk between linear polarisation and circular polarisation (and {\em vice versa}). Due to the relatively strong circular polarisation in the spectral lines of our targets, contamination of the significantly weaker linearly polarised profiles is potentially a serious problem.

\begin{figure}
\begin{center}
   \includegraphics[width=0.54\textwidth]{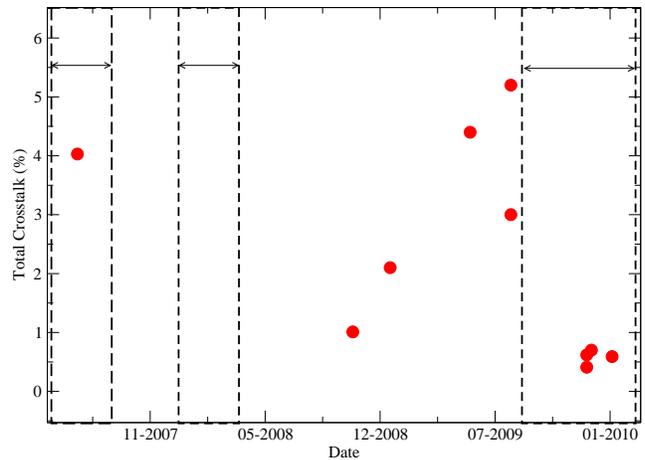}
  \caption{The total crosstalk of Stokes $V$ into Stokes $Q$ and $U$ of ESPaDonS as function of time, as reported by Barrick et al. (in preparation). The dashed boxes indicate periods during which observations were obtained as part of this investigation. }
\label{xtalk}
\end{center}
\end{figure}

The contributing optical component to this crosstalk identified initially was the collimating triplet within the polarimeter unit. This component was replaced in June 2006, resulting in a reduction of the crosstalk to the 5\% level (from an initial level of 10-15 \%). After replacing the triplet lens again in October 2008, the crosstalk appeared to have been reduced to 2-3\%, but still exhibited strong temporal changes. It was discovered in October 2009 that the atmospheric dispersion corrector (ADC) was an important and previously unrecognised source of crosstalk within the polarimetric unit.  The ADC was replaced in the fall of 2009, and since that time the crosstalk has been small and stable, with crosstalk from Stokes $V$ into Stokes $U$ at the 0.5\% level, and no measurable crosstalk from Stokes $V$ into Stokes $Q$ (i.e. below $\sim 0.1-0.2$\%). The procedure and results of the investigations of the crosstalk, are reported by Barrick et al. (in preparation).  The evolution of the ESPaDOnS crosstalk with time, based on the results of Barrick et al., is illustrated in Fig. \ref{xtalk}.

Less extensive crosstalk monitoring has been performed with NARVAL. Results from September 2009 indicated that with the ADC in place, the crosstalk was 3.1\% from Stokes $V$ to Stokes $Q$, and below 0.1-0.2\% from Stokes $V$ to Stokes $U$.  Without the ADC in place, the crosstalk was reduced to 2.1\% in Stokes $Q$, but increased to 1\% in Stokes $U$ (illustrating that the NARVAL ADC also introduces crosstalk, but again that it is not the sole source). It is important to note that at the current time, no tests of how this crosstalk changes with time have been made with NARVAL. 

Although the crosstalk is now below 1\% in ESPaDOnS and probably around 2\% in NARVAL,  observations for this project have been acquired over the last 4 years and during some of this time the crosstalk may have been higher (and almost certainly was for ESPaDOnS). It is important to understand how crosstalk could affect the Stokes $Q$ and $U$ signatures measured in spectral lines. A series of test was therefore performed to evaluate the importance of this crosstalk. 

The first test was to examine the Stokes profiles of the Fe~{\sc ii} $\lambda 6149$ line. This line has a relatively large Land\'e factor, but in the linear regime of the Zeeman effect it is predicted be purely circularly polarised as a consequence of the $\sigma$ and $\pi$ components of this line having the same strength and identical splitting.  In this case we would interpret any signal in Stokes $Q$ and $U$ as due to crosstalk from Stokes $V$. We have examined $\lambda 6149$ in our spectra, as well as spectra discussed by Barrick et al. (in preparation). In no case do we observe any significant signal in this line.

\begin{figure}
\begin{center}
   \includegraphics[width=0.55\textwidth]{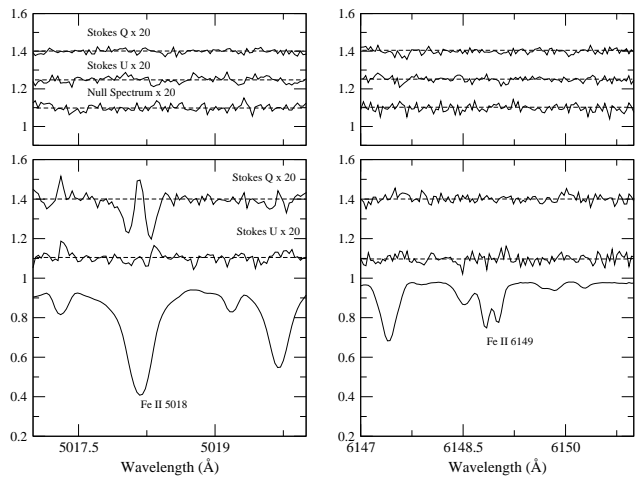}
        \caption{A comparison between the measured crosstalk in Stokes $Q$ and $U$ and the null spectrum for observations of $\gamma$ Equ taken with ESPaDOnS (16th July 2009) (top row),  with the actual observations shown below (lower frame). In the strong, magnetically sensitive line Fe II 5018 line, the crosstalk in Stokes $Q$ is below the noise,  where as the crosstalk in Stokes $U$ is slightly above the noise. }
\label{residual-xtalk}
\end{center}
\end{figure}

Then, we carefully examined one of the crosstalk diagnostic observations of the cool magnetic star $\gamma$ Equ obtained during CFHT engineering time. As described by Barrick et al. (in preparation), on-sky crosstalk diagnosis employs observations of slowly-rotating magnetic Ap stars observed in all Stokes parameters at two positions of the CFHT's Cassegrain bonnette to unambiguously measure the crosstalk into Stokes $Q$ and Stokes $U$. The diagnostic observation of $\gamma$ Equ (obtained in July 2009, with a peak S/N of over 1000) yielded relatively high crosstalk levels of $2.3\%$ in Stokes $Q$ and $5.1\%$ in Stokes $U$.  By comparing the observations with and without crosstalk, it was found that the crosstalk contributions to Stokes $Q$ is within the noise, whilst the crosstalk contributions to Stokes $U$ appears to be slightly above the noise, as illustrated by Fig. \ref{residual-xtalk}.   It is important to note that the removal of the crosstalk using techniques as described in Barrick et al. (in preparation) requires the acquisition of a second series of observations following a rotation of the Cassegrain bonnette,  a procedure which is not standard practice during regular observations.

$\gamma$ Equ is an extremely sharp lined star,  which means the crosstalk contribution will have a greater effect than it would in the broader lined stars studied in this paper.  Nevertheless,  the potential influence of the crosstalk on MDI will be investigated and discussed in a future paper  (Silvester et al. in preparation).

These comparisons demonstrate that the contribution of crosstalk to the Stokes profiles is below, or at most just above, the level of the noise of the best-quality observations of Ap stars acquired with these instruments.

\begin{figure*}
\begin{center}
   \includegraphics[width=0.95\textwidth]{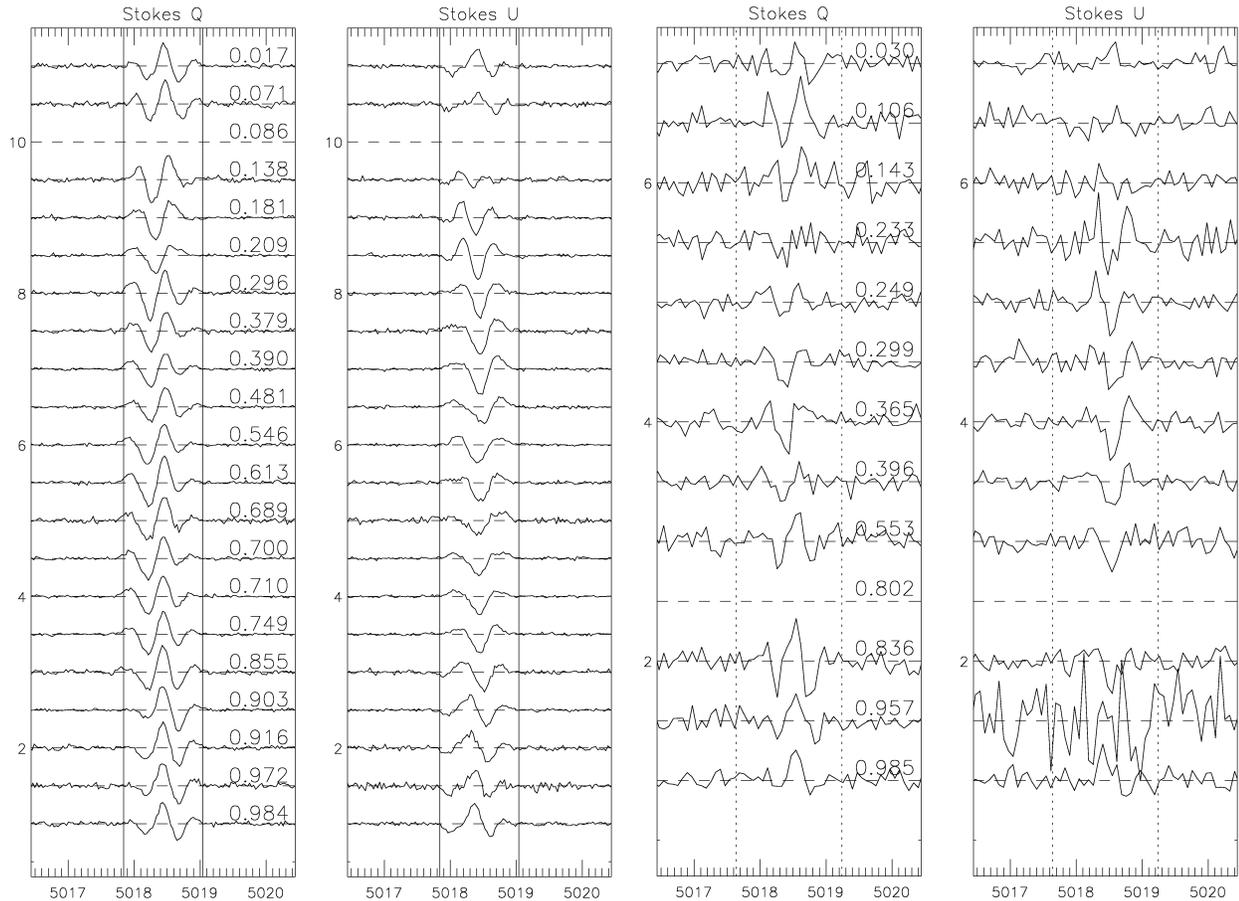}
  \caption{Comparison between Stokes $Q$ and $U$ profiles of HD 32633 in the 5018   \AA\ line for ESPaDOnS/NARVAL on the left and MuSiCoS on the right. Rotation phase for each observation are indicated.}
\label{emcomp}
\end{center}
\end{figure*}

\section{Longitudinal magnetic field and net linear polarisation} 
To examine the self-consistency of the new polarisation spectra, as well as to evaluate their consistency with published magnetic data for our targets, we have measured the mean longitudinal magnetic field and net linear polarization using Least-Squares Deconvolution (LSD). LSD  (Donati et al. 1997 and Kochukhov et al. 2010) is a multiline analysis method that produces mean Stokes $I$ and $V$ profiles using essentially all metallic  lines in the stellar spectrum. It assumes that the observed spectrum can be represented as the convolution of a single mean line profile with an underlying spectrum of unbroadened metal and helium lines of appropriate wavelength, depth and Land\'e factor (the ``line mask" computed using spectrum synthesis; e.g. Wade et al. 2000a). 

The LSD model allows the computation of single, average Stokes $I$ and $V$ line profiles, usually characterised by a signal-to-noise ratio significantly higher than that of individual spectral lines, scaling roughly as the square root of the number of lines used. 

\begin{figure*}
\begin{center}
   \includegraphics[width=0.80\textwidth]{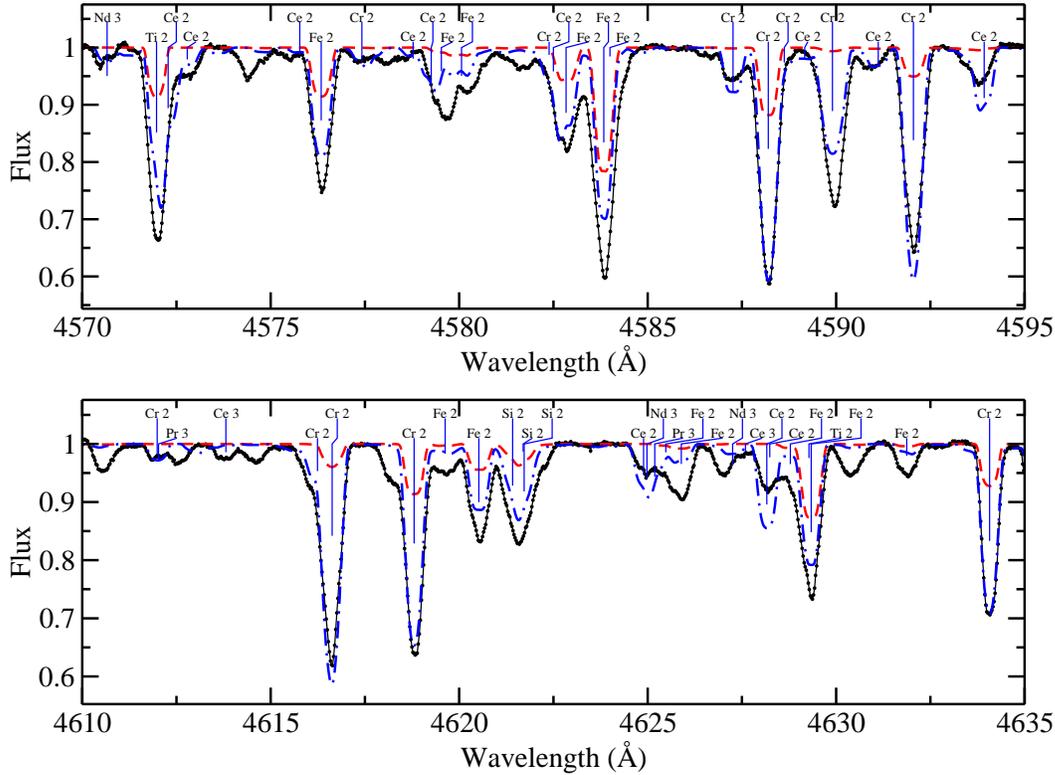}
 \caption{Example abundance fit for HD 32633, with a synthetic spectrum calculation of solar abundance show with a dashed line, and the spectrum computed with the inferred abundances used in the LSD mask shown with a dot-dashed line.}
\label{abunfit}
\end{center}
\end{figure*}

The projected rotational velocity $ v\sin i$ of each star was derived by fitting a selection of spectral lines (typically in the 4500   \AA\ region) with a synthesised spectrum created using the Synth3 spectrum synthesis code (Kochukhov, 2007).  Each line in the selection had $ v\sin i$ determined by using a $\chi^2$ fitting function which is included as part of the Binmag spectral visualization tool. The resulting mean (and standard deviation) values one reported in Table \ref{stars}. Line masks for this study were compiled using Vienna Atomic Line Database (VALD, Kupka et al. 1999) ``extract stellar" requests, with effective temperatures (adopted based on the literature) of each target. It should be noted when creating each line mask, a constant log g of 4.0 was used for all stars. Input chemical abundances were determined by a rough abundance analysis, based typically on lines around the 4500   \AA\ region.  The abundance analysis was performed by using the Synth3 spectrum synthesis code (Kochukhov 2007) with ATLAS9 solar abundance model atmospheres used to create synthetic spectra that were compared directly to the observations. The input abundances were then adjusted in the synthetic spectrum until a reasonable agreement was found with the observed spectrum.  A comparison between of two synthetic spectra, one computed using solar abundances and one using the final determined abundances, is shown in Fig \ref{abunfit} for HD 32633.  The final abundances were then adopted in the line mask creation. By doing so we ensure that any uncertainties caused by the mask when performing the LSD analysis are minimal. 

As discussed by Shorlin et al. (2002), the LSD S/N is only weakly sensitive to the line-depth cutoff employed to populate the mask. Following their results, we have chosen to employ a line-depth cutoff equal to 10\% of the continuum. Imposing such a cutoff has a related advantage: because weaker lines are less likely to have published experimental Land\'e factors (and to generally have more poorly-determined atomic data), we pre-filter our line list to (statistically) exclude those lines with the poorest data. In addition Balmer lines  are removed from the mask and the mask is restricted to the ESPaDOnS/NARVAL spectral range. Application of LSD to the data yields a set of mean profiles (Stokes $I$, Stokes $V$ and N) for each reduced spectrum.

The mean longitudinal magnetic field $B_\ell$ was computed from each LSD profile set. This quantity was evaluated by computing the first-order moment of the Stokes $V$ profile in velocity according to:

\begin{equation}
\displaystyle 
B_\ell =  -2.14\times 10^{11}  \frac {\int (v-v_O) V(v) dv} {\lambda g c \int [1-I(v)] dv}
\end{equation}

\noindent  (Mathys 1989, Donati et al. (1997), Wade et al. 2000a) where $v_O$ is the centre-of-gravity of the Stokes V profile, $g$ is the integrated mean Land\'e factor and $\lambda$  is the weighted mean wavelength of all the lines included in the mask. LSD profiles were locally re-normalised to a continuum level of 1.0 before evaluation of Eq. (1). Uncertainties associated with $B_\ell$ were computed by propagating the formal uncertainties of each LSD spectral pixel through Eq. (1).  LSD profiles are extracted for each star uniformly weighted to the same land\'e factor, line depth and wavelength. 

To determine the net linear polarisation (see e.g. Wade et al 2000a) the LSD Stokes $Q$ or $U$ profile was integrated to compute the normalized equivalent width of the line polarization using:

\begin{equation}
\displaystyle 
\frac{Q}{I} = \frac {\int Q(v) dv} {\int[1-I(v)] dv} .
\end{equation}

It was found that both net linear polarisation and longitudinal field measurements were sensitive to the integration range used to calculate Eqs. (1) and (2).  Integration ranges were carefully chosen to include the entire line profile, while avoiding including excess continuum outside of the profile (which contributes only noise). This was accomplished initially by selecting limits based on the apparent extent of the wings of the Stokes $I$ profile. We then evaluated visually if any significant polarised flux was located outside of the limits, adjusting the integration limits as necessary.

As was the case with Stokes $V$, LSD Stokes $I$ profiles associated with  Stokes $Q$ and $U$ profiles were locally re-normalised to a continuum level of 1.0 before evaluation of Eq. (2).

Longitudinal field measurements are phased according to the ephemerides given in Table. 1. We verified that all adopted periods were sufficiently precise that no significant relative phase uncertainties exist. A harmonic curve was fitted by least-squares to the phased longitudinal field data.   
The degree of the harmonic function which yielded the lowest reduced  $\chi^{2}$, while still providing a significant and systematic improvement to the fit, was chose as the ''best`` fit. The results of the fit are shown in Table \ref{curvefit}. It should be noted that whilst the fits to the longitudinal field variations help compare different data sets, quantify the dispersion of the data and the verify the accuracy of the error bars, their parameters and degree do not necessarily have any easily-interpretable physical meaning.

{\tiny
\begin{table}
\begin{tabular}{lllrr}
\hline
 & \multicolumn{2}{c}{Stokes $V$ Fit}  &  Null $N$ \\
  Star    & Degree & $\chi^{2}$  & $\chi^{2} $  \\
   \hline
   HD 4778     & 2 & 1.41 & 2.68 \\
HD 32633   & 3 & 1.95 & 5.42\\
HD 40312   & 2 & 1.96  &  0.51  \\
HD 62140   & 3 & 1.29  &  4.78 \\
HD 71866   & 4 & 1.02 & 3.70 \\
HD 112413 & 3 & 0.83  & 1.06 \\
HD 118022 & 1 & 1.92 &  1.12 \\
\hline\hline
  \end{tabular}
\caption{Reduced $\chi^{2}$s of harmonic fits to the longitudinal magnetic field phase curves determined from NARVAL and ESPaDOnS data}
\label{curvefit}
\end{table}
}

\begin{figure*}
\begin{center}
    \includegraphics[width=0.95\textwidth]{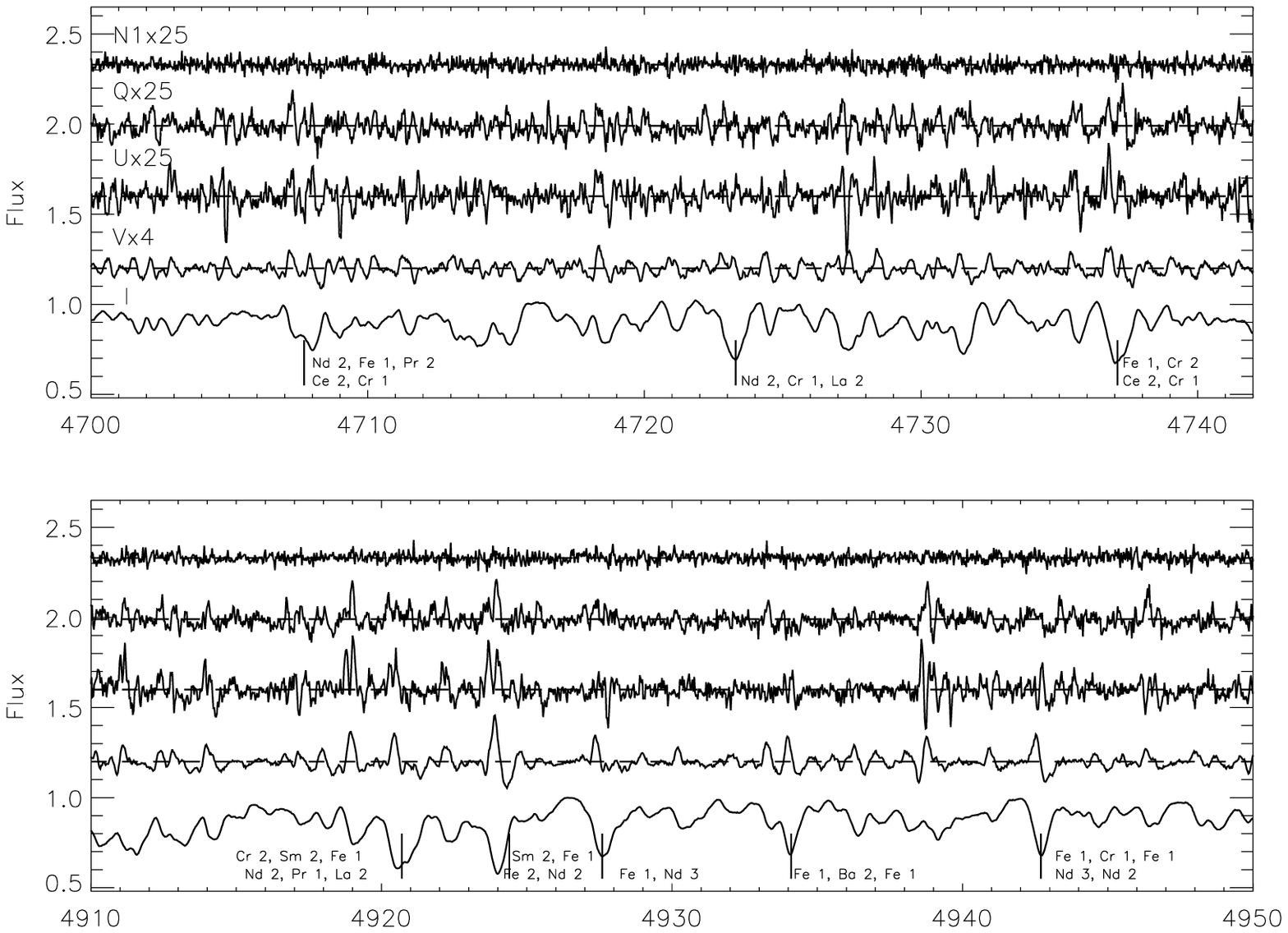}
    \includegraphics[width=0.95\textwidth]{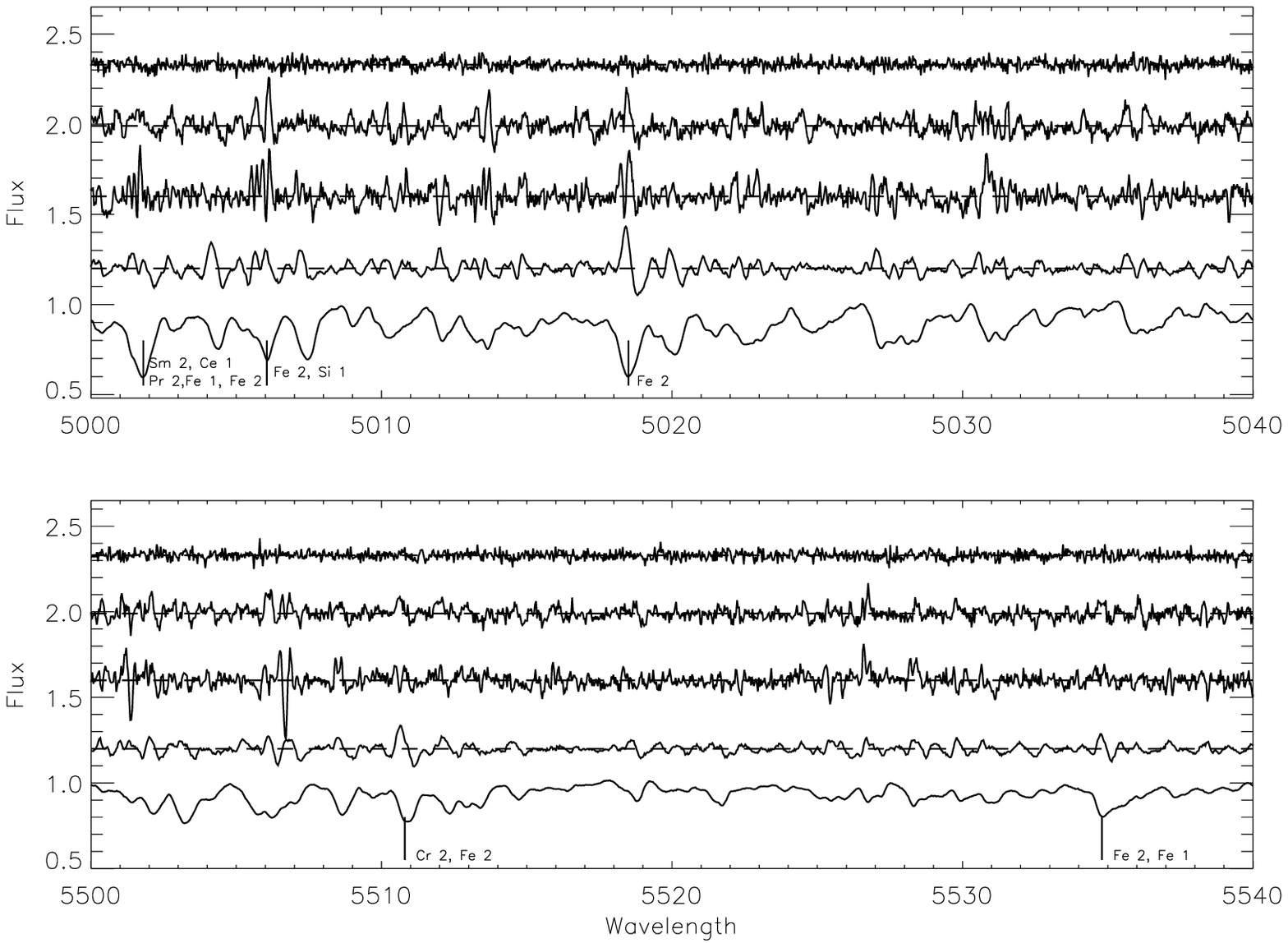}
 \caption{Selected regions of the Stokes $IVQU$ and the N spectra of HD 62140, with strongly contributing atomic lines labeled for reference}
\label{atlas}
\end{center}
\end{figure*}

To allow a consistent comparison between the MuSiCoS and ESPaDOnS/NARVAL data, the normalisation of the ESPaDOnS/NARVAL data had to be performed carefully.  Care also had to be taken with the integration limits when determining both the longitudinal field and net-linear polarization.  

When we compared the new longitudinal field measurements with those computed from MuSiCoS spectra, immediate agreement was found between the new measurements and those of Wade et al. (2000b),  except in the case of HD 71866 and $\alpha^2$ CVn which showed a slight discrepancy.  By re-computing the Stokes $V$ MuSiCoS LSD profiles for these stars with the new masks, the observations were brought into agreement.  It is possible that if both HD 71866 and $\alpha^2$ CVn have sufficiently peculiar abundances,  a general Ap star line mask  (with arbitrarily enhanced and depleted abundances of particular elements, as used in the original analysis of Wade et al. 2000b) may result in an underestimation of the longitudinal field.  Because each mask used in this study is ``tailor-made'' to the chemical abundances of the respective star, we are less susceptible to this problem and thus measure slightly increased longitudinal field values.

An example of the comparison between the new measurements, and those corresponding to the the two sets of MuSiCoS LSD profiles, are shown in Section 6.6 (Fig. 27) for $\alpha^2$ CVn. This figure demonstrates that overall the agreement between the new measurements and the re-computed MuSiCoS measurements are acceptable. Using the new masks for the other stars makes little difference to the longitudinal field measurements,  so the original measurements were kept for comparison.  In addition good agreement can be seen between LSD profiles obtained with MuSiCoS and ESPaDOnS/NARVAL observations. Fig. \ref{LSDespnavcomp} shows that LSD profiles both from MuSiCoS and the current observations are consistent for an identical phase of observation in the case of HD 32633 and HD 112413, with the new data showing much reduced noise levels.

\begin{figure}
\begin{center}
   \includegraphics[width=0.50\textwidth]{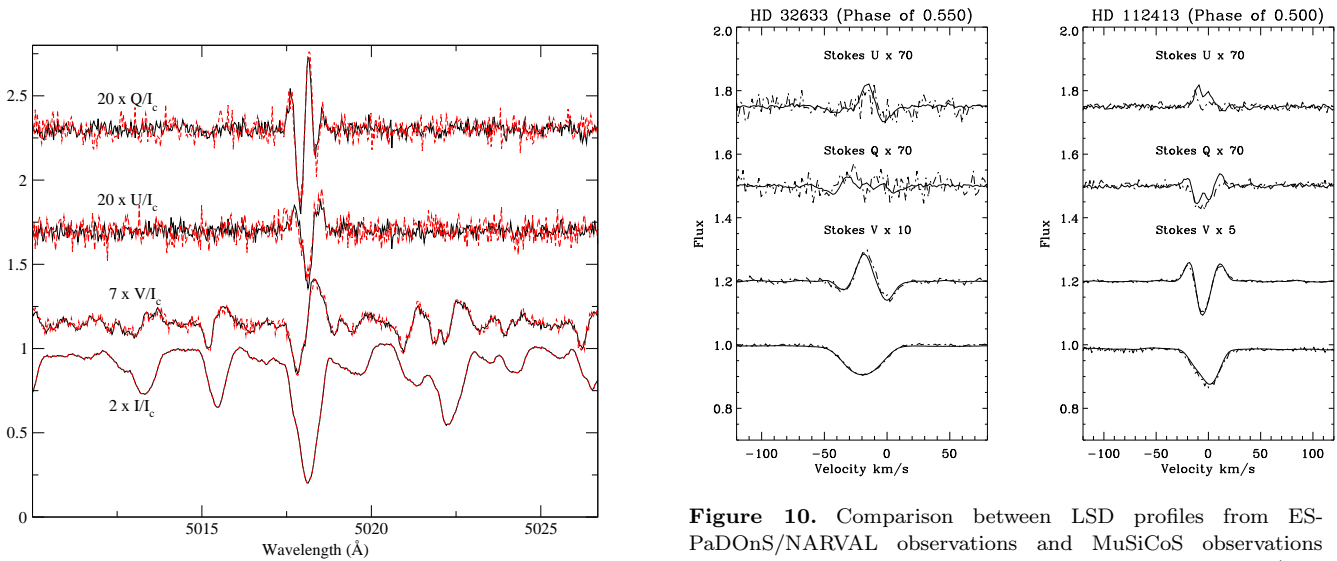}
 \caption{Two observations of HD 32633 obtained at close phases, one with NARVAL (Phase 0.689, taken 17 Dec 2006) and one with ESPaDOns (Phase 0.700, taken 26 Jan 2008).   The signal-to-noise ratio is $\approx$ 400 and $\approx$ 700 respectively.  Excellent agreement between the signatures can been seen, illustrating the consistency between the two instruments.  }
\label{phasecomp}
\end{center}
\end{figure}

\begin{figure}
\begin{center}
   \includegraphics[width=0.52\textwidth]{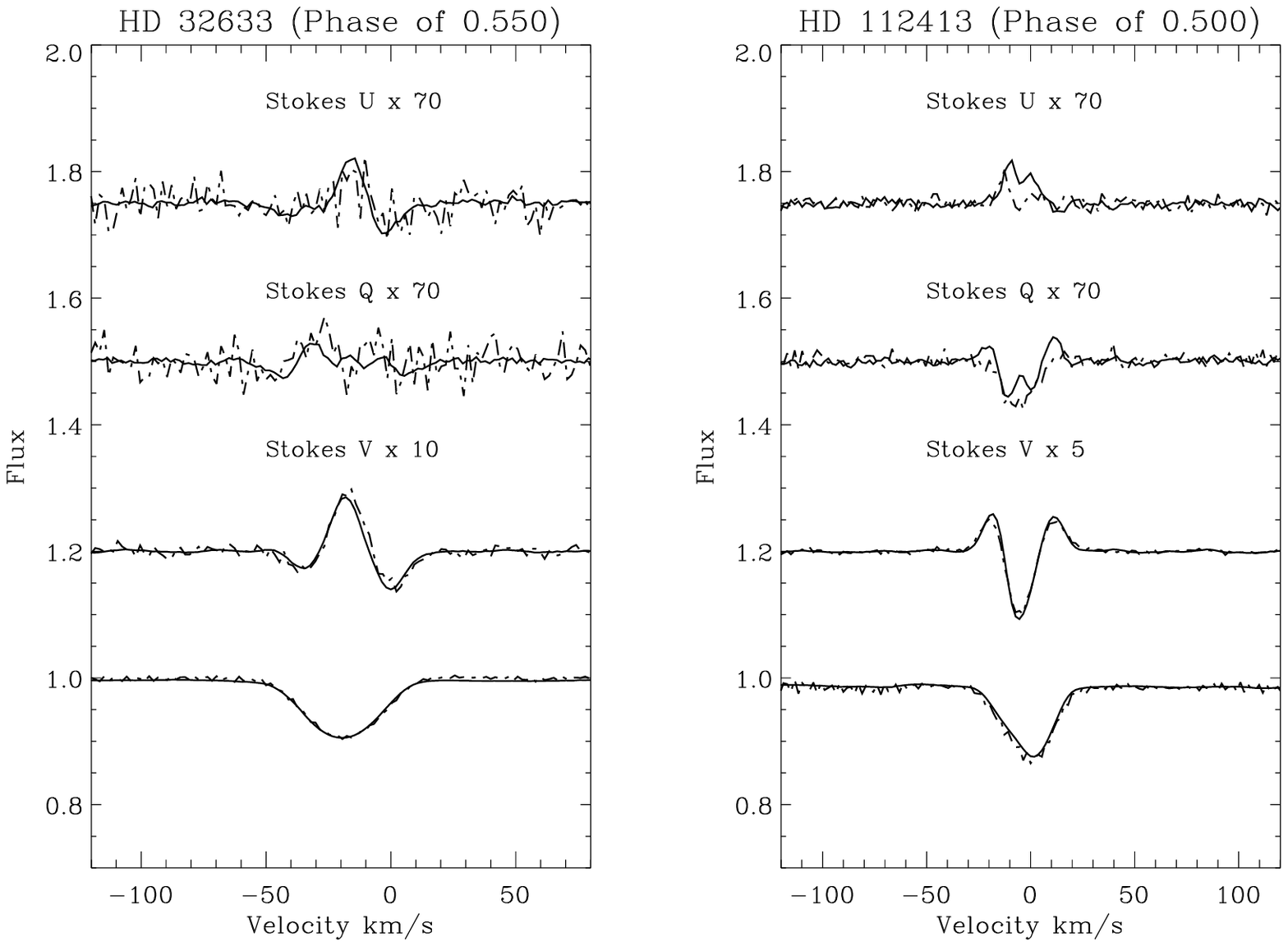}
  \caption{Comparison between LSD profiles from ESPaDOnS/NARVAL observations and MuSiCoS observations taken at near identical phases for HD 32633 and HD 112413 (MuSiCoS data dashed lines, ESPaDOnS/NARVAL solid lines). Good agreement can be seen.}
\label{LSDespnavcomp}
\end{center}
\end{figure}

It was quickly seen that both the ESPaDOnS and NARVAL data were consistent with one another. An excellent example of this can be seen in Fig. \ref{phasecomp},  where two observations of HD 32633 are shown, one at a phase of 0.689 (taken with NARVAL) and one with a phase of 0.700 (taken with ESPaDOnS). The detailed agreement between the Stokes profiles indicates that both instruments appear to be performing extremely consistently with one another.     Another test of consistency is described in Section 6.2 and 6.6 ,  where for certain phases, the ESPaDOnS/NARVAL data for  $\alpha^2$ CVn and HD 32633 have been convolved to the same approximate resolution as MuSiCoS (R=35000) and then compared to MuSiCoS observations of a similar phase.  Very good agreement can be seen in individual lines. 

As proposed by Landi Degl'Innocenti et al. (1982), assumed by Landolfi et al, (1993) and Leroy et al. (1993, 1995), and finally confirmed by Wade et al. (2000b), the phenomenon of broadband or net linear polarisation results from differential saturation of $\pi$ and $\sigma$ Zeeman components of individual spectral lines. Broadband polarisation measurements, such as those reported by Leroy (1995), effectively average the signal from the net polarised spectral lines with regions of continuum (which may be unpolarised, or polarised due to e.g. interstellar polarisation). In contrast, net polarisation measurements obtained using LSD measure only the net polarisation in lines. Therefore the measurements are not equivalent. In particular, Wade et al. (2000b) found that it was necessary to arbitrarily scale and shift the LSD measurements relative to the broadband measurements in order to bring them into agreement. Here we adopt a similar procedure.

Table \ref{curvefit} also shows the reduced $\chi^{2}$ of longitudinal field measurements taken from the null spectra.   A value close to 1 indicates that the null spectra and their error bars are consistent with zero longitudinal field and that the instrument is performing as expected.    We see for HD 4778, HD 32633, 49 Cam and HD71866, this is not the case.   These high values are a result of a handful of non-zero null measurements obtained during the December 2006 observations with NARVAL and very small number with obtained ESPaDonS, in which the null LSD profiles contained weak signatures (as discussed in Section 3).   

In addition, for these stars,  we find that the error bars on the Stokes $V$ longitudinal field measurements are larger than the null longitudinal field error bars, typically by a factor of 3-4. This is significant,  because it tells us that we begin to see that the Stokes $V$ longitudinal field measurements are no longer limited by photon noise. Rather, the larger error bars for the longitudinal field in Stokes $V$ are telling us that the LSD model is failing to fit the $V$ spectrum within the Libre-ESpRIT computed error bars, likely due to blends and the limitation of the weak-field approximation. This only happens when a significant number of individual lines show very strong Stokes $V$ signatures, and (as demonstrated by our ability to fit the longitudinal field curves within the error bars) is fully compensated for in the error bar calculation for Stokes V.  As discussed by Wade et al. (2000b) this compensation is achieved by scaling the photon noise statistical error bars of LSD profiles by a factor equal to the square root of the reduced $\chi^2$.  Wade et al. (2000b) describes that this scaling is almost always required for Stokes $V$ in Ap stars. If we recompute the reduced $\chi^{2}$ calculation again for the null spectra,  but use the Stokes $V$ error bars instead of the null error bars,  the resulting reduced $\chi^{2}$ is consistent with a zero longitudinal field (for example in the case of HD 32633,  the reduced $\chi^{2}$ goes from 5.42 to 0.36).

\section{Results for Individual Stars}
In the following section, the results for each star will be discussed,  including figures showing both the Stokes profile variation in selected individual lines and in the LSD profiles. Also the variation of the longitudinal field and in some cases the variation in net linear polarisation as a function of phase will be shown for each target.   Results are compared with those of Wade et al. (2000a) and Leroy (1995) where possible.  In the cases where the net linear polarisation was close to or consistent with zero at all phases (such as HD 32633 and HD 40312), these plots have been omitted. 

\subsection{HD 4778}

\begin{figure*}
\begin{center}
\includegraphics[width=0.85\textwidth]{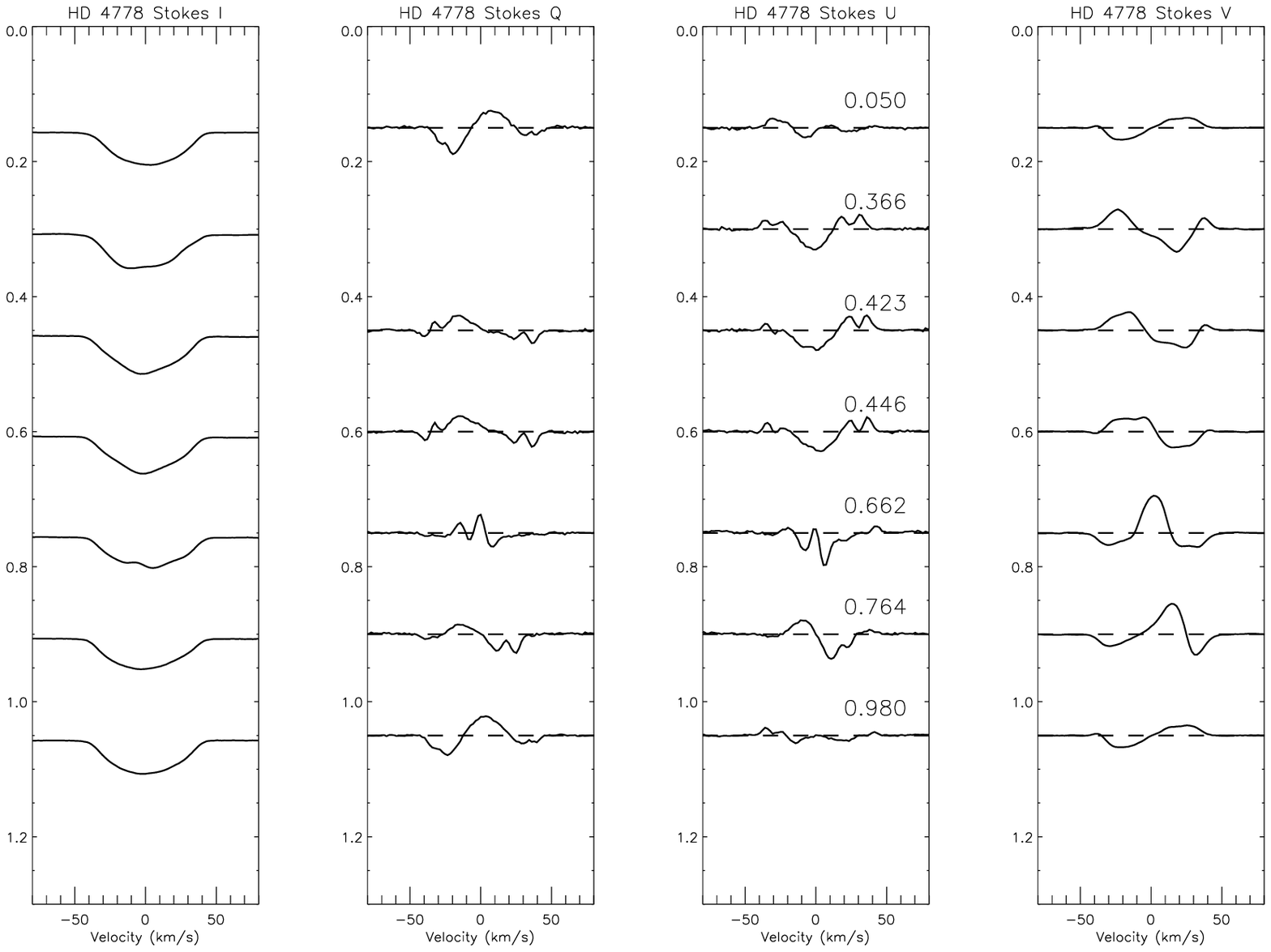}
   \includegraphics[width=0.60\textwidth, angle=90]{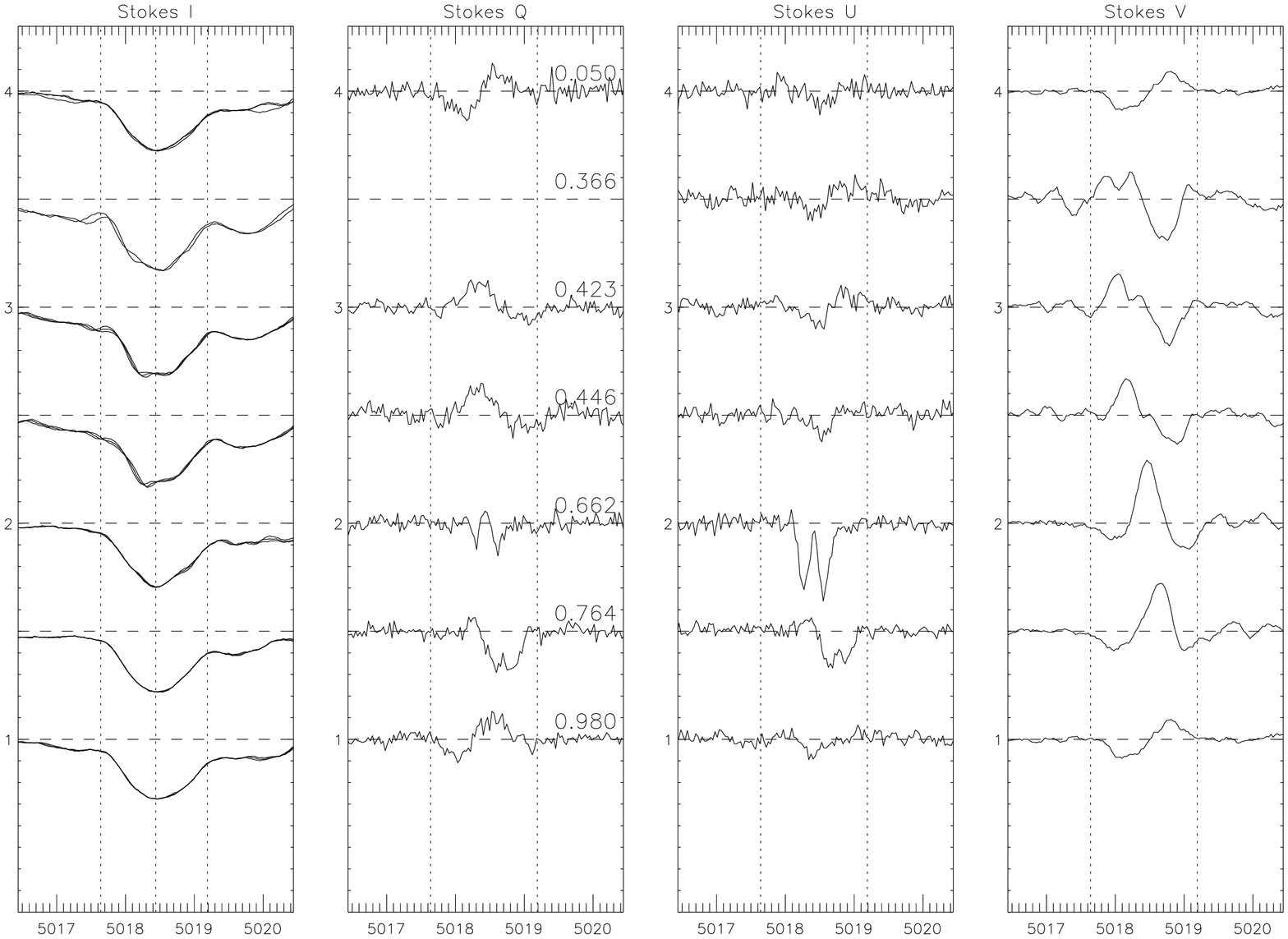}
 \caption{Top: Variation of Stokes $I, Q, U$ and $V$ LSD profiles for  HD 4778. Rotational phase increases from top to bottom, represented on the vertical axis. A scaling factor of 25 was used for both the Stokes $Q$ and $U$ observations and a factor of 5 was used for Stokes $V$. Bottom: Variation of Stokes $I, Q, U$ and $V$ profiles for HD 4778 in the Fe~{\sc ii}  $\lambda$ 5018 line (scaled by a factor of 20 for Stokes $Q$ and $U$ and a factor of 4 for Stokes $V$).}
\label{4778lines}
\end{center}
\end{figure*}

HD 4778 is the only star in our sample that was not studied by Wade et al. (2000a \& 2000b). It was classified by Renson and Manfroid (2009) as an A1CrSrEu star.  We obtained Stokes $V$, $Q$ and $U$ spectra for 7 rotational phases (although no Stokes $Q$ observation was obtained at phase 0.366).  Using the method described in Sect. 5, the projected rotational velocity ($v\sin i$) was determined to be $36 \pm 2$ ~\kms\,  which is a little larger than the Abt \& Morrell (1995) value of 33 ~\kms\ .  The data have been phased according to the ephemeris reported by Leone et al. (2000): 

\begin{equation}{\rm JD} = 2446674.006 + 2\fd56171\cdot {\rm E}.\end{equation}

The individual spectral lines show strong signatures in Stokes $V$,  and although variation in Stokes $Q$ and $U$ can be seen in individual lines in this star (an example is shown for the Fe~{\sc ii} $\lambda$ 5018 in Fig. \ref{4778lines}) the signatures in Stokes $Q$ and $U$ are relatively weak. Signatures are more prominent in the LSD profiles shown in Fig. \ref{4778lines}.   Stokes $I$ shows variability in the  Fe~{\sc ii} 4923, 5169 and 5018   \AA\  lines (as shown in Fig. \ref{4778lines}).

\begin{figure}
\begin{center}
   \includegraphics[width=0.55\textwidth]{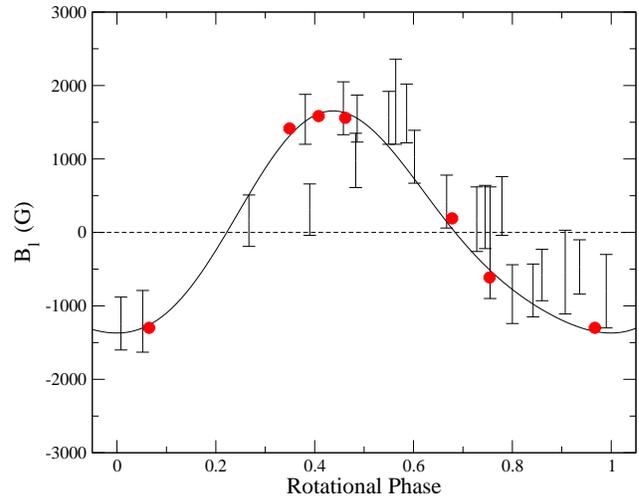}
 \caption{Longitudinal field measurements for HD 4778 obtained with ESPaDOnS/NARVAL (shown by filled circles), compared with those obtained by Bohlender (1989).  A  2nd order Fourier fit to the ESPaDOnS/NARVAL data is given by the solid curve.}
 \label{4778long}
\end{center}
\end{figure}

Both $B_\ell$ and net linear polarisation measurements have been obtained for each phase and are reported in Table \ref{bltable}.  The average uncertainly on the longitudinal field measurement is 20 gauss. The LSD profiles are illustrated in Fig.  \ref {4778lines}, and the longitudinal field curve shown in Fig.  \ref{4778long}. The 2nd order Fourier fit to this curve,  using only the new measurements gives a reduced $\chi^2$ of 1.41.  Also included in Fig. \ref{4778long} are measurements reported by Bolender (1989) for comparison. The slight apparent phase shift could be due to a cumulative error in phase over an almost 30 year span caused by the period uncertainty. 

\begin{figure}
\begin{center}
   \includegraphics[width=0.52\textwidth]{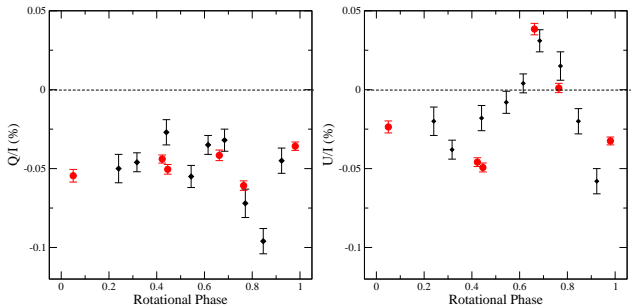}
   \caption{Net linear polarisation (Stokes $Q$ and $U$) measurements for HD 4778 obtained with ESPaDOnS/NARVAL (filled circles, scaled 0.04), and those obtained by Leroy (1995) (filled diamonds).  The improvement in data quality is evidenced by the smaller error bars associated with the ESPaDOnS/NARVAL measurements. }
\label{stokes4778}
\end{center}
\end{figure}

Leroy et al. (1995) obtained broadband linear polarisation measurements of HD 4778; their values are compared with the measurements obtained in this work in Fig.  \ref{stokes4778}. Reasonable agreement in Stokes $Q$ and $U$ can be seen between the two epochs of data. The ESPaDonS/NARVAL data had the mean subtracted and then were scaled by 0.08 to bring them into agreement with those of Leroy et al. (1995).

\subsection{HD 32633}
HD~32633 is a fairly broad-lined B9p star with a strong, non-sinusoidal longitudinal magnetic field variation. We obtained Stokes $V$, $Q$ and $U$ profiles for 20 rotational phases (Stokes $Q$ and $U$ observations are missing at phase 0.086).  Using the method described in Sect. 5, the projected rotational velocity ($v\sin i$) was determined to be $19 \pm 2$~~\kms\, which agrees within uncertainty with the value adopted by Wade et al. (2000b).

The data have been phased according to the ephemeris of Adelman (1997b): 

\begin{equation}{\rm JD}=2437635.200 + 6\fd 43000\cdot{\rm E}.\end{equation}

Very strong and variable Stokes $Q$ and $U$ polarisation signatures can be seen in individual lines in the spectra of this star,  an example of which is shown for the Fe~{\sc ii} 5018   \AA\  line in Fig. \ref{32633lines}. Interestingly, the shape of the Stokes $Q$ and $U$ profiles change relatively little as a function of phase (in individual lines, as well as LSD profiles of Fig. \ref{32633lines}). This could be an indication that we are "seeing" a limited part of the transverse geometry of the field due to the overall geometry of the star.  It should be noted in Fig. 15 that the shapes of the Stokes Q and U profiles of individual Fe lines look quite different from the LSD profiles. This may be a result of differences between the Stokes profiles of weak versus strong lines, or possibly differences in the Stokes profiles of lines of different chemical elements. As with all the stars with strong linear polarisation signatures in this study, HD 32633 has many more individual lines showing signatures in the new observations than in the MuSICoS observations.  Also Stokes $I$ shows variability in lines such as in the  Cr~{\sc ii} line 4588 \AA\ and  Fe~{\sc ii}4824 and 5018   \AA\ lines (as shown in Fig. \ref{32633lines}).

\begin{figure}
\begin{center}
   \includegraphics[width=0.55\textwidth]{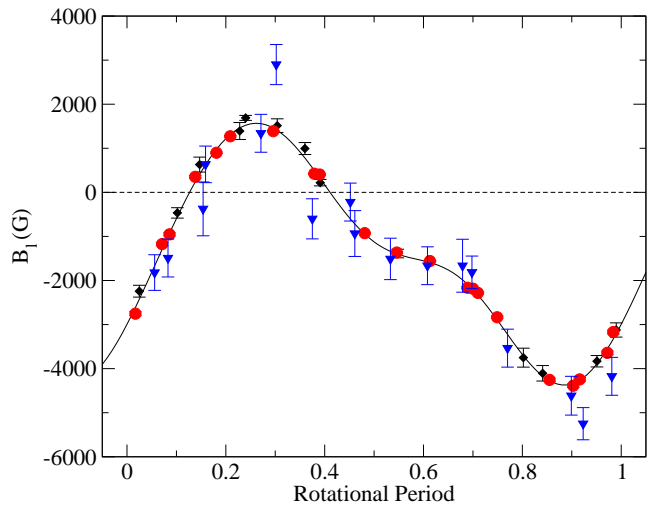}
 \caption{Longitudinal field measurements for HD 32633 obtained with ESPaDOnS/NARVAL (shown by filled circles), compared with those obtained by Wade et al. (2000b) with MuSiCoS (shown by filled diamonds) and measurements by Borra \& Landstreet (1980) (shown as triangles).  A good agreement can be seen, confirming consistency between the instruments and the improvement in data quality is evidenced by the smaller error bars associated with the ESPaDOnS/NARVAL measurements.  A  3rd order Fourier fit to the ESPaDOnS/NARVAL data is given by the solid curve.}
\label{32633long}
\end{center}
\end{figure}

\begin{figure*}
\begin{center}
   \includegraphics[width=0.85\textwidth]{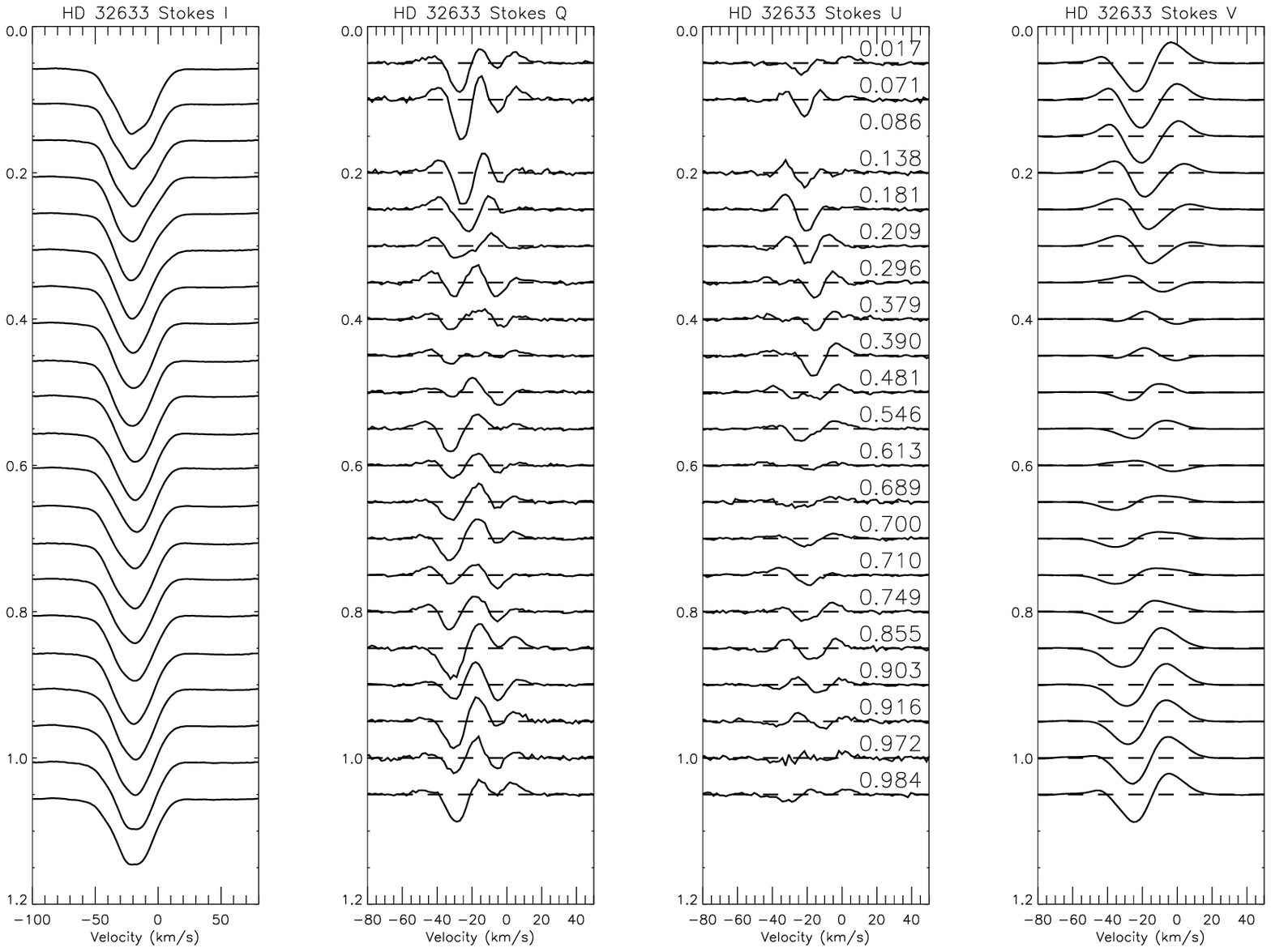}
      \includegraphics[width=0.60\textwidth, angle=90]{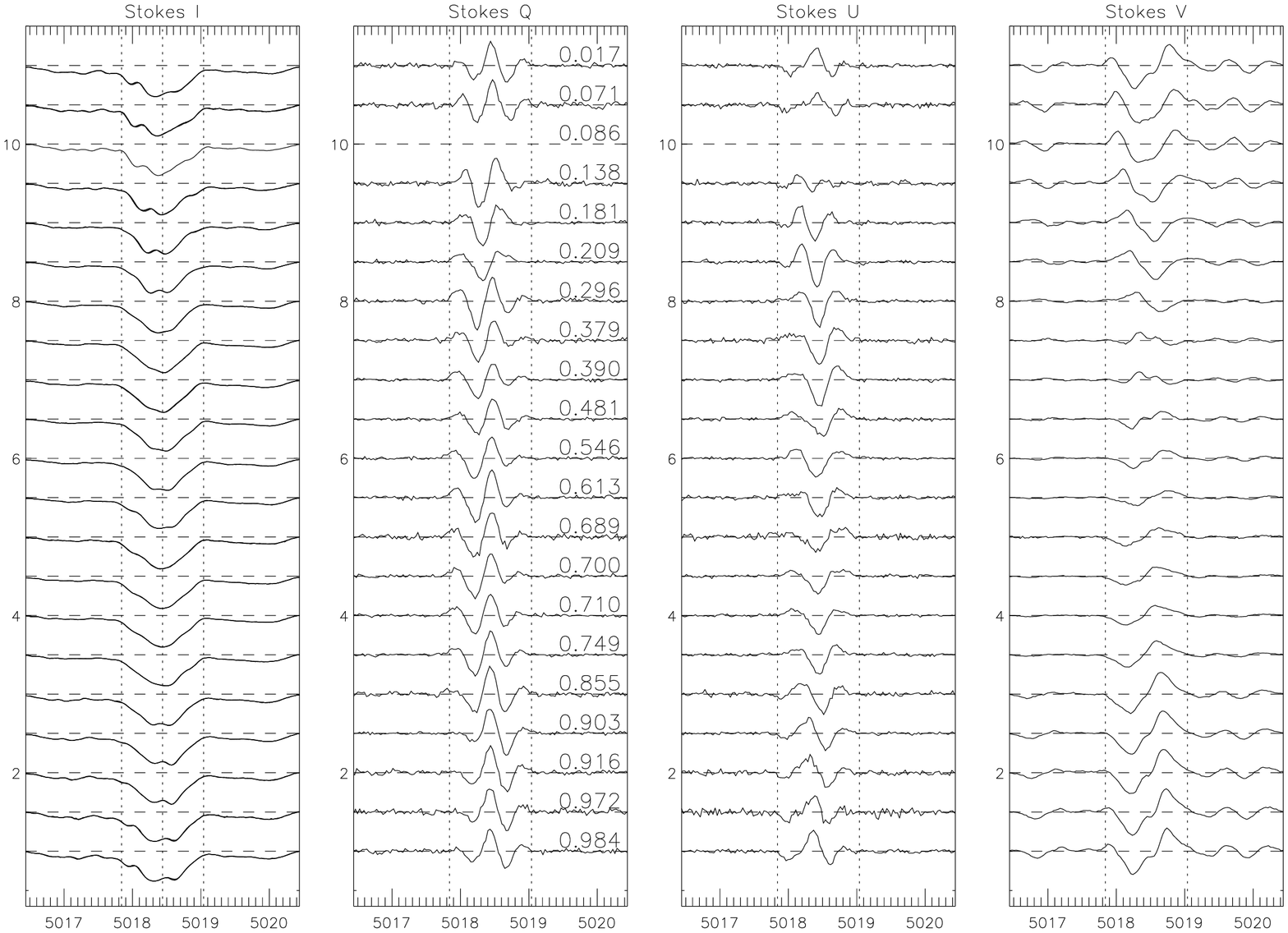}
 \caption{Top: Variation of Stokes $I,Q,U$ and $V$ LSD profiles for  HD 32633. Rotational phase is represented on the vertical axis. With a scaling factor of 20 was used both the Stokes $Q$ and $U$ observations. Bottom: Variation of Stokes $I,Q,U$ and $V$ profiles for  HD 32633 in the 5018 Fe~{\sc ii} line  (scaled by a factor of 10 for Stokes $Q$ and $U$ and 2 for Stokes $V$).}
\label{32633lines}
\end{center}
\end{figure*}

\begin{figure}
\begin{center}
   \includegraphics[width=0.50\textwidth]{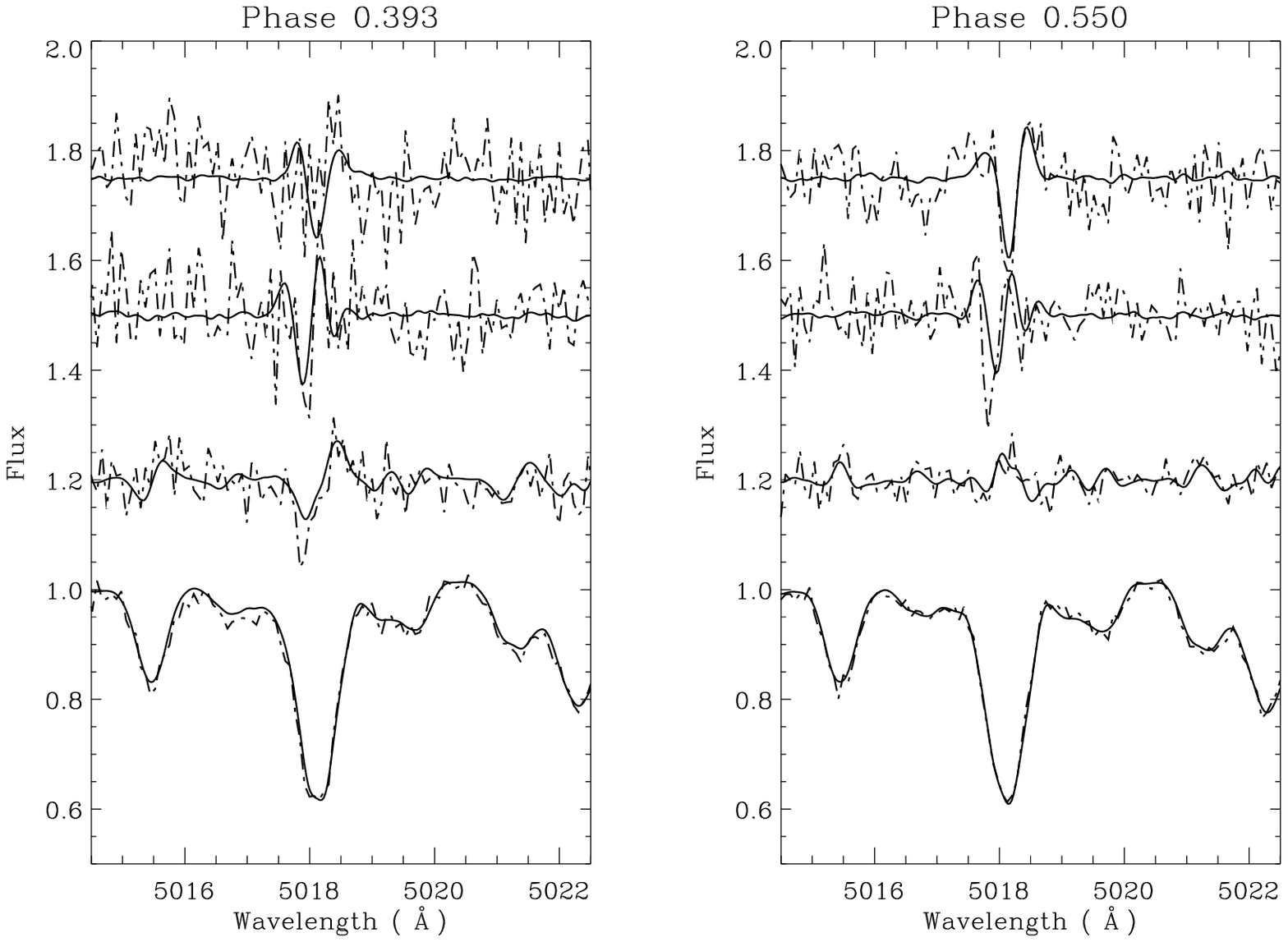}
 \caption{Comparison between two phases of the ESPaDOnS/NARVAL observations convolved to a resolution of 35000 and the MuSiCoS data at a similar phase for HD 32633 (MuSiCoS data dashed lines, ESPaDOnS/NARVAL solid lines). Good overall agreement can be seen between the two epochs of data.}
 \label{32633_35000}
\end{center}
\end{figure}

Both $B_\ell$ and net linear polarisation measurements have been obtained for each phase and are reported in Table \ref{bltable}.  The longitudinal field values obtained by Borra \& Landstreet (1980)  and Wade et al. (2000b) are compared to the new measurements in Fig.  \ref{32633long}. Very good agreement can be seen between the three epochs of data, with the clear reduction in uncertainties shown by the smaller error bars associated with the new data. The reduced $\chi^2$ of the Fourier fit to the field curve is 1.95, with an order of fit of 3. The average uncertainty of the longitudinal field measurements is 26 gauss.   The netlinear polarisation plot is not shown for HD 32633, due to the fact the measurements show little or no variation and are consistently null at all phases.  In addition for certain phases, the ESPaDOnS/NARVAL data for  HD 32633 have been convolved to the same approximate resolution as MuSiCoS (R=35000) and then compared to MuSiCoS observations of a similar phase.  Within the limits imposed by noise, very good agreement can be seen in individual lines between the two data sets in Fig. \ref{32633_35000}.

\subsection{HD 40312 - $\theta$ Aur}
$\theta$~Aur is a broader-lined A0p star with a weak magnetic field.  We obtained Stokes $V$, $Q$ and $U$ profiles for 7 rotational phases.  Using the method as described above, the projected rotational velocity ($v\sin i$) was determined to be $53 \pm 1$~~\kms\, which agrees within uncertainty with the value adopted by Wade et al. (2000b)

All magnetic measurements have been phased according to the ephemeris of Wade et al. (2000a): 

\begin{equation}
{\rm JD}=2450001.881 + 3\fd 61860\cdot{\rm E}.
\end{equation}

\begin{figure*}
\begin{center}
   \includegraphics[width=0.85\textwidth]{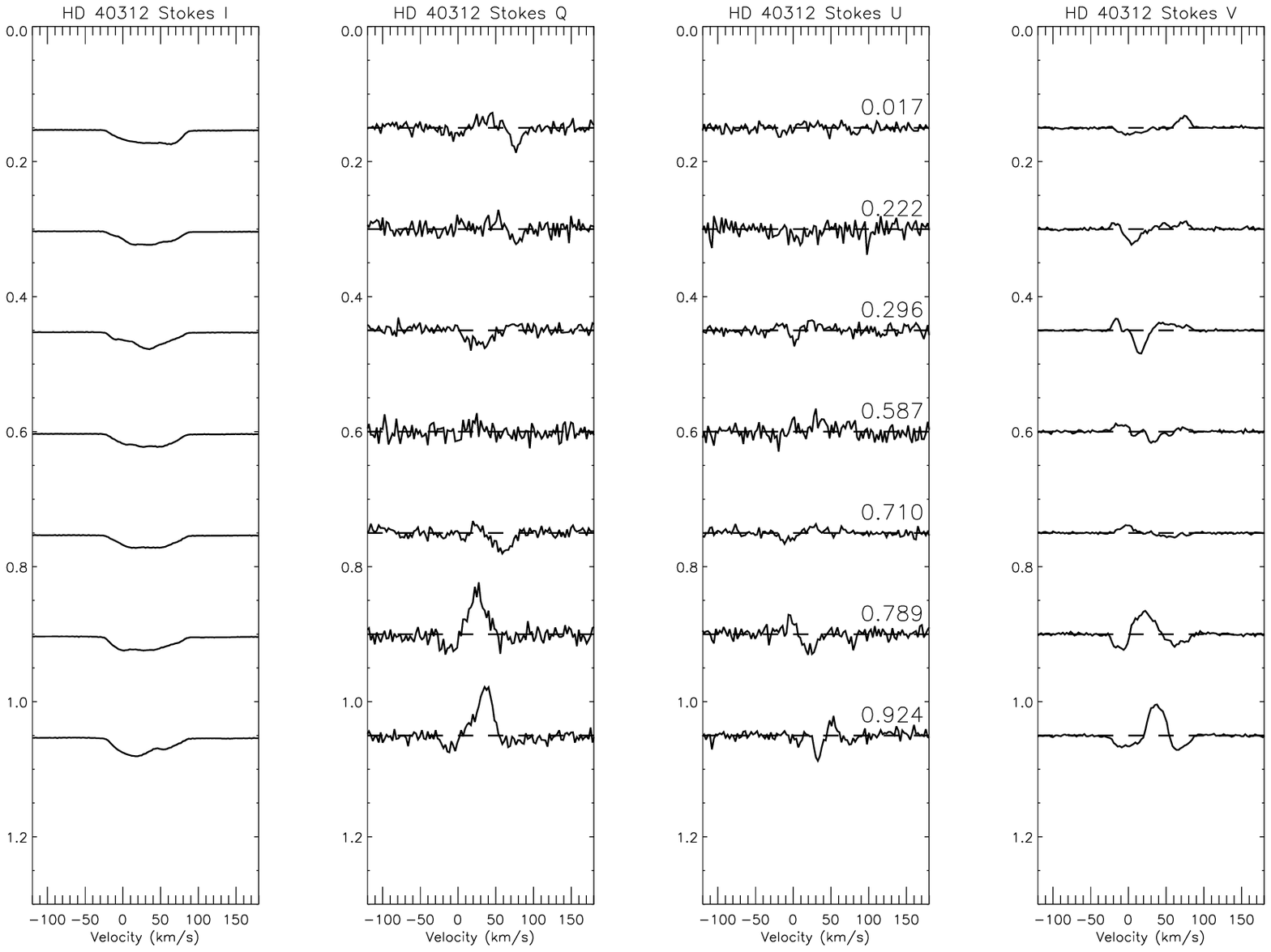}
    \includegraphics[width=0.60\textwidth, angle=90]{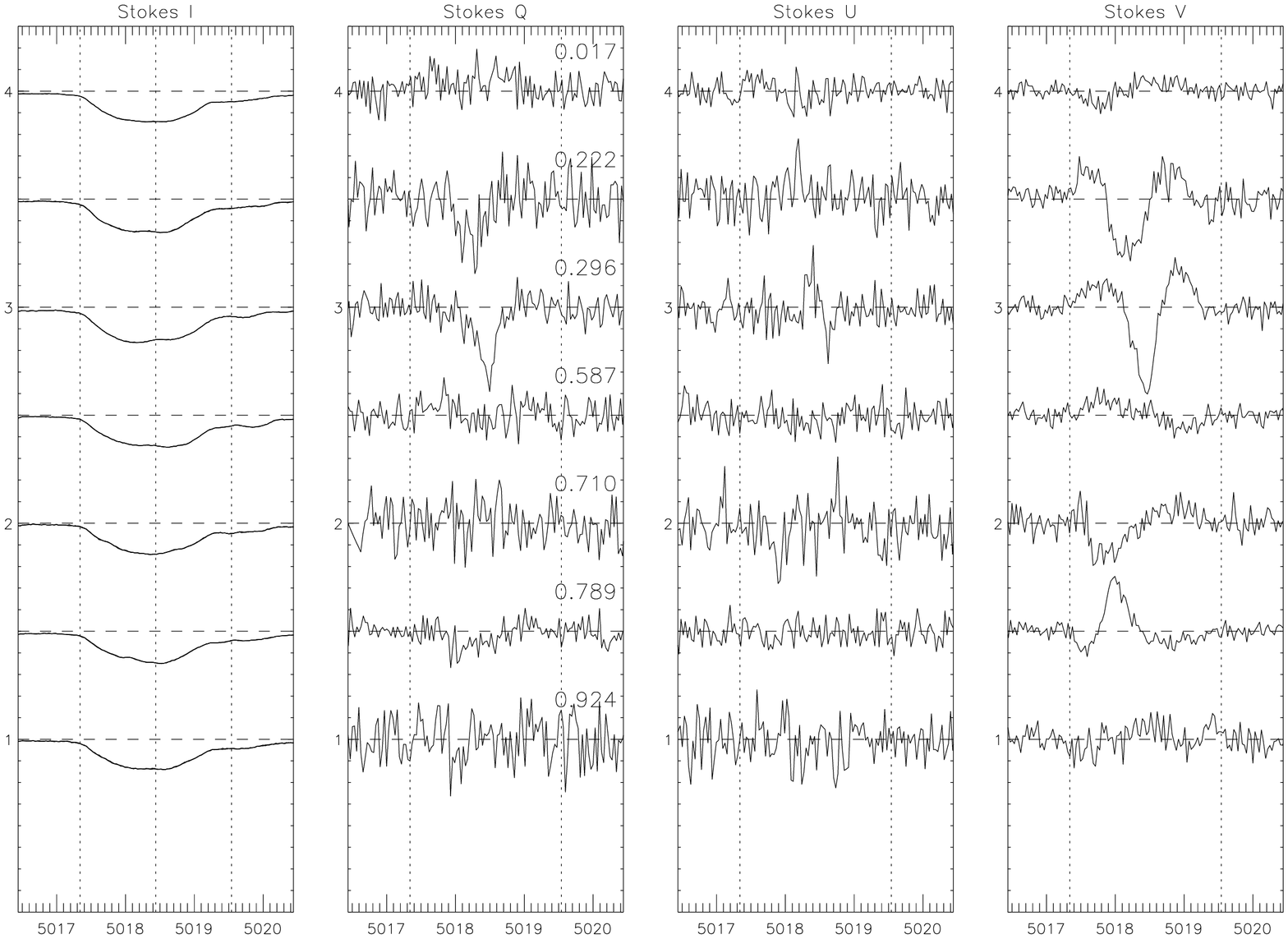}
 \caption{Top: Variation of Stokes $I, Q, U$ and $V$ LSD profiles for  HD 40312. Rotational phase is represented on the vertical axis. A scaling factor of 200 was used for both the Stokes $Q$ and $U$ observations, while a factor of 50 was used for Stokes $V$. Bottom:  Variation of Stokes $I, Q, U$ and $V$ profiles for HD 40312 in the Fe~{\sc ii} 5018 line (scaled by a factor of 80 for Stokes $Q$ and $U$ and 40 for Stokes $V$).  }
\label{40312lines}
\end{center}
\end{figure*}

The individual line profiles show weak signatures in Stokes $V$; marginal Stokes $Q$ and $U$ signatures are visible at phase 0.7-0.8 (an example for the Fe~{\sc ii} 5018   \AA\ line is shown in Fig. \ref{40312lines}). Signatures are slightly more prominent in the LSD profiles. Stokes $I$ exhibits small variations in the Fe~{\sc ii}  5018   \AA\ and very subtle variation in the  Fe~{\sc ii} line 4923 \AA\.   Because of the weak Stokes $Q$ and $U$ signatures in individual lines in the spectra of this star, it would be a challenging candidate for $IQUV$ mapping,  but would be very suitable for $IV$ mapping.

\begin{figure}
\begin{center}
   \includegraphics[width=0.55\textwidth]{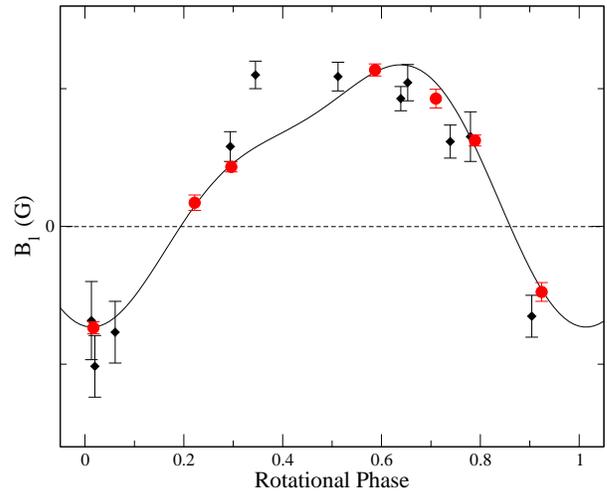}
 \caption{Longitudinal field measurements for HD 40312 obtained with ESPaDOnS/NARVAL (shown by filled circles), compared with those obtained by Wade et al.(2000b) with MuSiCoS (shown by filled diamonds).  A  2nd order Fourier fit to the ESPaDOnS/NARVAL data is given by the solid curve. }
\label{40312long}
\end{center}
\end{figure}

Both $B_\ell$ and net linear polarisation measurements have been obtained for each phase and are reported in Table \ref{bltable}.  The longitudinal field values obtained by Wade et al. (2000b) are compared to this work in Fig.  \ref{40312long}. Agreement can be seen between the two epochs of data. The average uncertainty on the new longitudinal field measurement is 14 gauss. A Fourier fit to the longitudinal field curve yields a reduced $\chi^2$ of 1.53 with an order of fit of 2.  Because of the limited phase coverage and the small net linear polarisation measurements,  a plot of net linear polarisation is not shown for HD 40312.

\subsection{HD 62140 - 49 Cam}
49~Cam is a fairly broad-lined F0p star with a moderately strong magnetic field.  We obtained Stokes $V$, $Q$ and $U$ profiles for 19 rotational phases.  The projected rotational velocity ($v\sin i$) was determined to be $24 \pm 2$~~\kms\, which agrees within uncertainty with the value adopted by Wade et al. (2000b).

All measurements have been phased according to the ephemeris of Adelman (1997a): 

\begin{equation} {\rm JD}=2441257.300 + 4\fd 28679\cdot{\rm E}.\end{equation}

\begin{figure*}
\begin{center}
   \includegraphics[width=0.85\textwidth]{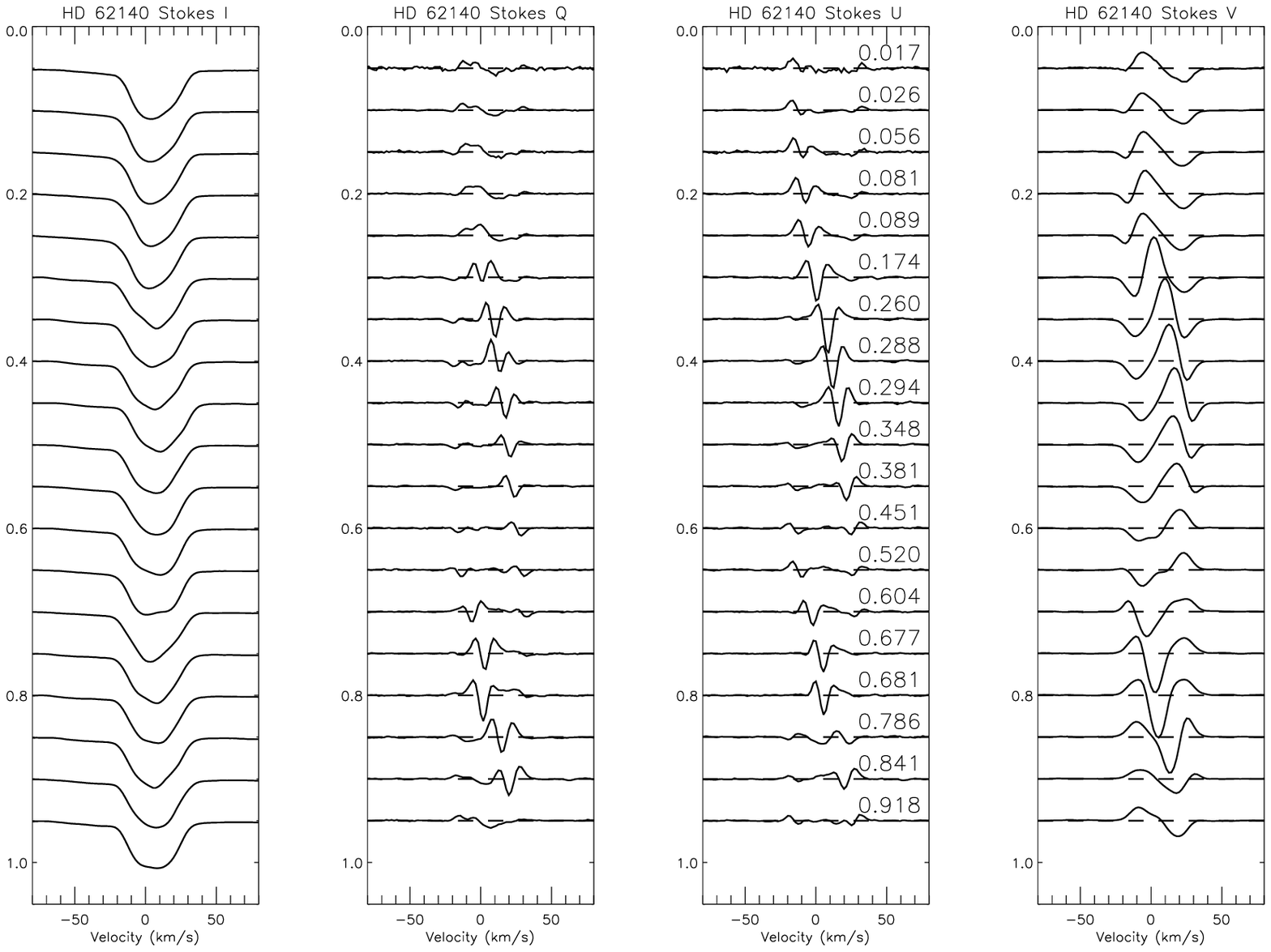}
     \includegraphics[width=0.60\textwidth, angle=90]{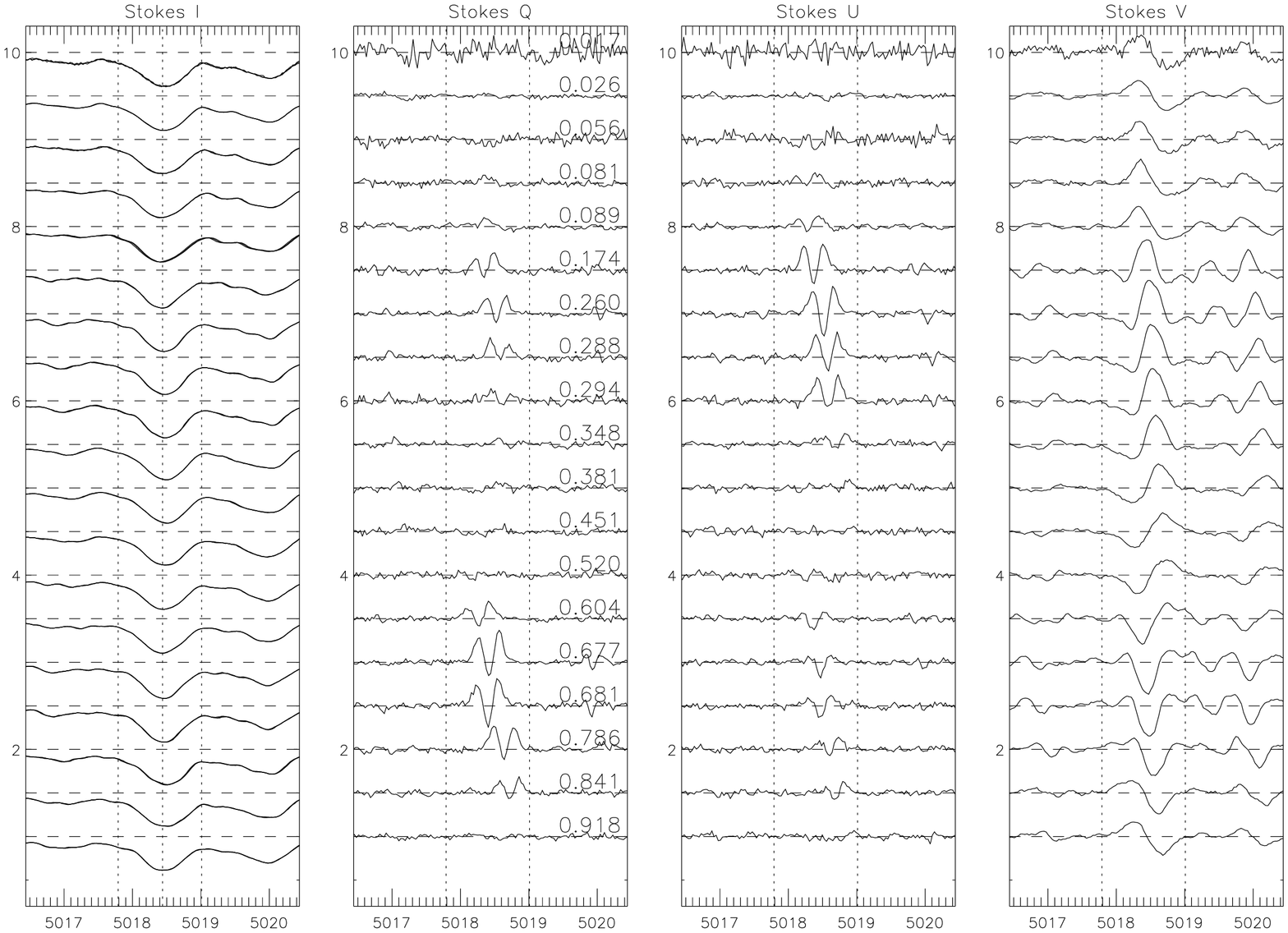}
 \caption{Top: Variation of Stokes $I,Q,U$ and $V$ LSD profiles for  HD 62140. Rotational phase represented on the vertical axis. With a scaling factor of 12 used both the Stokes $Q$ and $U$ observations and a scaling factor of 2 for Stokes V. Bottom: Variation of Stokes $I,Q,U$ and $V$ profiles for  HD 62140 in the 5018 Fe~{\sc ii} line  (scaled by a factor of 12 for Stokes $Q$ and $U$ and 4 for Stokes $V$).}
\label{62140lines}
\end{center}
\end{figure*}

Variable signatures can be seen in Stokes $Q$ and $U$ in individual lines in this star,  an example of which is shown for the Fe~{\sc ii} 5018   \AA\  line in Fig. \ref{62140lines}. The signatures are even clearer in the LSD profiles (Fig. \ref {62140lines}). The amplitudes of the Stokes $Q$ and $U$ profiles are small compared to HD 32633 (about 50 \% smaller), in both the individual lines and LSD profiles. In Stokes $I$,  49 Cam shows subtle variability in most lines, but as with all the stars with strong linear polarisation signatures in this study, many individual lines show signatures, as shown in Fig. \ref{atlas}.
 
\begin{figure}
\begin{center}
   \includegraphics[width=0.55\textwidth]{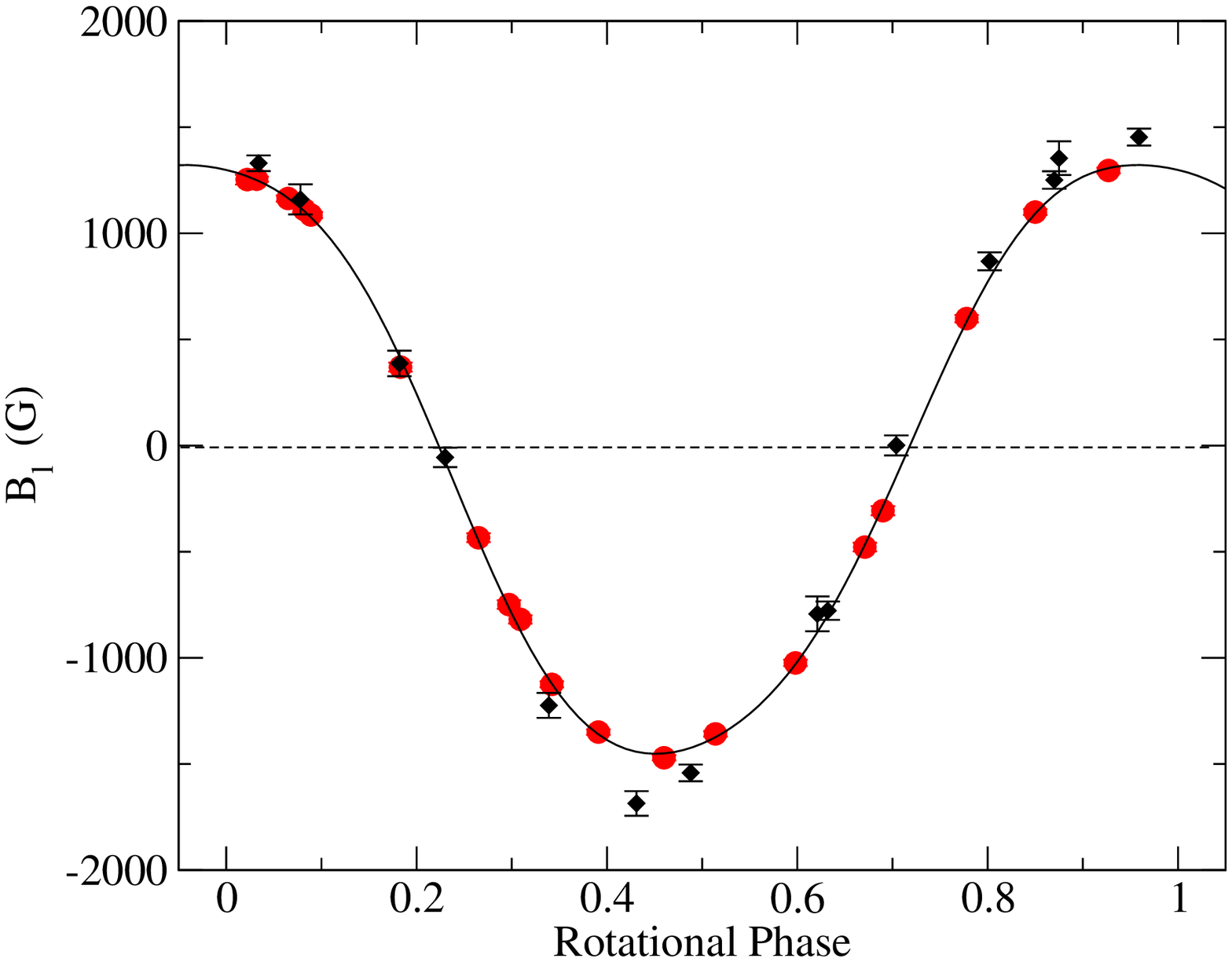}
 \caption{Longitudinal field measurements for HD 62140 obtained with ESPaDOnS/NARVAL (shown by filled circles), compared with those obtained by Wade et al. (2000b) with MuSiCoS (shown by filled diamonds).   A good agreement can be seen, confirming consistency between the instruments and the improvement in data quality is evidenced by the smaller error bars associated with the ESPaDOnS/NARVAL measurements.  A  3rd order Fourier fit to the ESPaDOnS/NARVAL data is given by the solid curve. }
\label{49long}
\end{center}
\end{figure}

\begin{figure}
\begin{center}
   \includegraphics[width=0.53\textwidth]{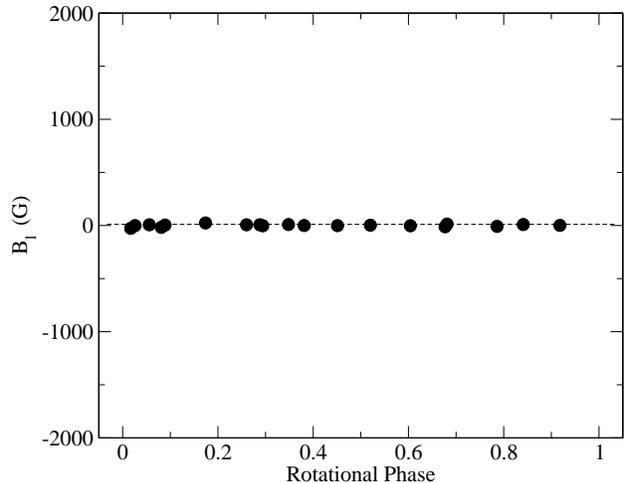}
 \caption{Longitudinal field measurements for HD 62140 obtained with ESPaDOnS/NARVAL for the null spectrum.}
\label{49null}
\end{center}
\end{figure}

Both $B_\ell$ and net linear polarisation measurements have been obtained for each phase and are reported in Table \ref{bltable}.  The longitudinal field values obtained by Wade et al. (2000b) are compared to those derived in this work in Fig.  \ref{49long}. Very good agreement can be seen between the two epochs of data. The average uncertainty on the new longitudinal field measurement is 14 gauss, and the Fourier fit to the new measurements gives a reduced $\chi^2$ of 1.29 with an order of fit of 3.  To further illustrate the quality of data, Fig. \ref{49null} shows the longitudinal measurements obtained from the null spectrum.  In the case of data free from any spurious signals (e.g caused by instrumental polarisation effects), the longitudinal field in the null spectrum should be consistent with zero and this is clearly the case with HD 62140 and is representative of most observations in this sample. 

\begin{figure}
\begin{center}
   \includegraphics[width=0.53\textwidth]{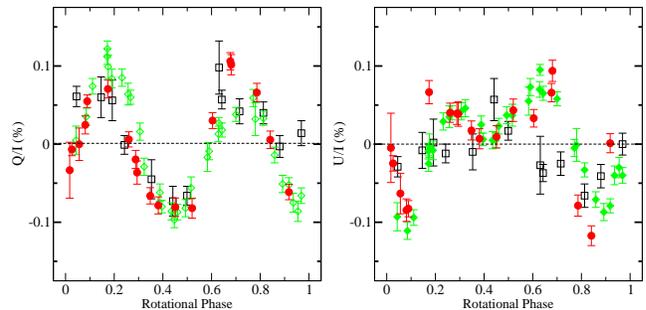}
  \caption{Net linear polarisation measurements (Stokes $Q$ and $U$) for 49 Cam obtained with ESPaDOnS/NARVAL (shown by filled circles, scaled by a factor of 0.28), compared with those obtained by Wade et al. (2000b) with MuSiCoS (shown by filled diamonds) and those obtained by Leroy (1995) (shown by open squares).  A good agreement can be seen, confirming consistency between the instruments and the improvement in data quality is evidenced by the smaller error bars associated with the ESPaDOnS/NARVAL measurements. }
\label{stokes49cam}
\end{center}
\end{figure}

The net linear polarisation as a function of phase is shown in Fig \ref{stokes49cam}. For both Stokes $Q$ and $U$ measurements there is reasonable agreement at most phases with the measurements reported by Wade et al. (2000b) and in most cases very good agreement with values reported by Leroy et al. (1995). The net linear polarisation was scaled by the same factor (0.28), double the scaling factor employed by Wade et al. (2000b). In addition the mean was subtracted from all measurements,  as prescribed by Wade et al. (2000b).

\subsection{HD 71866}

 \begin{figure*}
\begin{center}
   \includegraphics[width=0.85\textwidth]{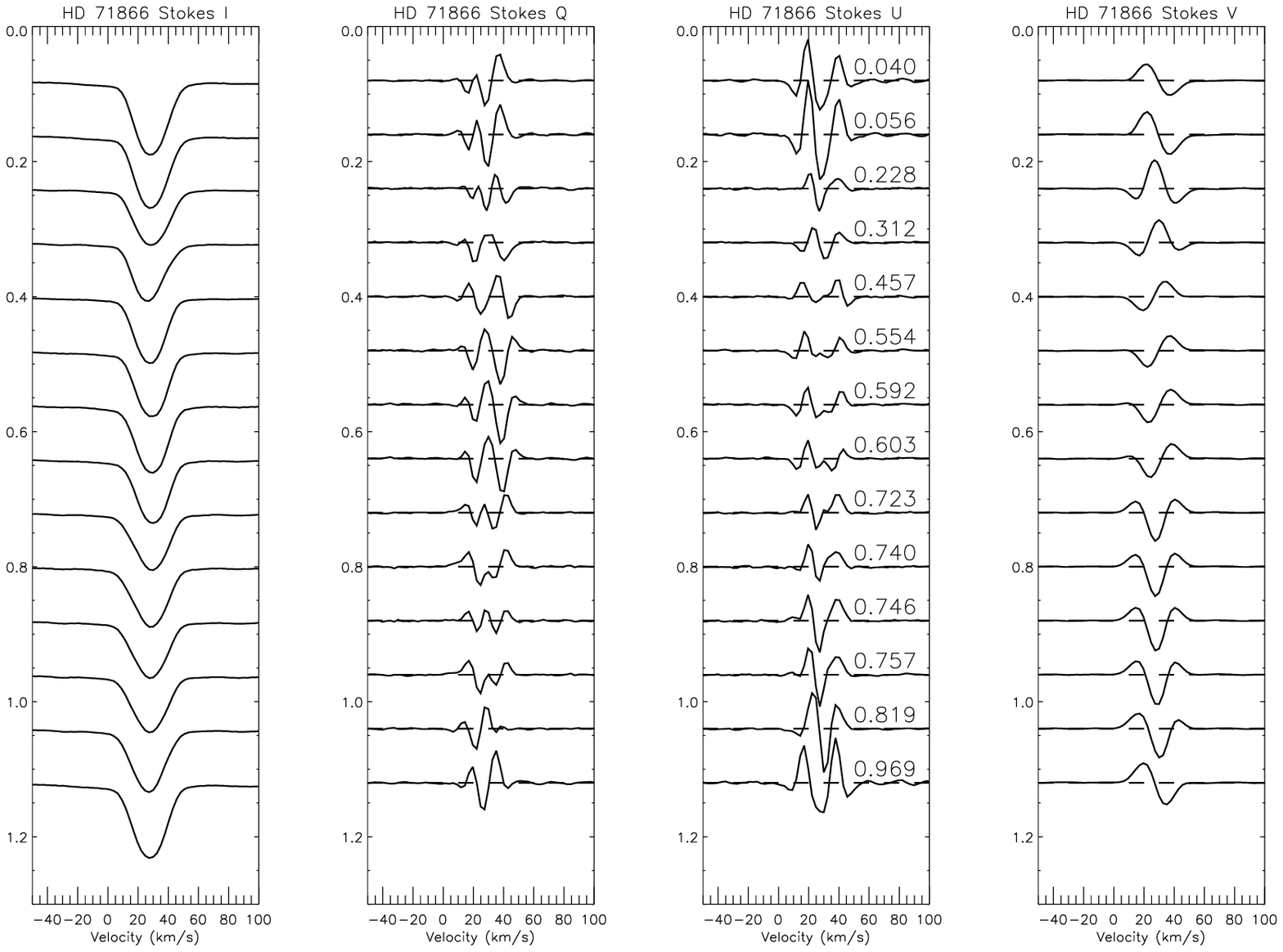}
     \includegraphics[width=0.60\textwidth, angle=90]{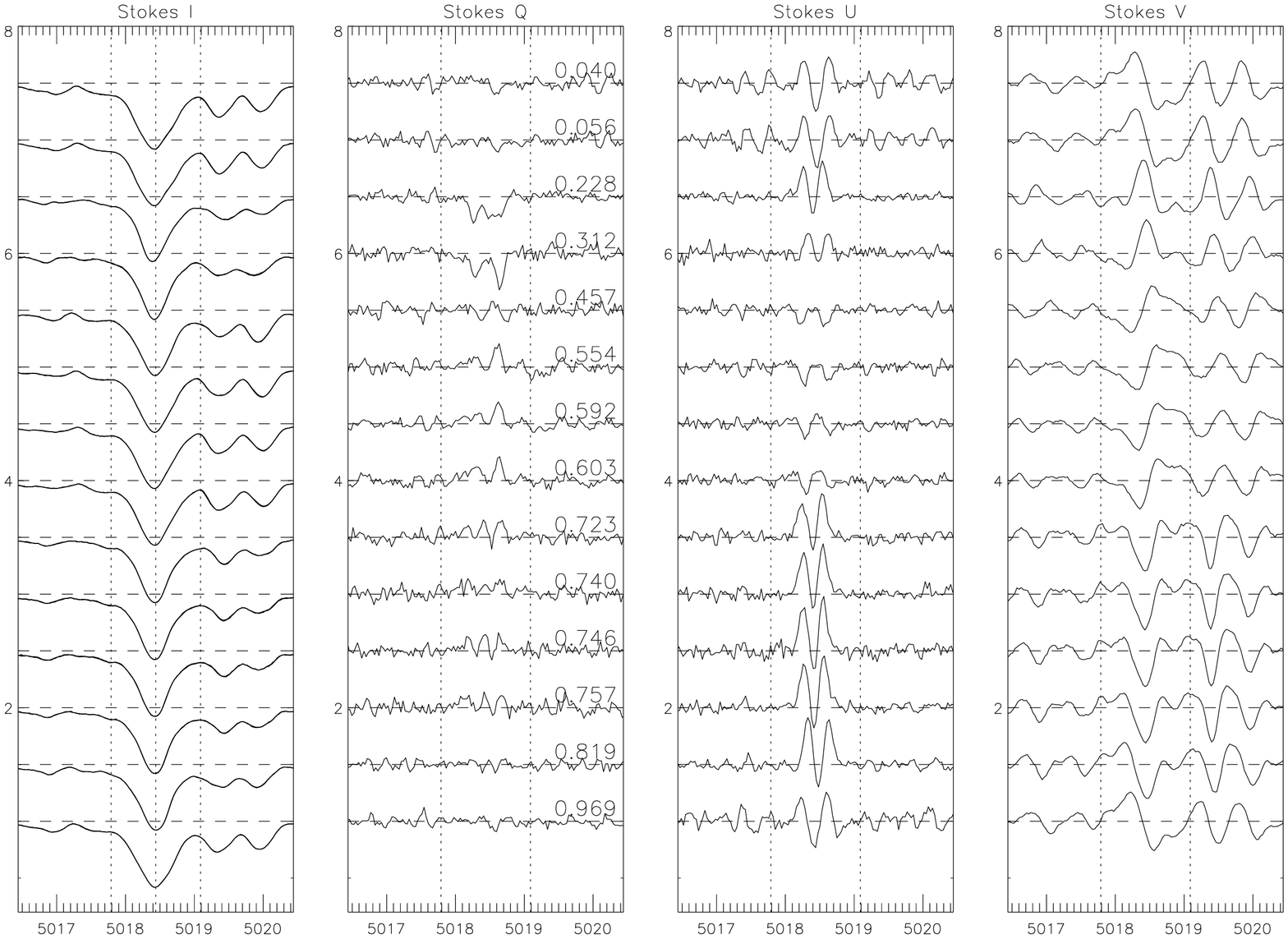}
 \caption{Top: Variation of Stokes $I,Q,U$ and $V$ LSD profiles for HD 71866. Rotational phase represented on the vertical axis. With a scaling factor of 15 used both the Stokes $Q$ and $U$ observations and a scaling factor of 1 for Stokes V. Bottom: Variation of Stokes $I,Q,U$ and $V$ profiles for  HD 71866 in the 5018 Fe~{\sc ii}  line  (scaled by a factor of 16 for Stokes Q, 16 for Stokes $U$ and 4 for Stokes $V$). }
\label{71866lsd}
\end{center}
\end{figure*}

HD~71866 is a sharp-lined A2p star with a moderately strong magnetic field.   We have obtained Stokes $V$, $Q$ and $U$ profiles for 14 rotational phases. The projected rotational velocity ($v \sin i$) we determine is $15 \pm 2$~~\kms\, which agrees within uncertainty with the value adopted by Wade et al. (2000b).

All measurements have been phased according to the ephemeris of Bagnulo et al. (1995): \begin{equation}{\rm JD}= 2438297.5 + 6\fd 80022\cdot{\rm E}. \end{equation}

Clear signatures and strong variability can be seen in Stokes $Q$ and $U$ in the individual lines of this star,  an example of which is shown for the Fe~{\sc ii} 5018 \AA\  line in Fig. \ref{71866lsd}. The amplitudes of the Stokes $Q$ and $U$ profiles are large compared to many of the other targets in this study (in both the individual line and LSD profiles, see Fig.  \ref {71866lsd}). Stokes $I$ appears to vary very slightly with phase,  with only small changes in line shape in the  Fe~{\sc ii} lines at 4923, 5018 and 5169 \AA\ . 

\begin{figure}
\begin{center}
   \includegraphics[width=0.55\textwidth]{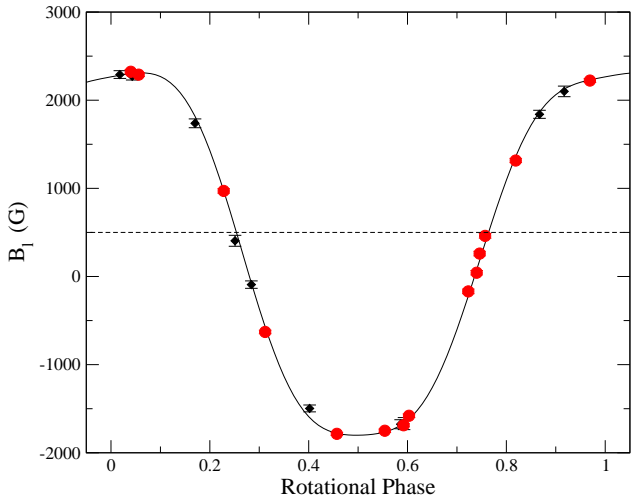}
 \caption{Longitudinal field measurements for HD 71866 obtained with ESPaDOnS/NARVAL (shown by filled circles), compared with those obtained by Wade et al.(2000b) with MuSiCoS (shown by filled diamonds).  A good agreement can be seen (with a re-anaylsis of the MuSiCoS observations),  confirming consistency between the instruments and the improvement in data quality is evidenced by the smaller error bars associated with the ESPaDOnS/NARVAL measurements.  A  4th order Fourier fit to the ESPaDOnS/NARVAL data is given by the solid curve. }
\label{71866long}
\end{center}
\end{figure}

Both $B_\ell$ and net linear polarisation measurements have been obtained for each phase and are reported in Table \ref{bltable}.   To get a good agreement between the two epochs of data required a re-analysis of the Wade et al. (2000b) data, using the updated and abundance specific line mask (as used for the ESPaDOnS/NARVAL data).  It is likely that this star is in a temperature regime where measurements are very sensitive to the line mask chosen or because of peculiar abundances.  The longitudinal field variation is illustrated in Fig.  \ref{71866long}. Good agreement can be seen between the two epochs of data. The average uncertainty of the longitudinal field measured from the ESPaDOnS and NARVAL spectra was 27 gauss.  A Fourier fit to the longitudinal field curve with a $\chi^2$ of 1.02, with order of fit of 4. 

\begin{figure}
\begin{center}
   \includegraphics[width=0.53\textwidth]{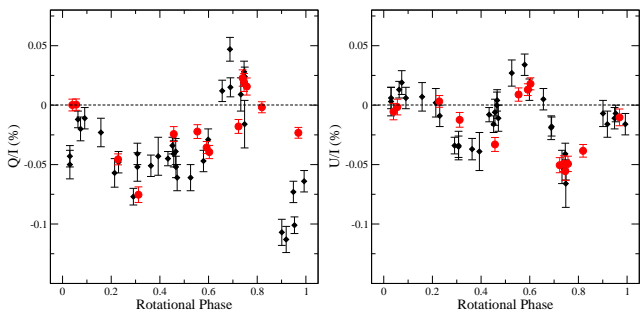}
  \caption{Net linear polarisation (Stokes $Q$ and $U$) measurements for HD 71866 obtained with ESPaDOnS/NARVAL (shown by filled circles and scaled by a factor of 0.10,  compared with those of Leroy et al. (1995) (shown by filled diamonds).}
\label{stokes71866}
\end{center}
\end{figure}

The net linear polarisation as a function of phase is shown in Fig \ref{stokes71866}. The Stokes $Q$ and $U$ measurements obtained in this study have been scaled by a factor of 0.07 to bring them into agreement with the measurements of Leroy et al. (1995) and this scaling is consistent with that of Wade et al. (2000b).  There is an approximate agreement at most phases.

HD 71866 is an ideal candidate for Magnetic Doppler Imaging. However, more observations will be required to complete phase coverage. As with all the stars with strong linear polarisation in this study many individual lines show clear signatures.

\subsection{HD 112413 - $\alpha^2$ CVn}

 $\alpha^2$~CVn is a fairly sharp-lined A0p star with a moderately strong magnetic field. We obtained Stokes $V, Q$ and $U$ profiles at 24 rotational phases.  The projected rotational velocity ($v \sin i$) was determined to be $17 \pm 1$ ~\kms\, which agrees within uncertainty with the values adopted by Wade et al. (2000b) and Kochukhov and Wade (2010).

 All measurements have been phased according to the ephemeris of Farnsworth (1932):
 
 \begin{equation}{\rm JD}=2419869.720 + 5\fd 46939\cdot{\rm E}. \end{equation}
 
\begin{figure*}
\begin{center}
   \includegraphics[width=0.85\textwidth]{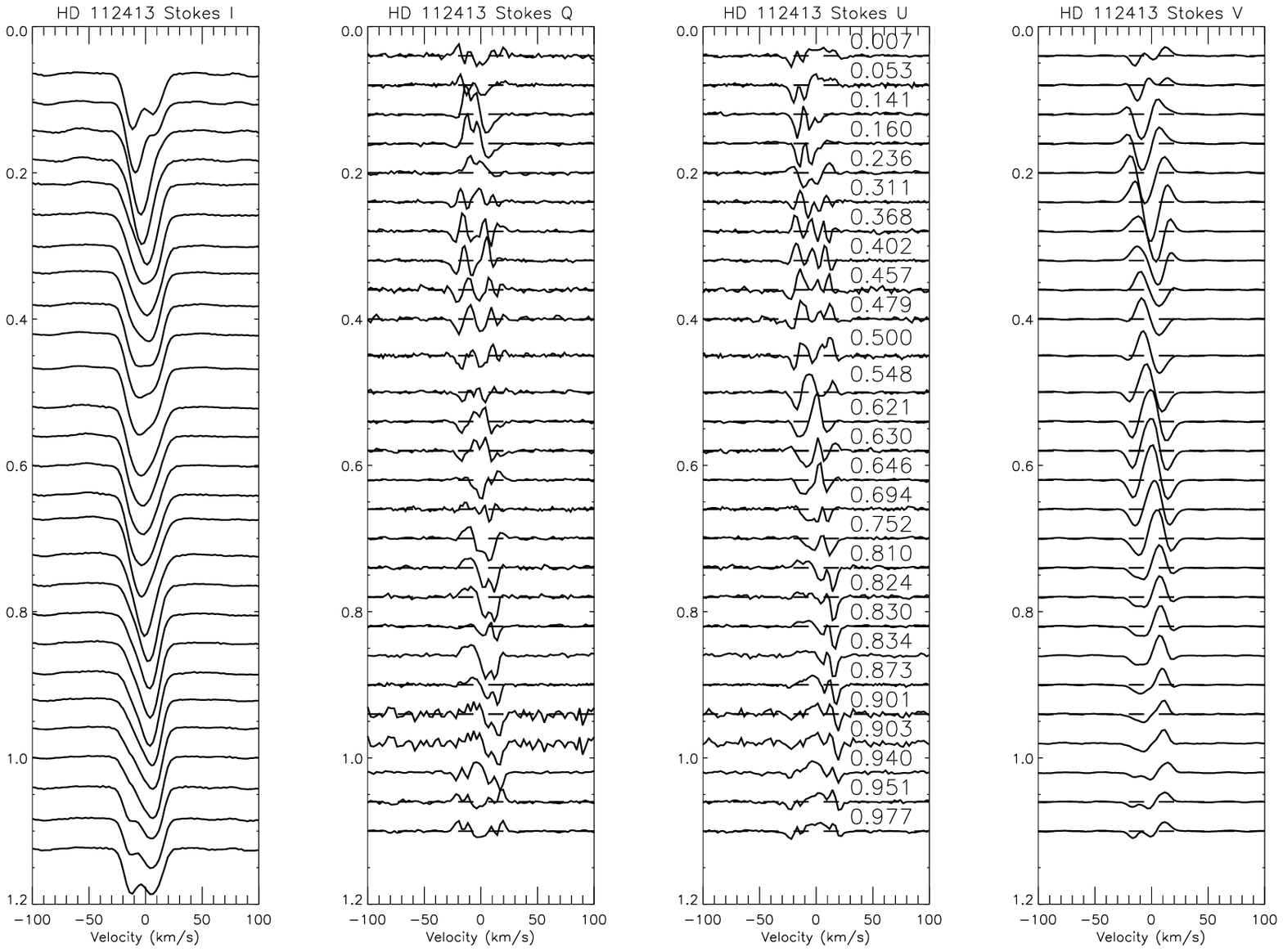}
      \includegraphics[width=0.60\textwidth, angle=90]{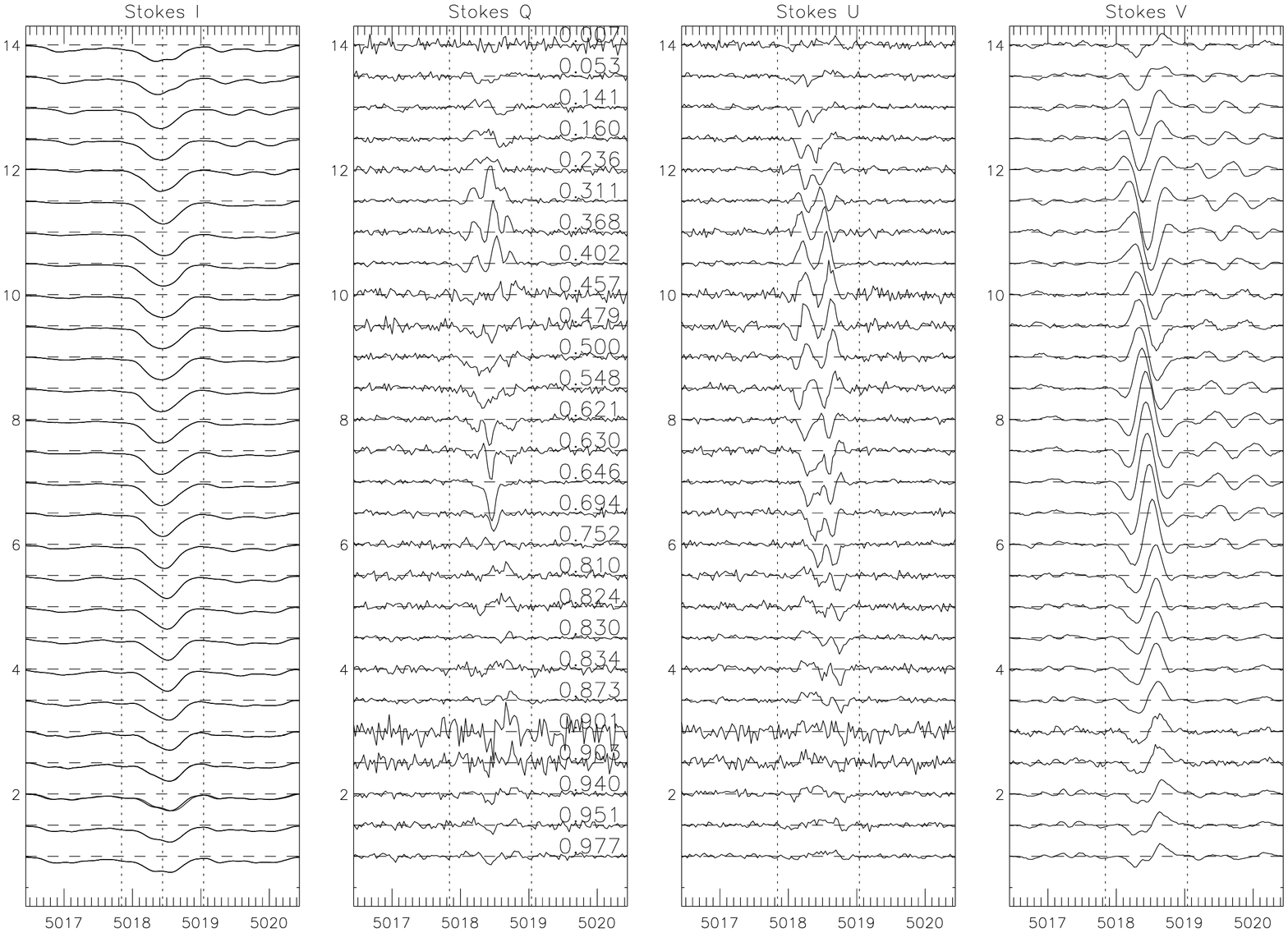}
 \caption{Top: Variation of Stokes $I, Q, U$ and $V$ LSD profiles for HD 112413. Rotational phase is represented on the vertical axis. A scaling factor of 25 was used for both the Stokes $Q$ and $U$ observations and a scaling factor of $2$ was used for Stokes $V$. Bottom:  Variation of Stokes $I, Q, U$ and $V$ profiles of HD 112413 in the Fe~{\sc ii} 5018 line  (scaled by a factor of 30 for Stokes $Q$ and $U$ and 7 for Stokes $V$). }
\label{112413lines}
\end{center}
\end{figure*}

\begin{figure}
\begin{center}
   \includegraphics[width=0.55\textwidth]{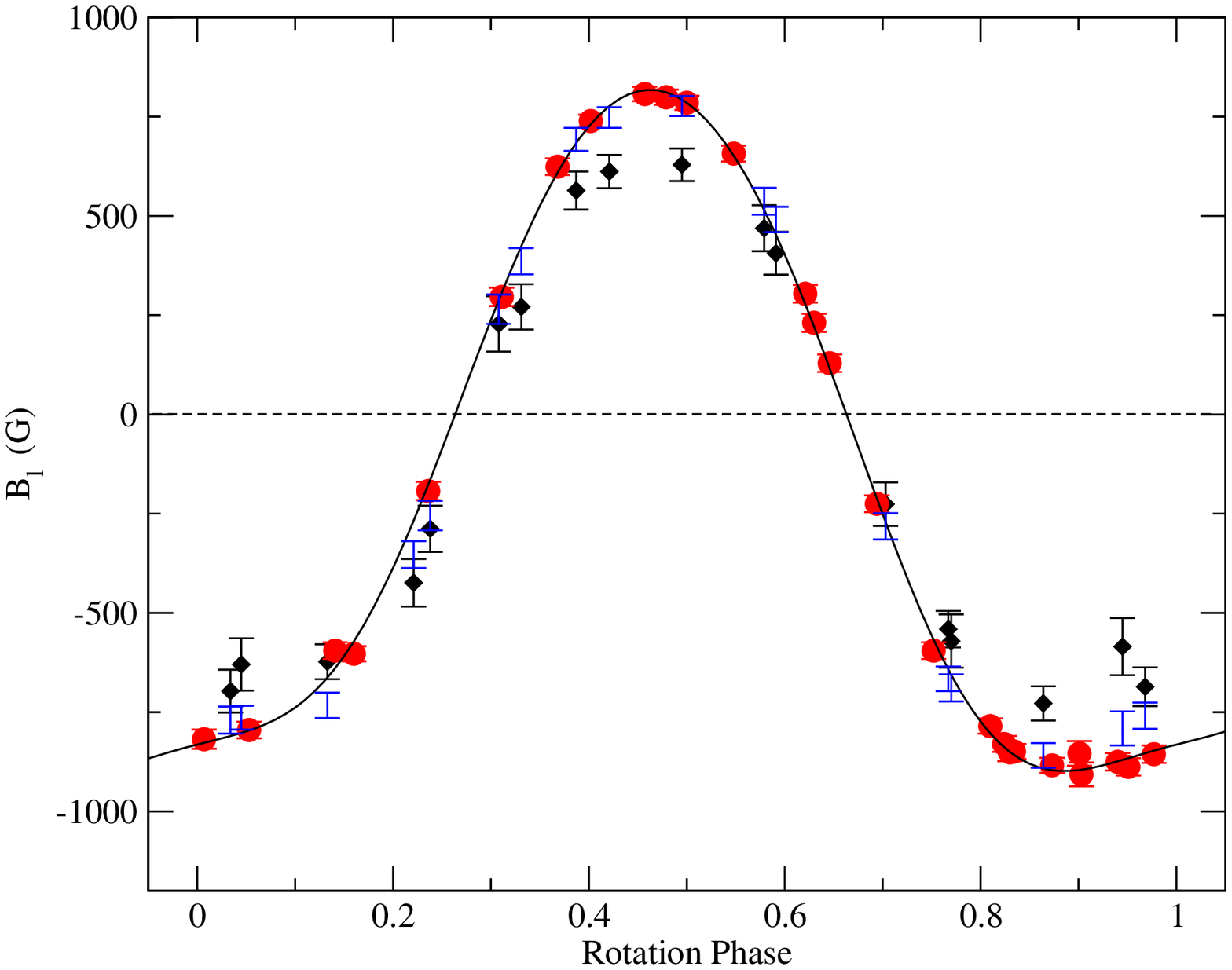}
 \caption{Longitudinal field measurements for HD 112413 obtained with ESPaDOnS/NARVAL (shown by filled circles), compared with those obtained by Wade et al. (2000b) with MuSiCoS (shown by filled diamonds). The measurements shown without symbols are the result of a re-anaylsis of the original Wade et al. (2000a) observations with the new line mask. The improvement in data quality is evidenced by the smaller error bars associated with the ESPaDOnS/NARVAL measurements. A 4th order Fourier fit to the ESPaDOnS/NARVAL data is given by the solid curve. }
 \label{2cvnlong}
\end{center}

\begin{center}
   \includegraphics[width=0.52\textwidth]{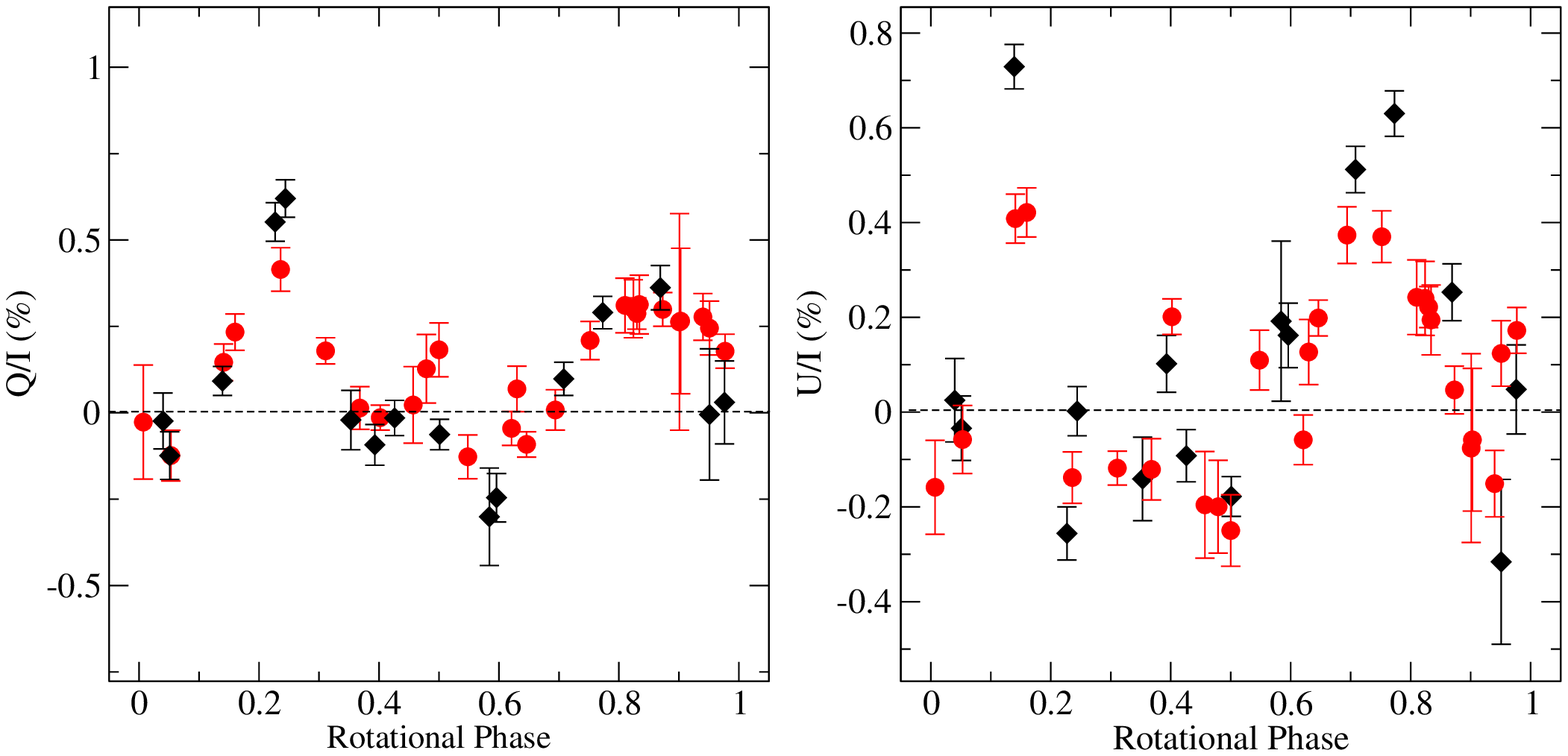}
   \caption{Net linear polarisation (Stokes $Q$ and $U$) measurements for HD 112413 obtained with ESPaDOnS/NARVAL (shown by filled circles), compared with those obtained by Wade et al. (2000b) with MuSiCoS (shown by filled diamonds).  A good agreement can be seen, confirming consistency between the instruments and the improvement in data quality is evidenced by the smaller error bars associated with the ESPaDOnS/NARVAL measurements. }
\label{stokes2cvn}
\end{center}
\end{figure}

Clear signatures and strong variability can be seen in Stokes $Q$ and $U$ in individual spectral lines of this star. Profile variations of the Fe~{\sc ii} $\lambda$ 5018 line and of LSD profiles are shown in  Fig. \ref{112413lines}.  Very strong variation in Stokes $I$ can be seen in the  Fe~{\sc ii} lines at 4923, 5018, 5169 \AA\.   . Both $B_{\ell}$ and net linear polarisation measurements have been obtained for each phase and are reported in Table \ref{bltable}.   The longitudinal field values  are compared to the results of Wade et al. (2000b) in Fig.  \ref{2cvnlong}.  To get a good agreement between the two epochs of data required a re-analysis of the Wade et al. (2000b) data, using the updated and abundance specific line mask (as used for the ESPaDOnS/NARVAL data). It is likely that this star is in a temperature regime where measurements are more sensitive to the line mask chosen or that the abundances are sufficiently peculiar to cause a difference. In addition the new measurements are much more precise. The average uncertainty on the longitudinal field measurement was 27 gauss, with a Fourier fit to the longitudinal field curve yielding a reduced $\chi^2$ of 0.83, with an order of fit of 3.

\begin{figure}
\begin{center}
   \includegraphics[width=0.50\textwidth]{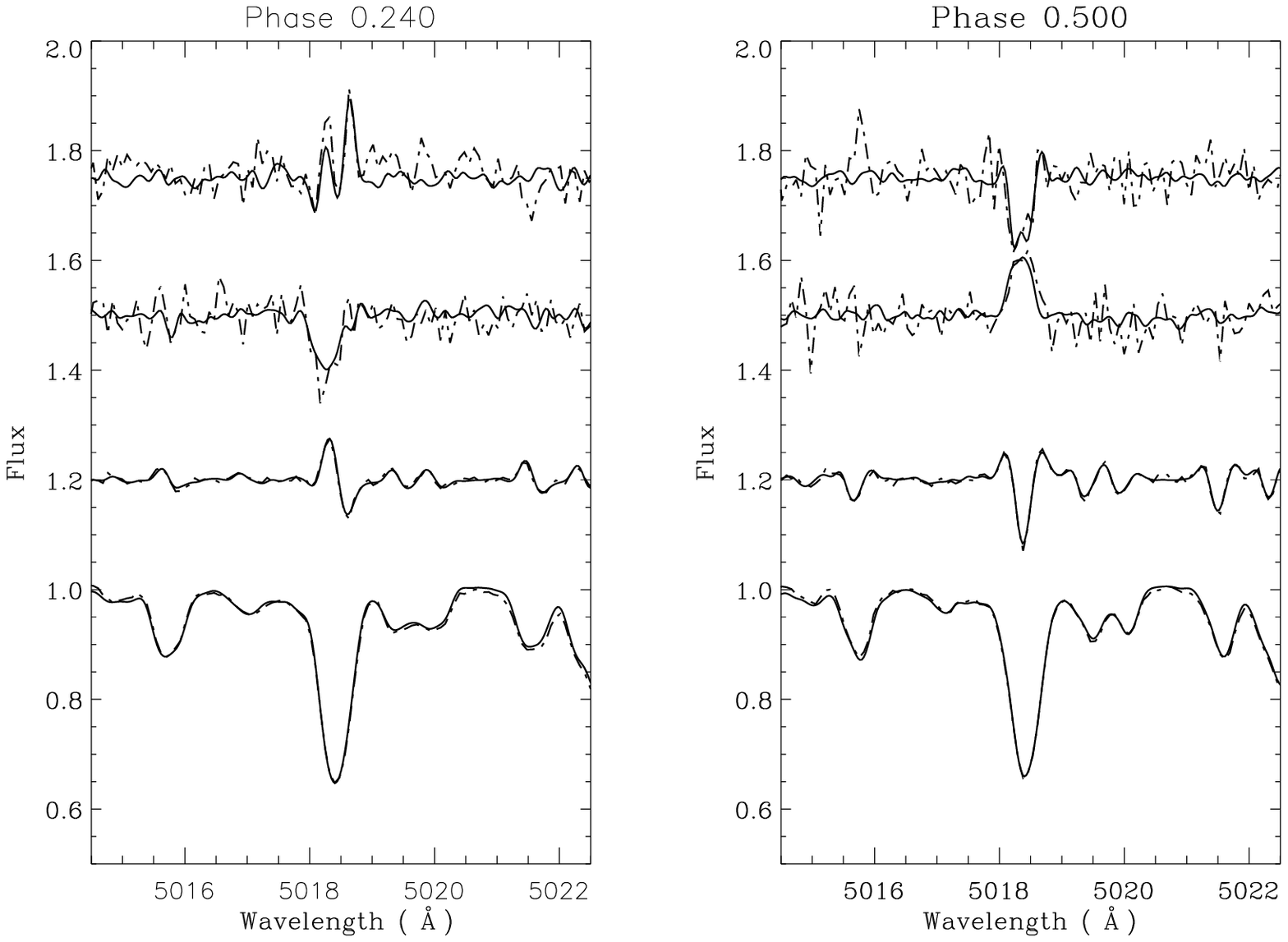}
 \caption{Comparison between two phases of the ESPaDOnS/NARVAL convolved to a resolution of R=35000 and the MuSiCoS data at a similar phase for HD 112413 (MuSiCoS data dashed lines, ESPaDOnS/NARVAL solid lines).}
 \label{2cvn35000}
\end{center}
\end{figure}

The net linear polarisation as a function of phase is shown in Fig \ref{stokes2cvn}. Both Stokes $Q$ and $U$ measurements were compared with the measurements of Wade et al. (2000b),  showing very good agreement at most phases. For certain phases, the ESPaDOnS/NARVAL data for  $\alpha^2$ CVn have been convolved to the same approximate resolution as MuSiCoS (R=35000) and then compared to MuSiCoS as described for HD 32633. Again good agreement can be seen in individual lines between the two data sets in Fig. \ref{2cvn35000}.

\subsection{HD 118022 - 78 Vir}

78 Vir is a sharp-lined A1p star with a moderately strong magnetic field. We obtained Stokes $V, Q$ and $U$ profiles for 5 rotational phases.  The projected rotational velocity ($v\sin i$) was determined to be $13 \pm 1$ ~\kms\, which agrees within uncertainty with the value adopted by Wade et al. (2000b).   All measurements have been phased according to the ephemeris of Preston (1969): \begin{equation}{\rm JD}=2434816.90 + 3\fd 7220\cdot{\rm E}.,\end{equation}

\begin{figure*}
\begin{center}
   \includegraphics[width=0.85\textwidth]{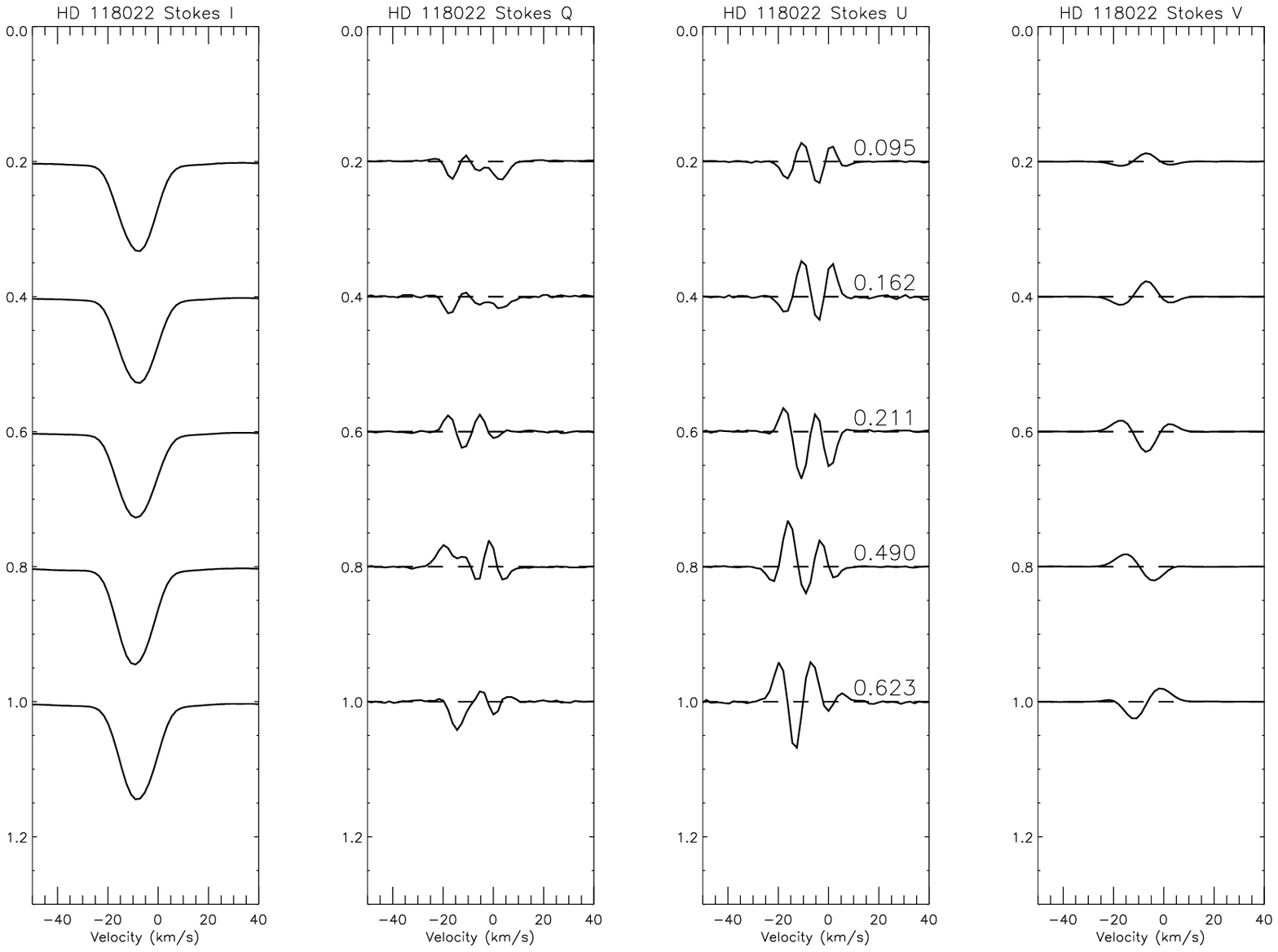}
    \includegraphics[width=0.60\textwidth, angle=90]{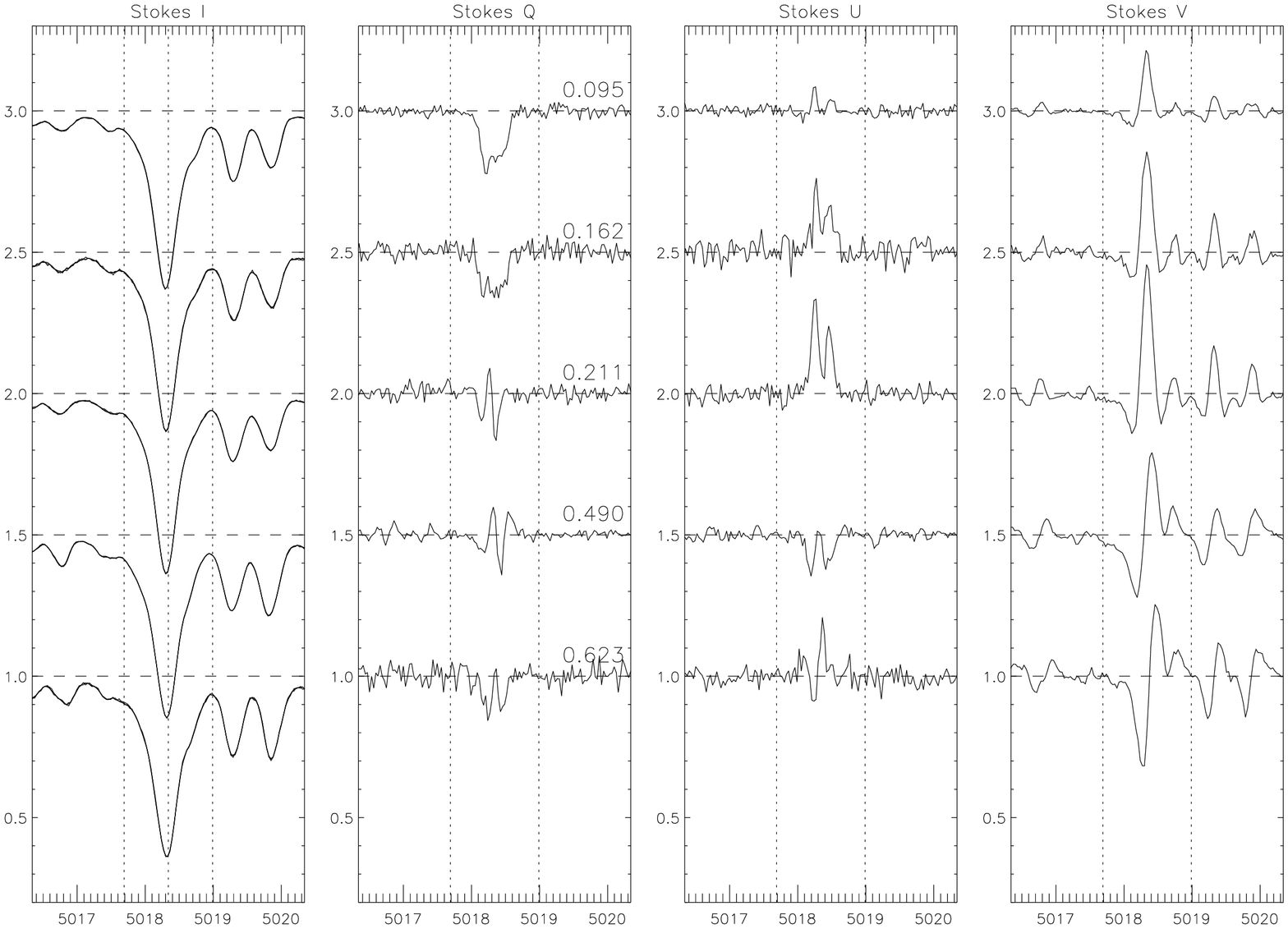}
 \caption{Top: Variation of Stokes $I,Q,U$ and $V$ LSD profiles for  HD 118022. Rotational phase represented on the vertical axis. With a scaling factor of 25 used for both the Stokes $Q$ and $U$ observations and no scaling on Stokes $V$ . Bottom: Variation of Stokes $I,Q,U$ and $V$ profiles for  HD 118022 in the 5018 Fe~{\sc ii} line (scaled by a factor of 10 for Stokes $Q$ and $U$ and 3 for Stokes $V$).}
\label{118022lsd}
\end{center}
\end{figure*}

Clear signatures and strong variability can be seen in Stokes $Q$ and $U$  in the individual lines of this star, as shown for  the Fe~{\sc ii} $\lambda$ 5018 line and LSD profiles in Fig. \ref {118022lsd}).  Stokes $I$ appears to vary only slightly between phases,  with only small changes in the  Fe~{\sc ii} lines at 4923, 5018 and 5169    \AA\  .

HD 118022 was studied by Khalack and Wade (2006) who constrained the global magnetic field of the star and determined the abundance distributions of titanium and chromium. This was performed using the magnetic charge distribution method (MCD) (Gerth et al. 1997; Khalack et al. 2001) and lower resolution MuSiCoS spectra, which limits the number of lines that could be modeled.  This star is an excellent target for MDI, and further observations are warranted to supplement the currently rather sparse phase coverage.

\begin{figure}
\begin{center}
   \includegraphics[width=0.55\textwidth]{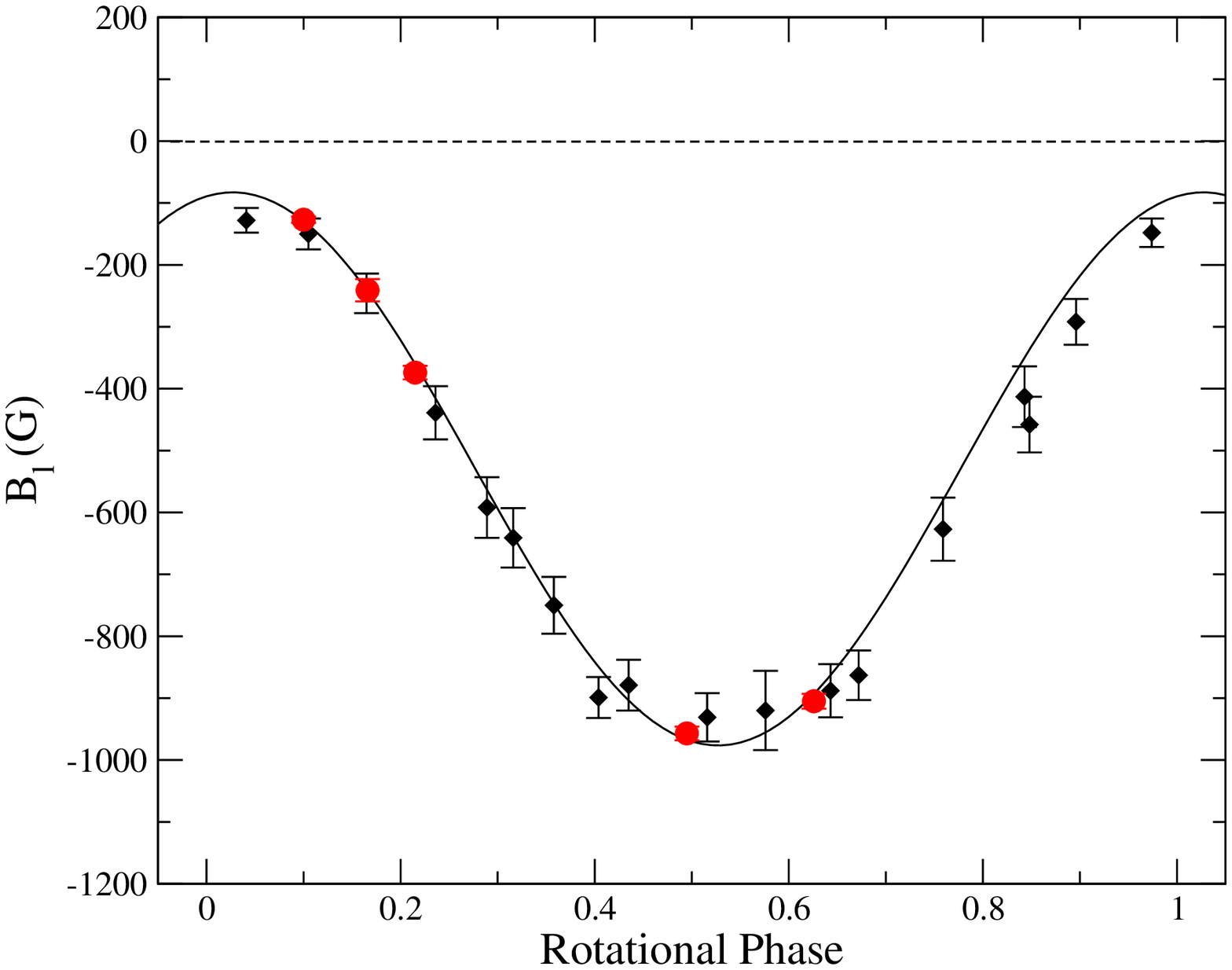}
 \caption{Longitudinal field measurements for HD 118022 obtained with ESPaDOnS/NARVAL (shown by filled circles), compared with those obtained by Wade et al. (2000b) with MuSiCoS (shown by filled diamonds). A  1st order Fourier fit to the ESPaDOnS/NARVAL data is given by the solid curve. }
\label{118022long}
\end{center}

\begin{center}
   \includegraphics[width=0.54\textwidth]{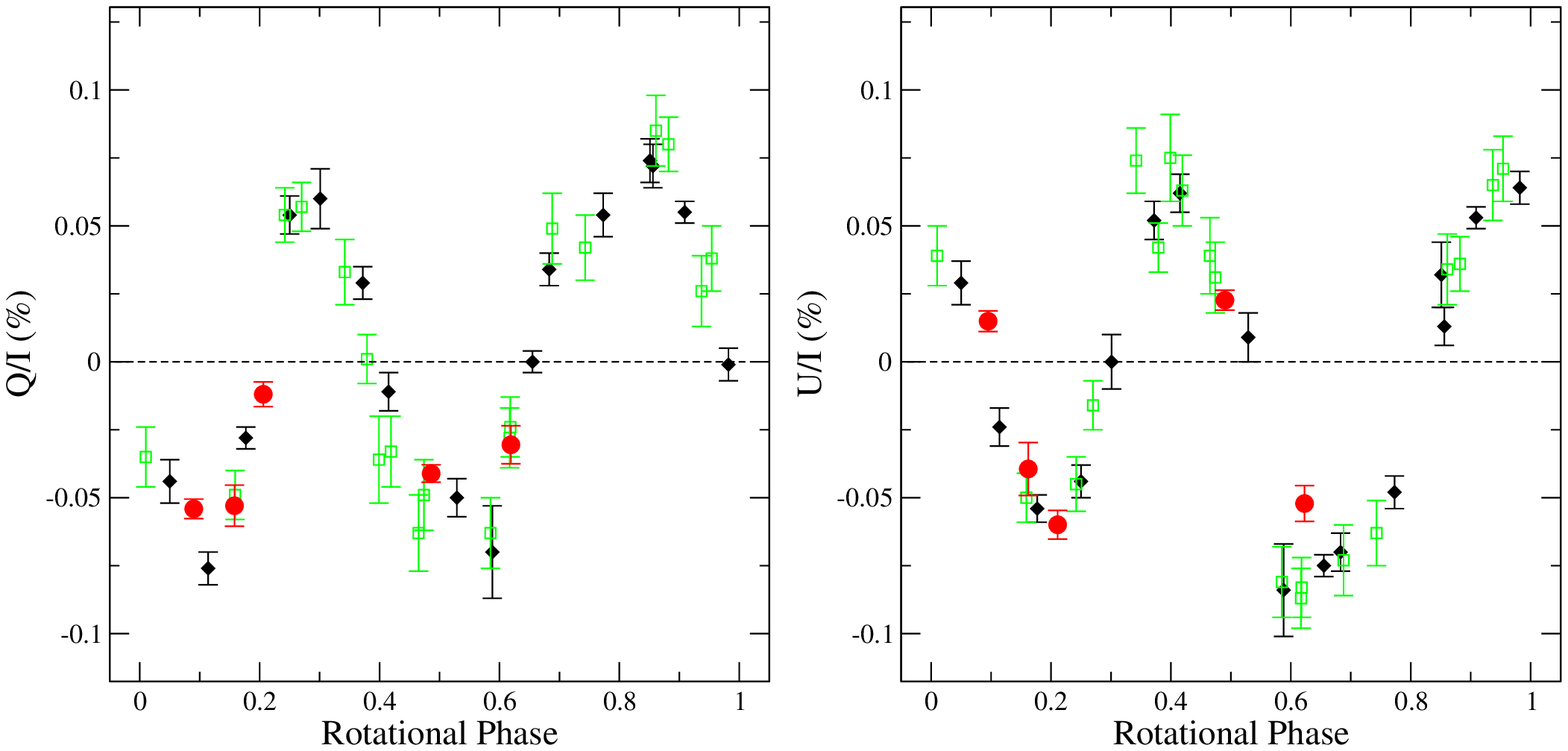}
   \caption{Net linear polarisation (Stokes $Q$ and $U$) measurements for HD 118022 obtained with ESPaDOnS/NARVAL (shown by filled circles, scaled by 0.10 consistent with Wade et al. (2000b)), compared with those obtained by Wade et al. (2000b) with MuSiCoS (shown by filled diamonds) and Leroy et al. (1995) broadband linear polarisation measurements (shown by open squares). Good agreement can be seen, confirming consistency between the instruments. The improvement in data quality is evidenced by the smaller error bars associated with the ESPaDOnS/NARVAL measurements.}
\label{stokes78vir}
\end{center}
\end{figure}

Both $B_\ell$ and net linear polarisation measurements have been obtained for each phase and are reported in Table \ref{bltable} .  The longitudinal field values obtained by Wade et al. (2000b) are compared to this work in Fig.  \ref{118022long}. The average uncertainty on the longitudinal field measurements was 12 gauss. The netlinear polarisation as a function of phase is shown in Fig \ref{stokes78vir} and compared with the measurements of Wade et al. (2000b) and Leroy (1995), with good agreement achieved at most phases. The netlinear polarisation measurements were scaled by the same factor (0.10) as reported by Wade et al. (2000b)

\section{Discussion and Conclusions}
The goal of this project was to obtain a new data set in all four Stokes parameters for a selection of well studied Ap stars, with the ultimate aim to map these stars using Magnetic Doppler Imaging. The target list contained stars which span a large part of the parameter space of interest, with sufficient signal-to-noise ratio to not only greatly improve on the previous observations, but to also be suitable for MDI mapping. The final selection was based primarily on stars already identified by Wade et al. (2000a) as promising candidates for such study. 

Early on in the project it was clear that both ESPaDOnS and NARVAL have greatly improved the level of detail at which Ap stars can be studied.  The resulting dataset obtained for this study is far superior to that obtained previously with MuSiCoS, and represents some of the highest resolution phase-resolved observations of Ap stars acquired to date. This data set has more individual lines showing variation, much improved signal-to-noise and smaller error bars associated with measurements of the longitudinal field and net linear polarisation.  The new data have been shown to be consistent with the previous observations of Wade et al. (2000a) and also those of Leroy at al. (1995), with most targets agreeing well between the different epochs. 

Surprisingly we found  that even when the data are of such high signal-to-noise and when the magnetic fields are strong,  the LSD analysis is sensitive to the normalisation and the measured magnetic field is rather sensitive to the integration ranges chosen, with variations of sometimes on the order of 100 gauss with very small changes of the integration range.  A key conclusion of this work is that even with such high-quality data,  extreme care must still be taken with all stages of analysis to ensure consistent results at this level of precision.  

Although crosstalk was originally a concern, through a series of experiments we have shown that it is at a level which should not have a significant impact on the results. By using observations of $\gamma$ Equ, we have seen that the highest level of crosstalk in Stokes $Q$ is still within the noise and slightly above the noise in Stokes $U$ (around the 5 \% level.). It should be noted that even at these levels, this effect will be less significant in the broader-lined stars studied here. 

Considering this, we believe the other uncertainties associated with the analysis techniques have a greater effect: normalisation, blending, line masks used for LSD  and the choice of integration ranges used for longitudinal field measurements.   But as regards the final impact of the crosstalk on MDI mapping,  this will be discussed and addressed in a future paper (Silvester et al. in preparation).  With these high quality observations, we suspect that the limitations for mapping will in fact come not from the data (with strong Stokes $Q$ and $U$ signatures seen in many individual lines),  but the ability to deal with line blends within the MDI code.

An important result of this study is the confirmed stability of the global properties of the magnetic fields of these Ap stars. Over multiple epochs of observations the fields have remained constant, with little variation in both longitudinal field and linear polarisation measurements. In some cases measurements separated by over a decade still agree with each other within the uncertainties. By comparing MDI maps produced from the new observations of $\alpha^2$ CVn with those of Kochukhov and Wade (2010), we can potentially test for evolution of the field geometry which may occur on small spatial scales.

Considering the longitudinal magnetic field measurements, agreement was found between the new measurements and those of Wade et al. (2000b),  with the exception of HD 71866 and $\alpha^2$ CVn which showed a slight discrepancy.  By re-reducing the MuSiCoS data with the new masks, the observations were brought into agreement.  Whilst the general shapes of the netlinear polarisation variations were in agreement, one needed to invoke free parameters such as scaling which is consistent with what was described in Wade et al. (2000b). 

For any study which requires observations over multiple semesters, it is imperative that the instrument is stable and consistent throughout the campaign. Both NARVAL and ESPaDOnS proved to be very stable instruments,  with resolution and signal-to-noise being constant over the 4 years of data. Indeed we have also shown the two instruments are consistent with one another with close to identical result from similar phases.  These facts demonstrate that ESPaDOnS and NARVAL are both very capable instruments, well suited to high-resolution four Stokes measurements of magnetic stars over multiple year time scales.  

\begin{figure*}
\begin{center}
   \includegraphics[width=0.85\textwidth]{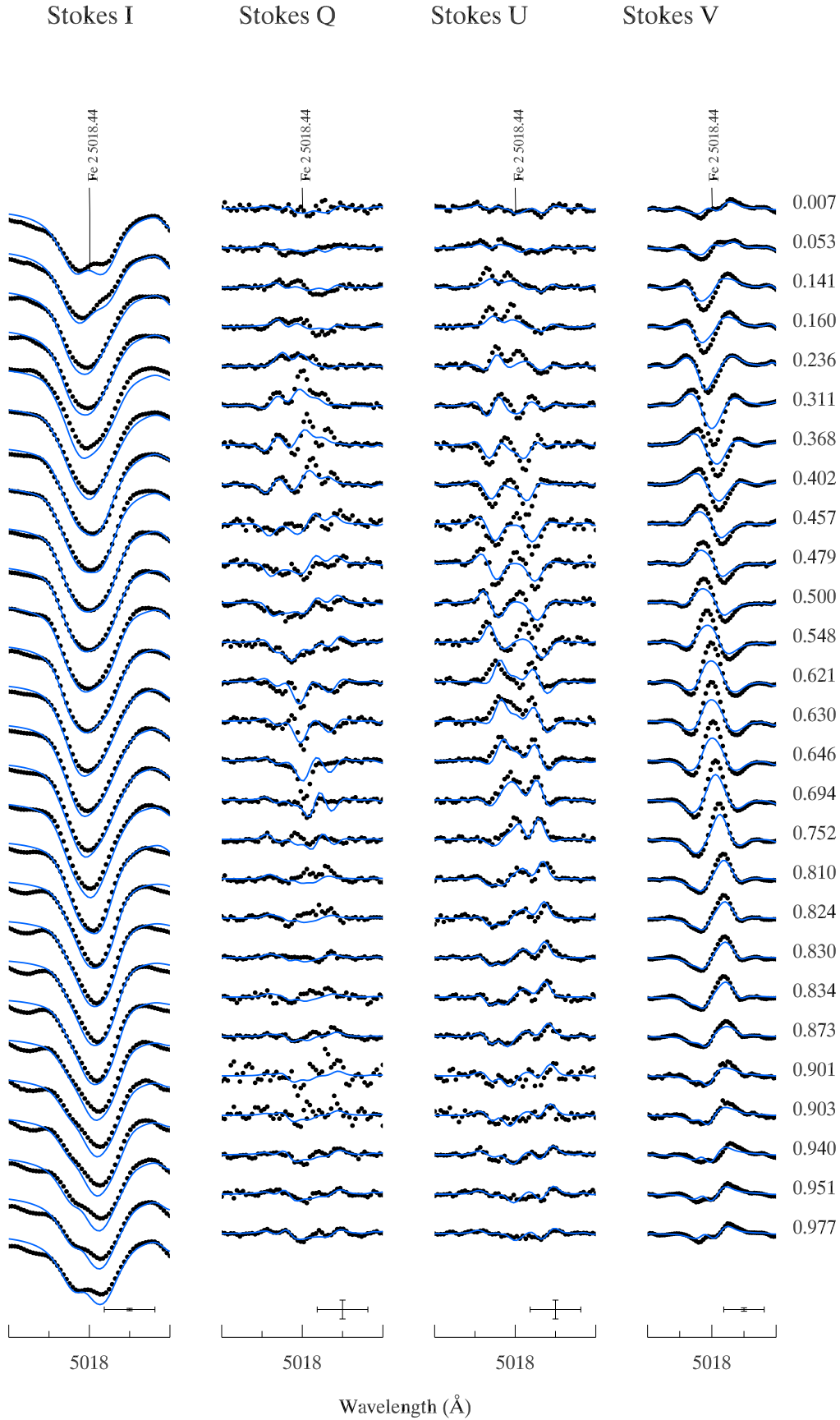}
 \caption{Comparison betwen the new ESPaDOnS/NARVAL data for the Fe~{\sc ii} 5018 line (shown by black points) with the final model profiles adopted in the mapping of HD 112413 (by solid blue lines) by Kochukhov \& Wade (2010) }
 \label{2cvnmapfit}
\end{center}
\end{figure*}

One of the targets ($\alpha^2$ CVn) has already been mapped with MDI using MuSiCoS data by Kochukhov \& Wade (2010).  HD 112413 is an ideal star for determining how much of an improvement the new polarimetric data could give to MDI mapping. To quantify this improvement and to further confirm consistency,  the new observations of $\alpha^2$ CVn were compared with the profiles predicted by the model of Kochukhov \& Wade (2010).  As shown in Fig. \ref{2cvnmapfit}, good general agreement between the new observations and the MuSiCoS-derived model is observed. However, the new profiles show more complexity than was present in the MuSiCoS data, which is not fully reproduced in the current model and would likely require a more complex magnetic field distribution.  In addition, the Stokes V profile amplitude is significantly underestimated by the model at a number of phases.

The complete phase coverage of both 49 Cam (Silvester et al. in preparation), $\alpha^2$ CVn and HD 32633 will allow the completion of  4 Stokes parameter MDI maps for these stars, doubling the number of Ap stars studied using this technique.  Out of the remaining targets  HD 4778, HD 71866 and HD 118022 would also be a worthwhile candidates for MDI; conversely HD 40312 has small linear polarisation signatures, relative to the noise in their spectra,  making them less suitable for MDI analysis.  The mapping of 49 Cam is well underway and the results will be presented in Paper II (Silvester et al. in preparation).

\section*{Acknowledgments} 
OK is a Royal Swedish Academy of Sciences Research Fellow supported by grants from the Knut and Alice Wallenberg Foundation and the Swedish Research Council.
GAW and DAH acknowledge support from the Natural Science and Engineering Research Council of Canada in the form of Discovery Grants.

\bsp

\begin{longtable}{rlcccccc}
\caption{Log of spectropolarmetric observations, where in the 2nd to last column, an observation is denoted by a star and a missing observation denoted by a dash. In the final column E $=$ ESPaDOnS and N $=$ NARVAL. }\\\\
\hline\hline
Object & Date          & JD           & Phase & $t_{exp}$ & S/N ratio      & Stokes & Instrument \\
       & (UT)          &  (2450000 +) &       &  (s)      &  (spectral pixel$^{-1}$)  &  $V$  $Q$ $U$  &                   \\
\endfirsthead
\caption{continued.}\\
\hline\hline
Object & Date          & JD           & Phase & $t_{exp}$ & S/N ratio      & Stokes & Instrument \\
       & (UT)          &  (2450000 +) &       &  (s)      &  (spectral pixel$^{-1}$)  &  $V$  $Q$ $U$  &                   \\ \hline
\endhead
\hline
\endfoot
\hline
\multicolumn{8}{c}{HD 4778}\\
\hline\noalign{\smallskip}
HD 4778 &  2006 Nov 30 & 4070.746 &  0.423   & 3200/3200/3200 &  152/709/638 & $\star \star \star$   & E  \\
        &  2006 Dec 04 & 4074.734 &  0.980   & 2600/2600/2600 &  281/667/858 & $\star \star \star$  & E \\
        &  2006 Dec 06 & 4076.744 &  0.764   & 2000/2000/2000 &  163/647/708 &  $\star \star \star$ & E\\
        &  2006 Dec 19 & 4089.251 &  0.662   & 3200/3200/3200 &  129/641/552 & $\star \star \star $ & N  \\
        &  2006 Dec 20 & 4090.244 &  0.050   & 3200/3200/3200 &  220/507/598 &  $\star \star \star$ & N \\
        &  2006 Dec 21 & 4091.260 &  0.446   & 3200/3200/3200 &  158/585/639 &  $\star \star \star$ & N \\
        &  2008 Jan 09 & 4475.394 &  0.366   & 2800/-/2800    & 137/-/585    &  $ \star - \star$ & N \\ 
 \hline
\multicolumn{8}{c}{$\theta$~Aur (HD 40132)}\\
\hline\noalign{\smallskip}
HD 40312 &  2006 Dec 04  & 4074.925 & 0.587  &  140/140/140 &  1231/1303/1350 & $\star \star \star$   & E  \\ 
         &  2006 Dec 05  & 4076.147 & 0.924  &     52/52/52 &  778/781/787 & $\star \star \star$   & E  \\
         &  2008 Jan 07  & 4473.417 & 0.710  &  240/240/240 &  783/835/836 & $\star \star \star$   & N  \\
         &  2008 Jan 09  & 4475.268 & 0.222  &  240/240/240 &  757/865/1056 & $\star \star \star$   & N  \\ 
         &  2008 Jan 22  & 4489.002 & 0.017  &  240/240/240 &  1377/1225/1601 & $\star \star \star$   & E  \\
         &  2008 Jan 23  & 4490.012 & 0.296  &  160/160/160 &  1296/906/1234 & $\star \star \star$   & E  \\
         &  2008 Jan 25  & 4491.795 & 0.789  &  160/160/160 &  1285/1501/1468 & $\star \star \star$   & E  \\
\hline
\multicolumn{8}{c}{HD 32633}\\
\hline\noalign{\smallskip}
HD 32633 &   2006 Nov 29       & 4070.098 & 0.972 & 3200/3200/3200 & 588/408/247  & $\star \star \star$   & E  \\  
       &   2006 Nov 30       & 4070.830 & 0.086 & 3200/-/-       & 646/- /-     & $\star - -$ & E  \\ 
       &   2006 Dec 04       & 4074.848 & 0.710 & 3200/3200/3200 & 789/806/776  & $\star \star \star$   & E  \\  
       &   2006 Dec 05       & 4075.094 & 0.749 & 3200/3200/3200 & 823/754/637  & $\star \star \star$   & E  \\ 
       &   2006 Dec 06       & 4076.085 & 0.903 & 3200/3200/3200 & 820/755/696  & $\star \star \star$   & E  \\ 
       &  2006 Dec 12        & 4082.577 & 0.916 & 1600/1600/1600 & 417/430/427 & $\star \star \star$   & N  \\ 
       &   2006 Dec 13       & 4083.558 & 0.071 & 3200/3200/3200 & 379/398/426 & $\star \star \star$   & N  \\ 
       &   2006 Dec 15       & 4085.537 & 0.379 & 3200/3200/3200 & 554/418/520 & $\star \star \star$   & N  \\ 
       &   2006 Dec 17       & 4087.528 & 0.689 & 3200/3200/3200 & 359/414/360 & $\star \star \star$   & N  \\ 
       &   2006 Dec 19       & 4089.430 & 0.984 & 3200/3200/3200 & 556/553/564 & $\star \star \star$   & N  \\ 
       &   2006 Dec 20       & 4090.419 & 0.138 & 3200/3200/3200 & 508/455/471 & $\star \star \star$   & N  \\ 
       &   2006 Dec 21       & 4091.434 & 0.296 & 3200/3200/3200 & 578/579/593 & $\star \star \star$   & N  \\      
       &   2008 Jan 09       & 4475.482 & 0.017 & 3200/3200/3200 & 550/538/527  & $\star \star \star$   & N  \\ 
        &   2008 Jan 24       & 4490.736 & 0.390 & 2400/2400/2400 & 763/788/785  & $\star \star \star$   & E  \\ 
       &   2008 Jan 25       & 4491.741 & 0.546 & 2600/2600/2600 & 755/766/745  & $\star \star \star$   & E  \\ 
       &   2008 Jan 26       & 4492.729 & 0.700 & 2000/2000/2000 & 672/752/726  & $\star \star \star$   & E  \\  
       &   2008 Jan 27       & 4493.730 & 0.855 & 2000/2000/2000 & 520/463/524  & $\star \star \star$   & E  \\ 
       &   2009 Jan 17       & 4849.432 & 0.181 & 3200/3200/3200 & 578/611/618  & $\star \star \star$   & N  \\ 
       &   2009 Oct 10       & 5115.055 & 0.481 & 2000/2000/2000 & 659/661/695  & $\star \star \star$   & E  \\ 
       &   2010 Feb 27       & 5254.743 & 0.209 & 2000/2000/2000 & 720/738/737  & $\star \star \star$   & E  \\ 
       &   2010 Mar 08       & 5263.774 & 0.613 & 2000/2000/2000 & 612/600/580  & $\star \star \star$   & E  \\
\hline
 \multicolumn{8}{c}{49 Cam (HD 62140)}\\
\hline\noalign{\smallskip}
HD 62140 &  2006 Dec 10       & 4080.681 &         0.381  & 2800/2800/2800 & 685/709/735 & $\star \star \star$   & N  \\ 
         &  2006 Dec 12       & 4082.653 &         0.841  & 3200/3200/3200 & 673/757/764 & $\star \star \star$   & N  \\ 
         &  2006 Dec 13       & 4083.684 &         0.081  & 3200/3200/3200 & 518/618/599 & $\star \star \star$   & N  \\ 
         &  2006 Dec 14       & 4084.595 &         0.294  & 3200/3200/3200 & 666/505/586 & $\star \star \star$   & N  \\ 
         &  2006 Dec 17       & 4087.715 &         0.017  & 1600/1600/1600 & 148/200/182 & $\star \star \star$   & N  \\ 
         &  2006 Dec 19       & 4089.557 &         0.451  & 3200/3200/3200 & 668/628/605 & $\star \star \star$   & N  \\ 
         &  2006 Dec 20       & 4090.542 &         0.681  & 3200/3200/3200 & 656/680/664 & $\star \star \star$   & N  \\ 
         &  2006 Dec 21       & 4091.557 &         0.918  & 3200/3200/3200 & 774/764/743 & $\star \star \star$   & N  \\ 
         &  2008 Jan 09       & 4475.706 &         0.520  & 2400/2000/2000 & 612/571/588 & $\star \star \star$   & N  \\ 
         &  2008 Jan 24       & 4491.003 &         0.089  & 2600/2600/2600 &957/1041/1029 & $\star \star \star$  & E  \\ 
         &  2008 Jan 27       & 4493.989 &         0.786  & 2600/2600/2600 & 709/836/692 & $\star \star \star$   & E  \\ 
         &  2008 Mar 24       & 4550.724 &         0.026  & 2000/2000/2000 & 951/929/956 & $\star \star \star$   & E  \\ 
         &  2008 Mar 25       & 4551.723 &         0.260  & 2000/2000/2000 & 939/944/936 & $\star \star \star$   & E  \\ 
         &  2008 Apr 24       & 4581.349 &         0.174  & 3200/3200/3200 & 605/740/655 & $\star \star \star$   & N  \\ 
         &  2009 Jan 14       & 4846.625 &         0.056  & 3200/3200/3200 & 465/363/308 & $\star \star \star$   & N  \\  
         &  2009 Jan 15       & 4847.620 &         0.288  & 3200/3200/3200 & 709/723/735 & $\star \star \star$   & N  \\ 
         &  2009 Sep 10       & 5085.125 &         0.677  & 2000/2000/2000 & 758/853/817 & $\star \star \star$   & E  \\ 
         &  2010 Feb 23       & 5250.873 &         0.348  & 2000/2000/2000 & 860/782/809 & $\star \star \star$   & E  \\ 
         &  2010 Feb 24       & 5251.972 &         0.604  & 2000/2000/2000 & 837/800/841 & $\star \star \star$   & E  \\ 
\hline
\multicolumn{8}{c}{HD 71866}\\
\hline\noalign{\smallskip}
HD 71866 & 2006 Dec 15  & 4085.756 & 0.723 & 2400/2400/2400 & 527/544/534 & $\star \star \star$   & N  \\ 
         & 2006 Dec 19  & 4089.756 & 0.312 & 2000/2200/2200 & 507/477/498 & $\star \star \star$   & N  \\ 
         & 2006 Dec 20  & 4090.751 & 0.457 & 2600/2600/2600 & 503/466/513 & $\star \star \star$   & N  \\ 
         & 2006 Dec 21  & 4091.748 & 0.603 & 2800/2800/2800 & 604/649/645 & $\star \star \star$   & N  \\ 
         & 2007 Mar 07  & 4167.957 & 0.819 & 2000/2000/2000 & 738/748/748 & $\star \star \star$   & E  \\ 
         & 2007 Mar 08  & 4168.975 & 0.969 & 2000/2000/2000 & 778/779/549 & $\star \star \star$   & E  \\ 
         & 2007 May 06  & 4227.424 & 0.554 & 3200/3200/3200 & 549/537/518 & $\star \star \star$   & N  \\ 
         & 2008 Jan 07  & 4473.530 & 0.746 & 3000/3000/3000 & 495/423/418 & $\star \star \star$   & N  \\ 
         & 2008 Jan 09  & 4475.566 & 0.056 & 2800/2800/2800 & 522/612/633 & $\star \star \star$   & N  \\ 
         & 2008 Jan 26  & 4492.827 & 0.592 & 2000/2000/2000 & 699/734/651 & $\star \star \star$   & E  \\ 
         & 2008 Jan 27  & 4493.829 & 0.740 & 2000/2000/2000 & 448/453/530 & $\star \star \star$   & E  \\ 
         & 2008 Apr 25  & 4582.411 & 0.757 & 3000/3000/3000 & 600/444/524 & $\star \star \star$   & N  \\ 
         & 2009 Jan 10  & 4842.734 & 0.040 & 2400/2400/2400 & 556/579/570 & $\star \star \star$   & N  \\ 
         & 2010 Jan 28  & 5224.799 & 0.228 & 2000/2000/2000 & 685/713/708 & $\star \star \star$   & E  \\ 
\hline
  \multicolumn{8}{c}{HD 112413}\\
\hline\noalign{\smallskip}
$\alpha^2$ CVn & 2006 Dec 04 & 4075.158 &         0.977 & 240/240/240 & 1536/1566/1546 & $\star \star \star$   & E  \\ 
       & 2006 Dec 05         & 4076.159 &         0.160  & 120/120/120 & 1187/1190/1184 & $\star \star \star$   & E  \\ 
       & 2006 Dec 12         & 4082.760 &         0.368  & 240/240/240 & 967/910/867 & $\star \star \star$   & N  \\ 
       & 2006 Dec 19         & 4089.663 &         0.630   & 240/240/240 & 811/867/828 & $\star \star \star$   & N  \\ 
       & 2006 Dec 20         & 4090.647 &         0.810  & 240/240/240 & 772/811/812 & $\star \star \star$   & N  \\ 
       & 2007 Mar 02         & 4163.087 &         0.053  & 120/120/120 & 1020/1038/1049 & $\star \star \star$   & E  \\ 
       & 2007 Apr 23         & 4214.514 &         0.457  & 240/240/240 & 515/572/580 & $\star \star \star$   & N  \\ 
       & 2007 Apr 23         & 4214.635 &         0.479  & 240/240/240 & 528/512/503 & $\star \star \star$   & N  \\ 
       & 2007 Apr 25         & 4216.523 &         0.824 & 240/240/240 & 796/781/841 & $\star \star \star$   & N  \\ 
       & 2007 May 06         & 4227.516 &         0.834 & 240/240/240 & 918/764/875 & $\star \star \star$   & N  \\ 
       & 2007 May 07         & 4228.467 &         0.007 & 400/400/400 & 589/494/795 & $\star \star \star$   & N  \\ 
       & 2008 Jan 22         & 4489.024 &         0.646 & 240/240/240 & 1482/1555/1557 & $\star \star \star$   & E  \\ 
       & 2008 Jan 23         & 4490.031 &         0.830 & 240/240/240 & 1564/1470/1569 & $\star \star \star$   & E  \\ 
       & 2008 Mar 23         & 4549.044 &         0.621 & 120/120/120 & 1086/1133/1088 & $\star \star \star$   & E  \\ 
       & 2008 Mar 24         & 4550.850 &         0.951 & 120/120/120 & 1011/947/1035 & $\star \star \star$   & E  \\ 
       & 2008 Mar 25         & 4551.891 &         0.141 & 120/120/120 & 1190/1181/1196 & $\star \star \star$   & E  \\ 
       & 2008 Apr 24         & 4581.460 &         0.548 & 240/240/240 & 931/896/906 & $\star \star \star$   & N  \\ 
       & 2008 Apr 25         & 4582.574 &         0.752 & 240/240/240 & 1082/1067/1077 & $\star \star \star$   & N  \\ 
       & 2008 Jul 26         & 4672.734 &         0.236 & 120/120/120 & 1010/877/1015 & $\star \star \star$   & E  \\ 
       & 2008 Aug 24         & 4703.724 &         0.901 & 120/120/120 & 370/226/357 & $\star \star \star$   & E  \\ 
       & 2008 Aug 24         & 4703.736 &         0.903 & 120/120/120 & 361/338/464 & $\star \star \star$   & E  \\ 
       & 2009 Jan 11         & 4843.728 &         0.500 & 240/240/240 & 767/797/825 & $\star \star \star$   & N  \\  
       & 2009 Jan 13         & 4846.143 &         0.940 & 120/120/120 & 1026/1085/1039 & $\star \star \star$   & E  \\ 
       & 2010 Jan 26         & 5223.163 &         0.873 & 240/240/240 & 1448/1458/1393 & $\star \star \star$   & E  \\ 
       & 2010 Jun 04         & 5351.852 &         0.402 & 240/240/240 & 1665/1637/1589 & $\star \star \star$   & E  \\ 
       & 2010 Jun 20         & 5367.761 &         0.311 & 240/240/240 & 1570/1557/1613 & $\star \star \star$   & E  \\ 
       & 2010 Jun 22         & 5369.856 &         0.694 & 240/240/240 & 1072/983/972 & $\star \star \star$   & E  \\ 
\hline
\multicolumn{8}{c}{78 Vir (HD 118022)}\\
\hline\noalign{\smallskip}
78 Vir &  2007 Apr 23        & 4214.570     & 0.623 & 800/800/800    & 425/369/395 & $\star \star \star$   & N  \\ 
       &  2007 Apr 25        & 4216.580     & 0.162 & 1000/1000/1000 & 327/378/346 & $\star \star \star$   & N  \\ 
       &  2007 May 06        & 4227.503     & 0.095 & 1400/1400/1400 & 681/821/741 & $\star \star \star$   & N  \\ 
       &  2008 Apr 24        & 4581.520     & 0.211 & 1200/1200/1200 & 587/634/588 & $\star \star \star$   & N  \\ 
       &  2008 Apr 25        & 4582.562     & 0.490 & 1200/1200/1200 & 891/928/944 & $\star \star \star$   & N  \\ 
\hline
\label{oblog}
\end{longtable}

\begin{longtable}{ccrrr}
\caption{Longitudinal magnetic field and net linear polarisation measurements of magnetic A and B stars, measured from LSD Stokes $V$ profiles and Stokes $Q$ and $U$ profiles.}\\
\hline\hline
JD - 2450000& Phase & $B_\ell\pm\sigma_B$ (G)  & $Q/I\pm\sigma_Q$ & $U/I\pm\sigma_U$ \\
\endfirsthead
\caption{continued.}\\
\hline\hline
\endhead
\hline
\endfoot
\hline
 \multicolumn{5}{c}{HD 4778}\\
\hline\noalign{\smallskip}
4070.746 & 0.423 & $ 1584\pm 22$ &	$ 0.002535\pm 0.000322$	& $ -0.002785\pm 0.000348$ \\
4074.734 & 0.980 & $ -1299\pm 15$ &	$ -0.000206\pm 0.000341$	& $ -0.001108\pm 0.000312$ \\
4076.744 & 0.764 & $ -613\pm 18$ &	$ -0.004080\pm 0.000390$	& $ -0.003208\pm 0.000370$ \\
4089.251 & 0.662 & $ 191\pm 21$ &	$ 0.001527\pm 0.000436$	& $ 0.007873\pm 0.000454$ \\
4090.244 & 0.050 & $ -1299\pm 15$ &	$ 0.003215\pm 0.000516$	& $ 0.000027\pm 0.000477$ \\
4091.260 & 0.446 & $ 1560\pm 22$ &	$ -0.003398\pm 0.000387$	& $ 0.003088\pm 0.000355$ \\
4475.394 & 0.366 & $ 1415\pm 24$ &	$ -$						& $ -0.000771\pm 0.000424$ \\
\hline
 \multicolumn{5}{c}{$\theta$~Aur (HD 40132)}\\
\hline\noalign{\smallskip}
4074.925 & 0.587 & $ 284\pm 11$ &	$ -0.000802\pm 0.000414$	& $ 0.000139\pm 0.000411$ \\
4076.147 & 0.924 & $ -119\pm 17$ &	$ -0.000159\pm 0.000810$	& $ 0.001076\pm 0.000817$ \\
4473.417 & 0.710 &$ 232\pm 17$ & 	$ 0.000107\pm 0.000736$	& $ 0.001391\pm 0.000738$ \\
4475.268 & 0.222 & $ 43 \pm 14 $ &		$ -0.001876\pm 0.000622$	& $ 0.000512\pm 0.000534$ \\
4489.002 & 0.017 & $ -184\pm 11$ &	$ -0.001646\pm 0.000456$	& $ 0.000054\pm 0.000389$ \\
4490.012 & 0.296 & $ 108\pm 9$ &	$ 0.001905\pm 0.000388$	& $ 0.000731\pm 0.000417$ \\
4491.795 & 0.789 & $ 156\pm 10$ &	$ -0.001068\pm 0.000446$	& $ -0.000547\pm 0.000460$ \\
\hline
 \multicolumn{5}{c}{HD 32633}\\
\hline\noalign{\smallskip}
4070.098 & 0.972 & $ -3641\pm 34$ &	$ 0.000794\pm 0.000830$	& $ -0.000148\pm 0.001156$ \\
4070.830 &   0.086 & $    -953\pm     31$ 	& $ - $ & $ -$ \\ 
4074.848 & 0.710 & $ -2281\pm 16$ &	$ 0.000290\pm 0.000383$	& $ 0.000450\pm 0.000384$ \\
4075.094 & 0.749 & $ -2835\pm 20$ &	$ 0.000576\pm 0.000406$	& $ 0.000601\pm 0.000457$ \\
4076.085 & 0.903 & $ -4387\pm 28$ &	$ 0.000267\pm 0.000384$	& $ 0.000267\pm 0.000401$ \\
4082.577 & 0.916 & $ -4245\pm 30$ &	$ -0.000343\pm 0.000688$	& $ -0.000122\pm 0.000696$ \\
4083.558 & 0.071 & $ -1174\pm 33$ &	$ 0.000590\pm 0.000790$	& $ -0.000113\pm 0.000743$ \\
4085.537 & 0.379 & $  423\pm 11$ &	$ -0.000766\pm 0.000632$	& $ -0.001007\pm 0.000632$ \\
4087.528 & 0.689 & $ -2165\pm 18$ &	$ 0.000202\pm 0.000788$	& $ -0.000962\pm 0.000907$ \\
4089.430 & 0.984 & $ -3170\pm 35$ &	$ -0.000083\pm 0.000536$	& $ -0.000059\pm 0.000525$ \\
4090.419 & 0.138 & $ 352\pm 32$ &	$ 0.000343\pm 0.000699$	& $ 0.000409\pm 0.000673$ \\
4091.434 & 0.296 & $ 1388\pm 18$ &	$ -0.001486\pm 0.000549$	& $ -0.000241\pm 0.000539$ \\
4475.482 & 0.017 & $ -2753\pm 46$ &	$ -0.000062\pm 0.000569$	& $ -0.000249\pm 0.000573$ \\
4490.736 & 0.390 & $ 404\pm 12$ &	$ 0.001182\pm 0.000388$	& $ 0.000729\pm 0.000395$ \\
4491.741 & 0.546 & $ -1366\pm 17$ &	$ 0.000510\pm 0.000441$	& $ 0.000231\pm 0.000436$ \\
4492.729 & 0.700 & $ -2185\pm 20$ &	$ 0.000384\pm 0.000449$	& $ 0.000414\pm 0.000449$ \\
4493.730 & 0.855 & $ -4256\pm 40$ &	$ 0.000263\pm 0.000642$	& $ 0.000013\pm 0.000558$ \\
4849.432 & 0.181 & $  898\pm 18$ &	$ 0.000194\pm 0.000515$	& $ -0.000166\pm 0.000509$ \\
5115.055 & 0.481 & $ -925\pm 13$ &	$ 0.000590\pm 0.000477$	& $ 0.000413\pm 0.000449$ \\
5254.743 & 0.209 & $ 1273\pm 19$ &	$ 0.000285\pm 0.000413$	& $ -0.000176\pm 0.000422$ \\
5263.774 & 0.613 & $ -1559\pm 15$ &	$ -0.000030\pm 0.000536$	& $ 0.000529\pm 0.000547$ \\
\hline
 \multicolumn{5}{c}{49 Cam (HD 62140)}\\
\hline\noalign{\smallskip}
4080.681 & 0.381 & $ -1350\pm 13$ 	& $ -0.001999\pm 0.000437$ 	& $ -0.000067\pm 0.000459$ \\
4082.653 & 0.841 & $ 1100\pm 14$ 		& $ 0.001359\pm 0.000447$ 	& $ -0.004501\pm 0.000452$ \\
4083.684 & 0.081 & $ 1086\pm 14$ 		& $ 0.002122\pm 0.000461$ 	& $ -0.003331\pm 0.000514$ \\
4084.595 & 0.294 & $ -819\pm 19$ 		& $ -0.000318\pm 0.000594$ 	& $ 0.001067\pm 0.000563$ \\
4087.715 & 0.017 & $ 1253\pm 23$ 		& $ -0.000198\pm 0.001432$ 	& $ -0.000483\pm 0.001584$ \\
4089.557 & 0.451 & $ -1471\pm 12$ 	& $ -0.002095\pm 0.000476$ 	& $ 0.000024\pm 0.000511$ \\
4090.542 & 0.681 & $ -307\pm 21$ 		& $ 0.005206\pm 0.000519$ 	& $ 0.003038\pm 0.000486$ \\
4091.557 & 0.918 & $ 1298\pm 13$ 	 	& $ -0.001324\pm 0.000393$ 	& $ -0.000273\pm 0.000438$ \\
4475.706 & 0.520 & $ -1358\pm 13$ 	& $ -0.002141\pm 0.000512$ 	& $ 0.001236\pm 0.000513$ \\
4491.003 & 0.089 & $ 1112\pm 13$ 		& $ 0.003333\pm 0.000323$ 	& $ -0.003265\pm 0.000370$ \\
4493.989 & 0.786 & $ 598\pm 17$ 		& $ 0.003778\pm 0.000473$ 	& $ -0.003123\pm 0.000484$ \\
4550.724 & 0.026 & $ 1255\pm 12$ 		& $ 0.000867\pm 0.000322$ 	& $ -0.001192\pm 0.000352$ \\
4551.723 & 0.260 & $ -434\pm 21$ 		& $ 0.001381\pm 0.000389$ 	& $ 0.001123\pm 0.000438$ \\
4581.349 & 0.174 & $ 370\pm 21$ 		& $ 0.003960\pm 0.000457$ 	& $ 0.002064\pm 0.000531$ \\
4846.625 & 0.056 & $ 1164\pm 14$ 		& $ 0.001132\pm 0.000830$ 	& $ -0.002572\pm 0.000928$ \\
4847.620 & 0.288 & $ -749\pm 20$ 		& $ 0.000353\pm 0.000463$  & $ 0.001085\pm 0.000497$ \\
5085.125 & 0.677 & $ -478\pm 20$ 		& $ 0.005382\pm 0.000443$ 	& $ 0.002041\pm 0.000419$ \\
5250.873 & 0.348 & $ -1125\pm 14$ 	& $ -0.001510\pm 0.000417$ 	& $ 0.000311\pm 0.000436$ \\
5251.972 & 0.604 & $ -1024\pm 13$ 	& $ 0.002344\pm 0.000407$ 	& $ 0.000870\pm 0.000401$ \\
\hline  
\multicolumn{5}{c}{HD 71866}\\
\hline\noalign{\smallskip}
4085.756 & 0.723 & $ -169\pm 26$ &	$ -0.001707\pm 0.000558$	& $ -0.004811\pm 0.000603$ \\
4089.756 & 0.312 & $ -629\pm 20$ &	$ -0.007183\pm 0.000630$	& $ -0.001153\pm 0.000592$ \\
4090.751 & 0.457 & $ -1785\pm 19$ &	$ -0.002275\pm 0.000598$	& $ -0.003174\pm 0.000559$ \\
4091.748 & 0.603 & $ -1580\pm 21$ &	$ -0.003833\pm 0.000532$	& $ 0.001778\pm 0.000484$ \\
4167.957 & 0.819 & $ 1316\pm 24$ &	$ -0.000151\pm 0.000441$	& $ -0.003645\pm 0.000506$ \\
4168.975 & 0.969 & $ 2223\pm 21$ &	$ -0.002215\pm 0.000431$	& $ -0.000973\pm 0.000672$ \\
4227.424 & 0.554 & $ -1750\pm 19$ &	$ -0.002216\pm 0.000554$	& $ 0.000866\pm 0.000535$ \\
4473.530 & 0.746 & $ 260\pm 26$ &	$ 0.001818\pm 0.000718$	        & $ -0.005288\pm 0.000736$ \\
4475.566 & 0.056 & $ 2290\pm 22$ &	$ 0.000009\pm 0.000456$	       & $ -0.000153\pm 0.000629$ \\
4492.827 & 0.592 & $ -1688\pm 18$ &	$ -0.003414\pm 0.000484$	& $ 0.001299\pm 0.000474$ \\
4493.829 & 0.740 & $ 43\pm 25$ &	$ 0.002172\pm 0.000653$	        & $ -0.004723\pm 0.000629$ \\
4582.411 & 0.757 & $ 461\pm 26$ &	$ 0.001491\pm 0.000686$	        & $ -0.004704\pm 0.000613$ \\
4842.734 & 0.040 & $ 2323\pm 22$ &	$ 0.000023\pm 0.000475$  	& $ -0.000549\pm 0.000659$ \\
5224.799 & 0.228 & $ 971\pm 25$ &	$ -0.004270\pm 0.000440$	& $ 0.000347\pm 0.000471$ \\
\hline
 \multicolumn{5}{c}{HD 112413}\\
\hline\noalign{\smallskip}
4075.158 & 0.977 & $ -856\pm 23$ 	& $ 0.001780\pm 0.000494$ & $ 0.001723\pm 0.000484$ \\
4076.159 & 0.160 & $ -603\pm 19$ 	& $ 0.002334\pm 0.000525$ & $ 0.004213\pm 0.000518$ \\
4082.760 & 0.368 & $ 624\pm 21$ 	& $ 0.000137\pm 0.000617$ & $ -0.001208\pm 0.000647$ \\
4089.663 & 0.630 & $ 231\pm 23$ 	& $ 0.000691\pm 0.000657$  & $ 0.001269\pm 0.000687$ \\
4090.647 & 0.810 & $ -785\pm 19$ 	& $ 0.003106\pm 0.000790$ & $ 0.002424\pm 0.000788$ \\
4163.087 & 0.053 & $  -795\pm 21$ & $ -0.001239\pm 0.000734$& $ -0.000578\pm 0.000718$ \\
4214.514 & 0.457 & $ 807\pm 18$ 	& $ 0.000226\pm 0.001107$ & $ -0.001956\pm 0.001125$ \\
4214.635 &0.479  &  $ 799\pm 19$ & $  0.001272\pm 0.000993$  & $ -0.001996\pm 0.000981$ \\
4216.523 & 0.824 & $ -830\pm 20$ 	& $ 0.003010\pm 0.000838$ & $ 0.002400\pm 0.000780$ \\
4227.516 & 0.834 & $ -849\pm 19$ 	& $ 0.003131\pm 0.000848$ & $ 0.001945\pm 0.000738$ \\
4228.467 & 0.007 & $ -818\pm 24$ 	& $ -0.000271\pm 0.001652$ & $ -0.001587\pm 0.000992$ \\
4489.024 & 0.646 & $ 129\pm 23$ 	& $ -0.000916\pm 0.000367$ & $ 0.001985\pm 0.000378$ \\
4490.031 & 0.830 & $ -851\pm 22$ 	& $ 0.002872\pm 0.000452$ & $ 0.002218\pm 0.000433$ \\
4549.044 & 0.621 & $ 304\pm 22$ 	& $ -0.000453\pm 0.000489$ & $ -0.000584\pm 0.000524$ \\
4550.850 & 0.951 & $ -888\pm 22$ 	& $ 0.002450\pm 0.000779$ & $ 0.001239\pm 0.000691$ \\
4551.891 & 0.141 & $ -595\pm 21$ 	& $ 0.001459\pm 0.000530$ & $ 0.004084\pm 0.000518$ \\
4581.460 & 0.548 & $ 657\pm 20$ 	& $ -0.001276\pm 0.000634$ & $ 0.001097\pm 0.000631$ \\
4582.574 & 0.752 & $ -595\pm 21$ 	& $ 0.002091\pm 0.000553$ & $ 0.003702\pm 0.000545$ \\
4672.734 & 0.236 & $ -193\pm 23$ 	& $ 0.004147\pm 0.000629$ & $ -0.001381\pm 0.000543$ \\
4703.724 & 0.901 & $ -854\pm 31$ 	& $ 0.002628\pm 0.003134$ & $ -0.000760\pm 0.001992$ \\
4703.736 & 0.903 & $ -907\pm 30$ 	& $ 0.002653\pm 0.002106$ & $ -0.000584\pm 0.001505$ \\
4843.728 & 0.500 & $ -785\pm 18$  & $  0.001821\pm 0.000779$& $ -0.002499\pm 0.000754$ \\
4846.143 & 0.940 & $ -875\pm 22$ 	& $ 0.002773\pm 0.000675$ & $ -0.001509\pm 0.000701$ \\
5223.163 & 0.873 & $ -884\pm 19$ 	& $ 0.002989\pm 0.000488$ & $ 0.000467\pm 0.000503$ \\
5351.852 & 0.402 & $ 739\pm 16$ 	& $ -0.000142\pm 0.000361$ & $ 0.002013\pm 0.000375$ \\
5367.761 & 0.311 & $ 296\pm 23$ 	& $ 0.001792\pm 0.000379$ & $ -0.001181\pm 0.000359$ \\
5369.856 & 0.694 & $ -225\pm 21$ 	& $ 0.000081\pm 0.000584$ & $ 0.003735\pm 0.000598$ \\
     
\hline
\multicolumn{5}{c}{78 Vir (HD 118022)}\\
\hline\noalign{\smallskip}
4214.570 & 0.623 & $ -905\pm 12$ &	$ -0.005410\pm 0.000359$	& $ 0.001493\pm 0.000380$ \\
4216.580 & 0.162 & $ -241\pm 10$ &	$ -0.003051\pm 0.000700$	& $ -0.005213\pm 0.000658$ \\
4227.503 & 0.095 & $ -127\pm 5$ &	            $ -0.001195\pm 0.000454$ & $ -0.005993\pm 0.000528$ \\
4581.520 & 0.211 & $ -374\pm 11$ &	$ -0.005294\pm 0.000757$	& $ -0.003943\pm 0.000973$ \\
4582.562 & 0.490 & $ -957\pm 10$ &	$ -0.004111\pm 0.000320$ & $ 0.002266\pm 0.000365$ \\
\hline
\label{bltable}
\end{longtable}

\label{lastpage}


\clearpage





\clearpage

\begin{thebibliography}{}
\bibitem{adel} Adelman S.J., 1997a, PASP 109, 9
\bibitem{adel2} Adelman S.J., 1997b, A\&AS 122, 249
\bibitem{aur} Auri\`erie, M.,  Wade, G. A., Silvester, J., et al. , 2007, A\&A, 475, 1053
\bibitem{babcc} Babcock, H. W., 1947, ApJ , 105, 105
\bibitem{babc} Babcock, H. W., 1949, The Observatory, 69, 191
\bibitem{babf} Babcock, H. W., 1951, ApJ, 114, 1
\bibitem{babd} Babcock, H. W., 1956, ApJ, 124, 489
\bibitem{babl} Babel, J., 1992, A\&A,  258, 449
\bibitem{bag3} Bagnulo S., Landi Degl'Innocenti E., Landolfi M., Leroy J.-L., 1995, A\&A 295, 459
\bibitem{bag} Bagnulo, S., Wade, G. A., Donati, J.-F., Landstreet, J. D., Leone, F., Monin, D. N., Stift, M. J., 2001, A\&A, 369, 889
\bibitem{bag2} Bagnulo, S., Landolfi, M., Landstreet, J. D., Landi Degl'Innocenti, E., Fossati, L., Sterzik, M., 2009, PASP, 121, 993
\bibitem{boh} Bohlender, D. A., 1989, A\&A, 220, 215
\bibitem{bor} Borra, E. F., Landstreet, J. D., 1977, ApJ, 212, 141
\bibitem{bor2} Borra, E. F., Landstreet, J. D., 1980, ApJS, 42. 421
\bibitem{bra} Braithwaite, J., \& Spruit, H.C., 2004, Nature, 431, 819
\bibitem{bra2} Braithwaite, J.; Nordlund, A., 2006, A\&A, 450, 1077
\bibitem{don} Donati J.F., Semel M., Carter B.D., Rees D.E., Cameron A.C., 1997, MNRAS 291, 658 
\bibitem{don2} Donati, J.F., Collier Cameron, A., Semel, M., et al.,  2003, MNRAS, 345, 1145
\bibitem{farn} Farnsworth G., 1932, ApJ 76, 313
\bibitem{gre} Grevesse, N., Asplund, M., \& Sauval, A. J., 2005, in the proceedings of '{\it Element Stratification in Stars: 40 Years of Atomic Diffusion}', EAS 17, 21
\bibitem{ger} Gerth, E., Glagolevskij, Yu. V., \& Scholz, G. 1997, in Stellar Magnetic Fields, ed. Yu. V. Glagolevskij, \& I. I. Romanyuk, Moscow, 67
\bibitem{koch2} Kochukhov O., Piskunov N., Ilyin I., Ilyina S., Tuominen I., 2002, A\&A 389, 420
\bibitem{koch} Kochukhov, O., Bagnulo, S., Wade, G. A., Sangalli, L., Piskunov, N., Landstreet, J. D., Petit, P., Sigut, T. A. A.,  2004, A\&A, 414, 613
\bibitem{koch3} Kochukhov O. \& Piskunov N., 2002, A\&A 288, 868
\bibitem{kb} Kochukhov, O., Bagnulo, S., 2006, A\&A, 450, 763
\bibitem{ksy} Kochukhov, O, ''Spectrum synthesis for magnetic, chemically stratified stellar atmospheres'', 2007, in Magnetic Stars 2006, eds. I.I. Romanyuk and D. O. Kudryavtsev, in press (astro-ph/0701084).
\bibitem{kaw} Kochukhov, O., \& Wade, G. A., 2010, A\&A, 513, 13
\bibitem{kha} Khalack, V. R., Wade, G. A., 2006, A\&A, 450.1157
\bibitem{kha2} Khalack, V. R., Khalack, J. N., Shavrina, A.V., \& Polosukhina, N. S., 2001, Astron. Reports, 45, 564
\bibitem{kup} Kupka, F., Piskunov, N.,  Ryabchikova, T. A., Stempels, H. C., Weiss, W. W., 1999, A\&A, 138, 119
\bibitem{lan1} Landi Degl'Innocenti, M., Calamai, G., Landi Degl'Innocenti, E., and Patriarchi, P. 1981, ApJ, 249, 228
\bibitem{lan} Landolfi, M., Landi Degl'Innocenti, E.,  Landi Degl'Innocenti, M.,  Leroy, J. L., 1993, A\&A, 272, 285	
\bibitem{landa} Landstreet, J.D.,1970, ApJ, 159, 1001
\bibitem{landb} Landstreet, J.D.,  1988, ApJ, 326, 967
\bibitem{lanc} Landstreet, J. D., Barker, P. K., Bohlender, D. A., Jewison, M. S.	1989, ApJ, 344, 876
\bibitem{landm} Landstreet, J.D., \& Mathys, G., 2000, A\&A, 359, 213 
\bibitem{land} Landstreet, J.D, Bagnulo, S., Andretta, V., Fossati, L., Mason, E., Silaj, J., Wade, G. A., 2007, A\&A, in press 
\bibitem{ler4} Leroy, J. L., Landolfi, M., Landi Degl'Innocenti, E., 1993, A\&A, 270, 335
\bibitem{ler2} Leroy, J. L., Landstreet, J. D., Bagnulo, S., 1994, A\&A, 284, 491
\bibitem{ler1} Leroy, J. L., 1995, A\&AS, 114, 79
\bibitem{ler} Leroy, J. L., Landolfi, M., Landi degl'Innocenti, E., 1996, A\&A, 311, 513
\bibitem{leon} Leone, F., Catanzaro, G.,  and Catalano, S., 2000, A\&A, 355, 315
\bibitem{luft} Luftinger, T., Kochukhov, O., Ryabchikova, T., Piskunov, N., Weiss, W. W., Ilyin, I., 2010, A\&A, 509, 71
\bibitem{mak} Makaganiuk, V., Kochukhov, O., Piskunov, N., Jeffers, S. V., Johns-Krull, C. M., Keller, C. U., Rodenhuis, M.,  Snik, F., Stempels, H. C., Valenti, J. A., 2011, A\&A...525, 97 
\bibitem{mat} Mathys, G., 1989, Fund. Cosmic Phys. 13, 143 
\bibitem{mat2} Mathys, G., 1993, in the proceedings of '{\it Peculiar Versus Normal Phenomena in A-type and Related Stars,}', ASP Conference Series, Vol. 44.
\bibitem{met} Mestel, L., 2003, in the proceedings of "{\it Magnetic Fields in O, B and A Stars: Origin and Connection to Pulsation, Rotation and Mass Loss, Proceedings of the Conference held 27 November - 1 December, 2002}'  305, 3
\bibitem{mic} Michaud, G., 1970, ApJ, 160, 641
\bibitem{mos} Moss, D., 2004, in the proceedings of '{\it The A-Star Puzzle, held in Poprad, Slovakia, July 8-13, 2004}', IAUS 224, 245
\bibitem{pas}Pasinetti Fracassini, L. E., Pastori, L., Covino, S., Pozzi, A.,2001, A\&A, 367, 521
\bibitem{pis} Piskunov N. \& Kochukhov O., 2002, A\&A 381, 736
\bibitem{pis2} Piskunov N., 1985, SvAL, 11, 18
\bibitem{pre3} Preston, G. W.,  Sturch C., 1967,  {\it Magnetic and Related Stars, Proceedings of the AAS-NASA Symposium on the Magnetic and other Peculiar and Metallic-Line A Stars, held at Greenbelt, Maryland, November 8-10} 
\bibitem{pre} Preston, G. W., 1969, ApJ, 156, 967
\bibitem{pre2} Preston G.W., 1969, ApJ 158, 243
\bibitem{pre4} Preston G.W., 1970, ApJ, 160, 1059
\bibitem{rens} Renson, P., Manfroid, J., 2009, A\&A, 498, 961
\bibitem{ric} Rice, J. B., Wehlau, W. H., 1994, A\&A, 291, 825
\bibitem{ric2} Rice, J. B., Holmgren, D. E., Bohlender, D. A., 	2004, A\&A, 424, 237
\bibitem{ric2} Richer, J., Michaud, G., Turcotte, S., 2000, ApJ, 529, 338
\bibitem{roy} Roby, S.W., \& Lambert, D.L., 1990, ApJS, 73, 67
\bibitem{sho} Shorlin, S. L. S., Wade, G. A., Donati, J.-F., Landstreet, J. D., Petit, P., Sigut, T. A. A., Strasser, S., 2002, A\&A, 392, 637
\bibitem{ste}  Stepie\'n, K., 1989, A\&A, 220, 105 
\bibitem{ste2} Stepie\'n, K., 2000, A\&A, 353, 227
\bibitem{sti} Stibbs, D. W. N., 1950, MNRAS, 110, 395
\bibitem{sem} Semel, M., Donati, J.-F., Rees, D. E., 1993, A\&A, 278, 231 
\bibitem{wad} Wade, G. A., 1997, A\&A, 325, 1063
\bibitem{wad1} Wade, G. A., Elkin, V. G., Landstreet, J. D., Leroy, J.-L., Mathys, G., Romanyuk, I. I., 1996, A\&A, 313, 209
\bibitem{wad3} Wade G.A., Donati J.-F., Landstreet J.D., Shorlin S.L.S., 2000a, MNRAS 313, 823
\bibitem{wad2} Wade, G. A., Donati, J.-F., Landstreet, J. D., Shorlin, S. L. S., 2000b, MNRAS, 313, 851 
\end{thebibliography}
\end{document}